\documentclass[%
  reprint,
  superscriptaddress,
  prmaterials,
  amsmath,amssymb,
  aps,
]{revtex4-2}

\usepackage{graphicx}
\usepackage{dcolumn}
\usepackage{bm}
\usepackage{siunitx}
\usepackage{color}
\usepackage[T1]{fontenc}
\usepackage{lmodern}
\usepackage[utf8]{inputenc}
\usepackage{textcomp}
\usepackage{newunicodechar}
\usepackage{pifont}
\usepackage{amssymb}
\newunicodechar{−}{\textminus}

\usepackage[version=3]{mhchem}
\usepackage{chemformula}
\usepackage{xspace}

\newcommand{\FeNi}{Fe--Ni\xspace}

\usepackage[pdfencoding=auto]{hyperref}

\begin{document}

\preprint{APS/123--QED}

\title{Can MACE Potentials Accurately Describe Magnetism and Phase Stability in \FeNi Alloys? A Systematic Benchmark}

\author{Kushal Ramakrishna}
\email{k.ramakrishna@hzdr.de}
\affiliation{Helmholtz--Zentrum Dresden--Rossendorf (HZDR), D--01328 Dresden, Germany}

\author{Mani Lokamani}
\affiliation{Helmholtz--Zentrum Dresden--Rossendorf (HZDR), D--01328 Dresden, Germany}

\author{Attila Cangi}
\email{a.cangi@hzdr.de}
\affiliation{Helmholtz--Zentrum Dresden--Rossendorf (HZDR), D--01328 Dresden, Germany}
\affiliation{Center for Advanced Systems Understanding (CASUS), D--02826 G\"orlitz, Germany}

\date{\today}

\begin{abstract}
We present a systematic benchmark of MACE potentials for iron--nickel alloys, focusing on structural, elastic, magnetic, and finite-temperature properties relevant to phase stability. The reference dataset comprises spin-polarized PBE density functional theory (DFT) calculations for chemically disordered special quasirandom structures (SQS), spanning compositions, bcc and fcc crystal structures, and volumetric and shear deformations. A system-specific MACE--sqs model trained on this dataset achieves validation errors of 2.0~meV/atom for energies and 24.3~meV/\AA\ for forces. Compared with several MACE foundation models, including models trained with Hubbard $U$ corrections, MACE--sqs gives the most consistent agreement with DFT and experiment for equations of state, equilibrium volumes, elastic constants, and thermal expansion trends in bcc and fcc \FeNi alloys. For the bcc--to--hcp transition, MACE--sqs predicts a pure-Fe transition pressure closer to experiment than the tested foundation models, but all models predict an incorrect increase of transition pressure with Ni content. This failure indicates that high-pressure magnetic collapse and composition-dependent magnetoelastic effects are not yet fully captured. Overall, targeted SQS-based training substantially improves the accuracy of MACE potentials for \FeNi alloys, while phase stability under magnetic collapse remains a key limitation for future model development.
\end{abstract}

\maketitle

\section{\label{sec:level1}Introduction}
Iron--nickel (\FeNi) alloys exhibit a variety of phases as a function of composition and temperature, and their phase diagram provides a foundation for understanding phase stability and transformations~\cite{KUWAYAMA2008379,MAO2006146}. Beyond their structural diversity, \FeNi alloys are a canonical system for studying composition--dependent magnetism and magnetoelastic effects, including Invar--like behavior~\cite{Schilfgaarde:1999aa,PhysRevLett.86.4851}. Control of phase fractions and chemical order through composition and thermal treatment enables tuning of microstructure and functional properties, which is central to materials engineering and alloy design~\cite{MUSHNIKOV2022118330}. Beyond industrial metallurgy, \FeNi alloys serve as reference materials in planetary science (e.g., Earth--core studies)~\cite{AHLES2021465}. The same coupling between crystal structure, chemical disorder, and magnetism makes \FeNi a stringent test system for machine--learned interatomic potentials (MLIPs).

Magnetism is strongly coupled to compression and structural stability in Fe and Fe-rich \FeNi alloys. Iron is ferromagnetic under ambient conditions, but its magnetic moment collapses under sufficient pressure; nickel is also ferromagnetic and moderates the pressure dependence of magnetic properties in \FeNi alloys. The pressure associated with magnetic collapse depends on nickel content~\cite{huang1988,akahama2020pressure}. Under compression, changes in magnetic order affect magnetic susceptibility and related properties, with implications for Earth's core and the magnetic behavior of planetary interiors~\cite{PhysRevB.107.115131}. At pressures above 13~GPa, iron undergoes a transition from the body--centered cubic structure to the hexagonal close--packed structure~\cite{takahashi1964,PhysRevLett.93.255503,Ramakrishna_2023}. Similar behavior is observed in \FeNi alloys, where nickel content shifts the transition pressure; this composition-dependent phase stability provides a stringent benchmark for the potentials considered below (Fig.~\ref{fig:transition_bcc}). 

Accurate prediction of material properties is a central challenge in computational materials science because predictive accuracy typically requires high--fidelity electronic--structure methods, whereas accessing realistic length and time scales demands far less expensive interatomic models. MLIPs address this accuracy--efficiency trade--off by learning the potential--energy surface from first--principles reference data, enabling near--DFT fidelity in large--scale atomistic simulations at substantially reduced computational cost~\cite{PhysRevLett.98.146401,doi:10.1021/acs.chemrev.0c00868}. Several MLIP frameworks have been developed for transition metals and alloys, including the spectral neighbor analysis potential~\cite{thompson2015spectral,doi:10.1021/acs.jpca.9b08723,Nikolov2021,Nikolov2022,Nikolov2024}, the Gaussian approximation potential~\cite{bartok2010gaussian,PhysRevMaterials.2.013808,PhysRevMaterials.8.033804,PhysRevMaterials.9.053807}, the moment tensor potential~\cite{Shapeev2016,Novikov2022,10.1063/5.0280935}, the atomic cluster expansion~\cite{drautz2019atomic,Rinaldi2024,Owen2024}, deep learning--based models~\cite{zhang2018deep,ZHANG2020107206,Gong2024,913y-p6qf}, NequIP~\cite{batzner2022e3equivariant}, and MACE~\cite{batatia2022mace}. MACE foundation models trained on broad materials datasets provide a useful starting point for property prediction across chemically diverse systems~\cite{Deng2023,Schmidt2022,barrosoluque2024openmaterials2024omat24,kaplan2025foundationalpotentialenergysurface}, but their reliability for chemically disordered magnetic transition--metal alloys remains a system-dependent question. This uncertainty is particularly important for \FeNi alloys, where the same local chemical environments that determine alloy energetics also influence magnetic moments, magnetovolume coupling, and phase stability.

In this work, we benchmark several MACE foundation models against a system-specific MACE--sqs model trained on spin-polarized DFT data for chemically disordered \FeNi SQS configurations. The models considered, their training data, and the reference quantities available for \FeNi structures are summarized in Table~\ref{MACE_table}. We evaluate equations of state, equilibrium volumes, elastic properties, magnetic moments, pressure-induced bcc--to--hcp phase stability, and finite-temperature lattice expansion across composition. The targeted MACE--sqs model gives the most consistent agreement with DFT and experiment for bcc and fcc equations of state, volumes, elastic constants, and thermal expansion trends. At the same time, the bcc--to--hcp transition benchmark exposes a remaining limitation: all tested models predict an increase in transition pressure with Ni content, contrary to experiment. The benchmark therefore identifies both the accuracy gains obtained from SQS-based training and the missing physics associated with high-pressure magnetic collapse and composition-dependent magnetoelastic coupling.

\section{\label{comp}Computational Methods}

\subsection{Special Quasirandom Structures and Short--Range Order}

\begin{figure*}[t]
  \centering
  \includegraphics[width=2.0\columnwidth]{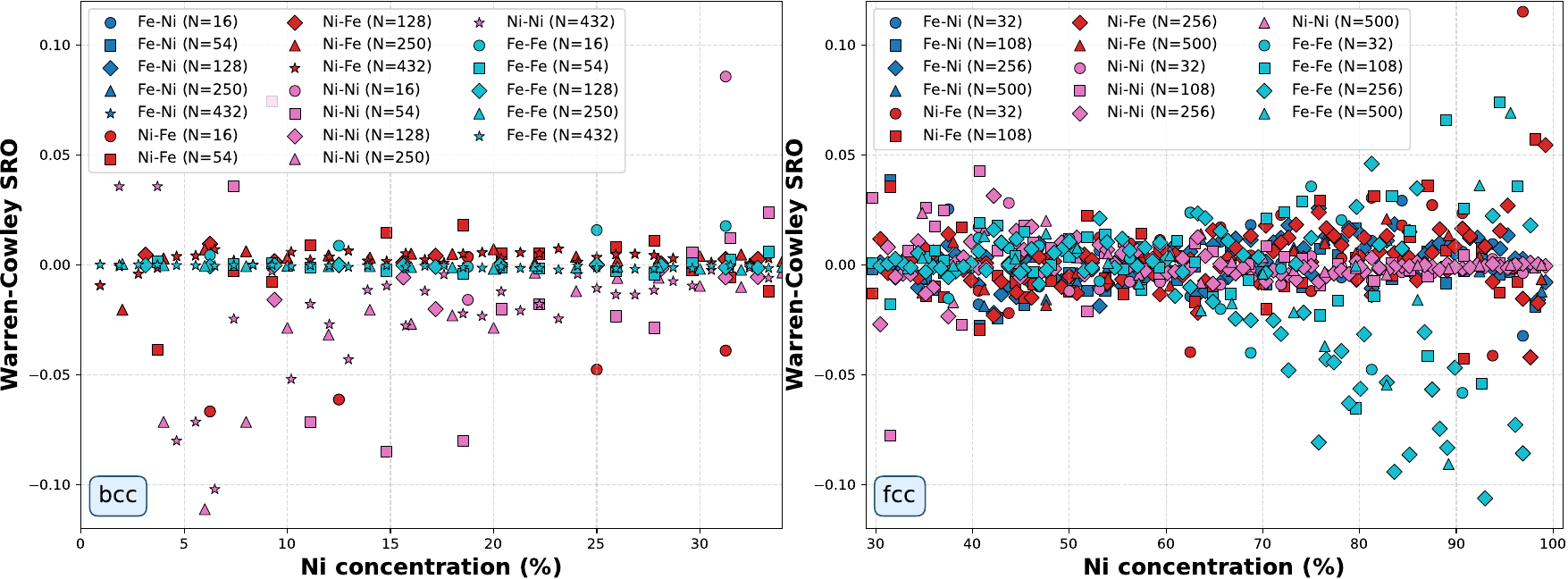}
  \caption{
    Warren--Cowley short--range order (SRO) parameters for \FeNi alloys as a function of nickel concentration for bcc (left) and fcc (right) structures. Marker color denotes the system size $N$, and marker shape denotes the SRO type (\FeNi, Ni--Fe, Ni--Ni, Fe--Fe). Larger system sizes yield SRO values closer to zero, indicating a more random chemical distribution, whereas deviations reflect clustering or ordering tendencies.}
  \label{fig:sro_fe_ni}
\end{figure*}

Disordered substitutional \FeNi alloys are most naturally described by an ensemble of random atomic configurations; however, direct ensemble averaging with first--principles calculations is computationally prohibitive. SQS provides an efficient representation of chemical disorder in a finite supercell by matching the relevant pair and, when desired, higher--order correlation functions of an ideal random alloy. This approach reduces configuration--to--configuration noise in computed energies, forces, magnetic moments, and mechanical properties at fixed composition, while retaining a physically meaningful distribution of local chemical environments that is important for magnetism and magnetoelastic response. Here, random \FeNi alloys at specific concentrations are represented using SQS generated with the Monte Carlo (MCSQS) program included in the Alloy Theoretic Automated Toolkit (ATAT) package~\cite{avdw:atat,VANDEWALLE201313}.

To quantify how closely an SQS reproduces the statistics of an ideal random \FeNi alloy, we use the Warren--Cowley short--range order (SRO) parameter~\cite{PhysRev.138.A1384,PhysRev.77.669}, $\alpha_{ij}$, as a convenient, neighbor--shell--resolved metric of residual ordering or clustering. It is defined as

\begin{equation}
  \alpha_{ij} = 1 - \frac{P(j|i)}{c_j},
  \label{sro_eqn}
\end{equation}

where $P(j|i)$ is the conditional probability that a neighbor of an atom of type $i$ is of type $j$, and $c_j$ is the concentration of species $j$. Values of $\alpha_{ij}$ close to zero indicate near--random mixing for the corresponding pair type and coordination shell, whereas systematic deviations indicate finite--size artifacts or incomplete matching of the target correlation functions. In a perfectly random alloy, $P(j|i) = c_j$ and thus $\alpha_{ij} = 0$; positive (negative) values indicate clustering (chemical ordering).

We evaluated SRO parameters for \FeNi alloys across a range of system sizes ($N$) and compositions, considering both fcc and bcc structures. Specifically, we report $\alpha_{\mathrm{Fe-Ni}}$ (Ni neighbors around Fe atoms), $\alpha_{\mathrm{Ni-Fe}}$ (Fe neighbors around Ni atoms), $\alpha_{\mathrm{Ni-Ni}}$ (Ni neighbors around Ni atoms), and $\alpha_{\mathrm{Fe-Fe}}$ (Fe neighbors around Fe atoms). The SRO values were extracted from supercells with sizes ranging from $N=16$ to $N=500$ atoms. The resulting SRO values are presented in Fig.~\ref{fig:sro_fe_ni} for \FeNi alloys in bcc and fcc structures as a function of nickel concentration. In these plots, the marker color denotes the system size $N$, and the marker shape distinguishes the SRO type. For most compositions and system sizes, the SRO values are close to zero, indicating a high degree of chemical randomness in the generated structures. However, deviations from zero are observed for certain compositions (Fe--rich or Ni--rich) and smaller system sizes, reflecting local clustering or ordering tendencies. As expected, larger system sizes yield SRO values closer to zero, consistent with improved statistical sampling of the random-alloy ensemble.

\subsection{DFT Calculations}

To parameterize and validate interatomic potentials and to obtain reliable reference data for the equation of state, elastic response, and magnetism of \FeNi\ alloys, we perform first--principles electronic--structure calculations on SQS configurations. The dataset includes both fully relaxed SQS structures and strained configurations, the latter comprising isotropic strains of up to $\pm 15\%$ and shear strains of up to $\pm 2\%$. The strained configurations systematically sample the relevant regions of the potential energy surface to ensure robust MLIP transferability across elastic deformation modes. Spin--polarized DFT provides a consistent framework for computing total energies, forces, and magnetic moments across compositions and supercell sizes, which is essential both for assessing finite--size and chemical--disorder effects and for constructing a training dataset for the MLIP. Spin--polarized DFT calculations for the SQS configurations were performed using the Vienna Ab initio Simulation Package (VASP)~\cite{PhysRevB.47.558,PhysRevB.59.1758,KRESSE199615,PhysRevB.54.11169}. We employed projector augmented-wave (PAW) pseudopotentials (PAW\_PBE\_Fe\_pv\_02Aug2007 and PAW\_PBE\_Ni\_pv\_06Aug2000) with the Perdew--Burke--Ernzerhof (PBE) exchange--correlation (XC) functional~\cite{perdew1996generalized}. To ensure representative random-alloy behavior, SQS configurations were selected based on near-zero SRO parameters (Eq.~\ref{sro_eqn}), retaining only structures with negligible residual chemical ordering. The availability of configurations meeting this criterion varies significantly with composition: abundant options exist for intermediate compositions around Fe$_{0.5}$Ni$_{0.5}$, whereas Fe-rich and Ni-rich compositions offer fewer suitable SQS structures. Convergence tests established the computational parameters: a plane-wave energy cutoff of 620~eV and Monkhorst--Pack $k$-point meshes of $4\times4\times4$ for system sizes with $N \leq 108$ atoms and $2\times2\times2$ for larger supercells.

We use spin-polarized PBE, without Hubbard corrections, as the reference level for the SQS dataset. This choice is motivated by the itinerant and metallic character of Fe and Ni $3d$ states, for which the definition of a localized correlated subspace and a transferable Hubbard $U$ is less clear than in many transition--metal oxides~\cite{PhysRevB.105.195153,Ramakrishna_2023}. In chemically disordered \FeNi alloys, the local environment, composition, and magnetic configuration can further affect any effective on-site correction, making a single fixed $U$ difficult to justify across the dataset. Metallic screening and double-counting ambiguities can also modify magnetic moments and relative phase energetics in ways that are difficult to separate from the physical magnetovolume response~\cite{Hausoel2017}. Similarly, inconsistent application of DFT+$U$ in the construction of MACE foundation models can result in incompatible potential-energy surfaces~\cite{warford2026betteruimpactselective}. For this reason, DFT+$U$-based foundation models are treated here as benchmark models rather than as the reference level for training MACE--sqs.

\subsection{MLIP Training}
The MACE--sqs model was trained to reproduce spin-polarized DFT energies and forces for chemically disordered \FeNi configurations over variations in composition, volume, and strain. Magnetic effects enter this model implicitly through the spin-polarized reference calculations, meaning the model encodes the magnetic ground state through the reference energetics and cannot represent changes in magnetic order driven by pressure or temperature. The MACE models compared in this work, together with their reference datasets and available training quantities, are summarized in Table~\ref{MACE_table}.

\begin{table*}
  \caption{Details of the MACE models and their training datasets.}
  \begin{ruledtabular}
    \begin{tabular}{lccccc}
      Model & Dataset & XC & E & F & S  \\
      \hline
      mpa0--medium & MPTrj+Alex~\cite{Deng2023,Schmidt2022} & PBE+\textit{U} & \ding{51} & \ding{51} & \ding{51} \\
      omat & OMAT24~\cite{barrosoluque2024openmaterials2024omat24} & PBE+\textit{U} & \ding{51} & \ding{53}$^{a}$ & \ding{53}$^{a}$ \\
      matpes--pbe & MatPES~\cite{kaplan2025foundationalpotentialenergysurface} & PBE & \ding{51} & \ding{53}$^{a}$ & \ding{53}$^{a}$ \\
      matpes--r2SCAN & MatPES~\cite{kaplan2025foundationalpotentialenergysurface} & r2SCAN & \ding{51} & \ding{53}$^{a}$ & \ding{53}$^{a}$ \\
      sqs  & SQS--based & PBE & \ding{51} & \ding{51} & \ding{53} \\
    \end{tabular}
  \end{ruledtabular}
  \vspace{0.25em}
  \raggedright
{\footnotesize E: energy, F: forces, S: stress. The Materials Project Trajectory (MPtrj), Alexandria (Alex), OMAT24, and MatPES datasets all contain E, F, and S. $^{a}$F and S are available in the OMAT24 and MatPES datasets but absent for the \FeNi structures.\par}
\label{MACE_table}
\end{table*}

\begin{figure*}[t]
\centering
\includegraphics[width=2.0\columnwidth]{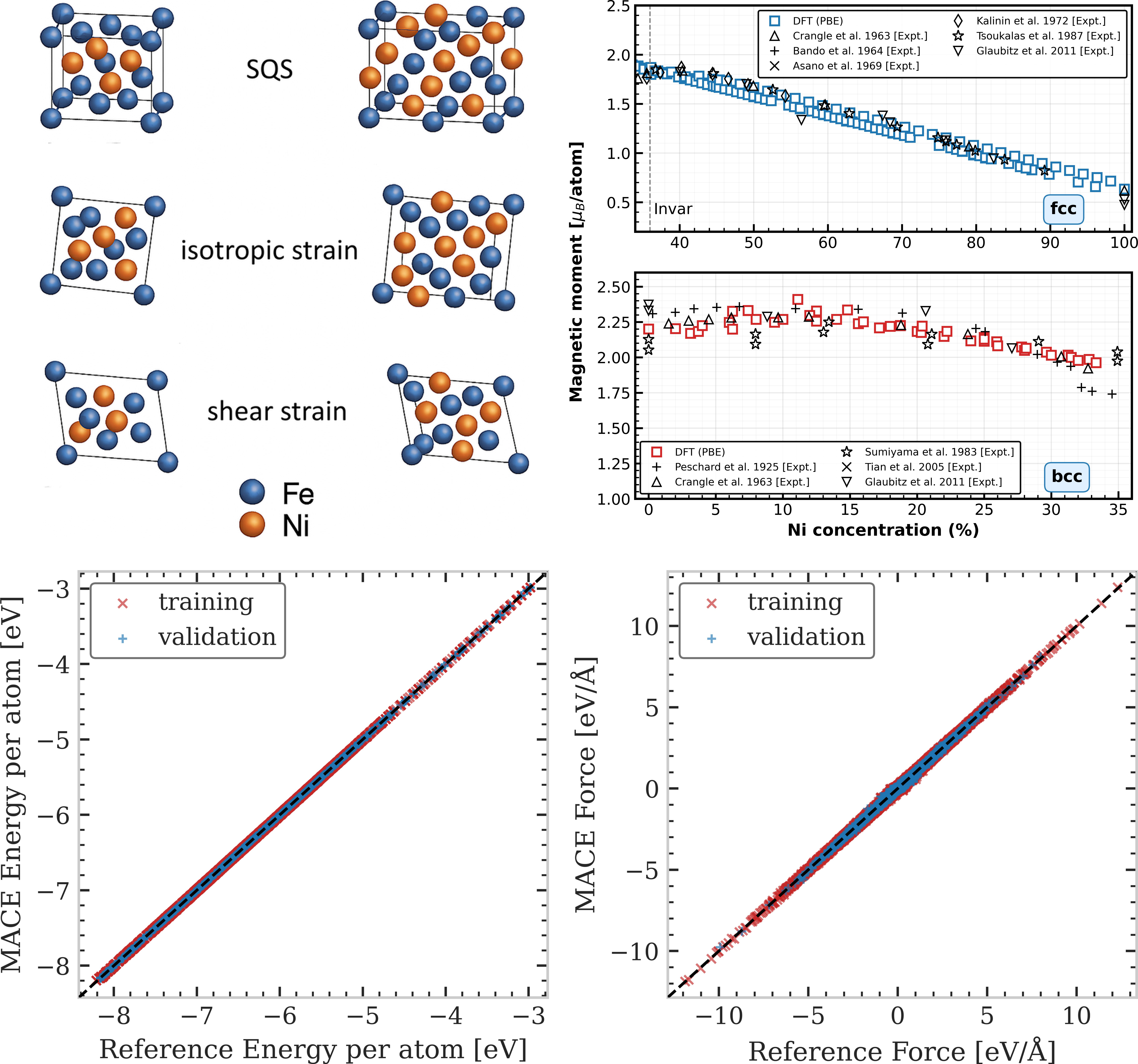}
  \caption{Overview of the training dataset and MACE model performance.
  Top left: Representative SQS supercells of \FeNi alloys from the training dataset, shown under three deformation conditions: undeformed reference (top), isotropic volumetric strain (middle), and shear strain (bottom). Blue and orange spheres represent Fe and Ni atoms, respectively.
  Top right: Mean magnetic moment per atom as a function of
  nickel concentration for fcc (upper) and bcc (lower) \FeNi alloys,
  computed from collinear spin-polarized DFT with the PBE
  XC functional and compared against experimental measurements~\cite{xiong2011magnetic,peschard1925contribution,crangle1963magnetization,PhysRevLett.94.137210,sumiyama1983metastable,asano1969magnetism,bando1964magnetization,Glaubitz_2011,Tsoukalas1987,Kalinin1972En}.
  Bottom row: Parity plots of MACE predicted energies (left,
  RMSE\,=\,2.2/2.0\,meV/atom) and forces (right,
  RMSE\,=\,22.0/24.3\,meV/\AA) against DFT reference values for the
  training and validation sets, respectively, confirming high model
  fidelity and absence of overfitting.}
  \label{fig:training}
\end{figure*}

The MLIP was trained on an SQS-based DFT dataset spanning variations in volume, lattice distortions, composition, and local chemical environment. Hyperparameter tuning and automated logging of training results were handled using Weights \& Biases (W\&B)~\cite{wandb}. For model development, the dataset contained 6389 DFT-computed structures, partitioned into 5115 training structures and 1274 validation structures. The final MACE--sqs model achieves RMSE values of 2.2~meV/atom for energies and 22.0~meV/\AA\ for forces on the training set. On the validation set, the corresponding RMSE values are 2.0~meV/atom and 24.3~meV/\AA. These comparable training and validation errors indicate no evident overfitting within the chosen split and are competitive with recent equivariant MLIPs~\cite{riebesell2025}. Figure~\ref{fig:training} gives an overview of the reference data and model performance. The top-left panel shows representative SQS supercells in the undeformed reference state and under isotropic and shear deformations, illustrating the structural diversity used for training. The top-right panels show the mean magnetic moment per atom from the spin-polarized DFT reference calculations as a function of nickel concentration for fcc and bcc \FeNi alloys, compared with experimental measurements~\cite{xiong2011magnetic,peschard1925contribution,crangle1963magnetization,PhysRevLett.94.137210,sumiyama1983metastable,asano1969magnetism,bando1964magnetization,Glaubitz_2011,Tsoukalas1987,Kalinin1972En}. The bottom panels show MACE--sqs predictions against DFT references for energies and forces, demonstrating that the model reproduces the training and validation data with similar accuracy.

Atomistic simulations were performed using ASE~\cite{Hjorth_Larsen_2017} together with MACE calculators, with TorchSim~\cite{Cohen_2025} used for GPU-accelerated batched force evaluations where applicable, the former handling structure generation, relaxation workflows, and analysis of relaxed configurations, while the latter enabled efficient evaluation of large batches of MACE forces on NVIDIA A100 and H100 GPUs. This workflow was used for the EOS, elastic-property, phase-stability, and finite-temperature calculations reported below.

\section{\label{results}{Results}}

\subsection{Equation of State}

\begin{figure*}[!htb]
  \centering
  \includegraphics[width=\textwidth]{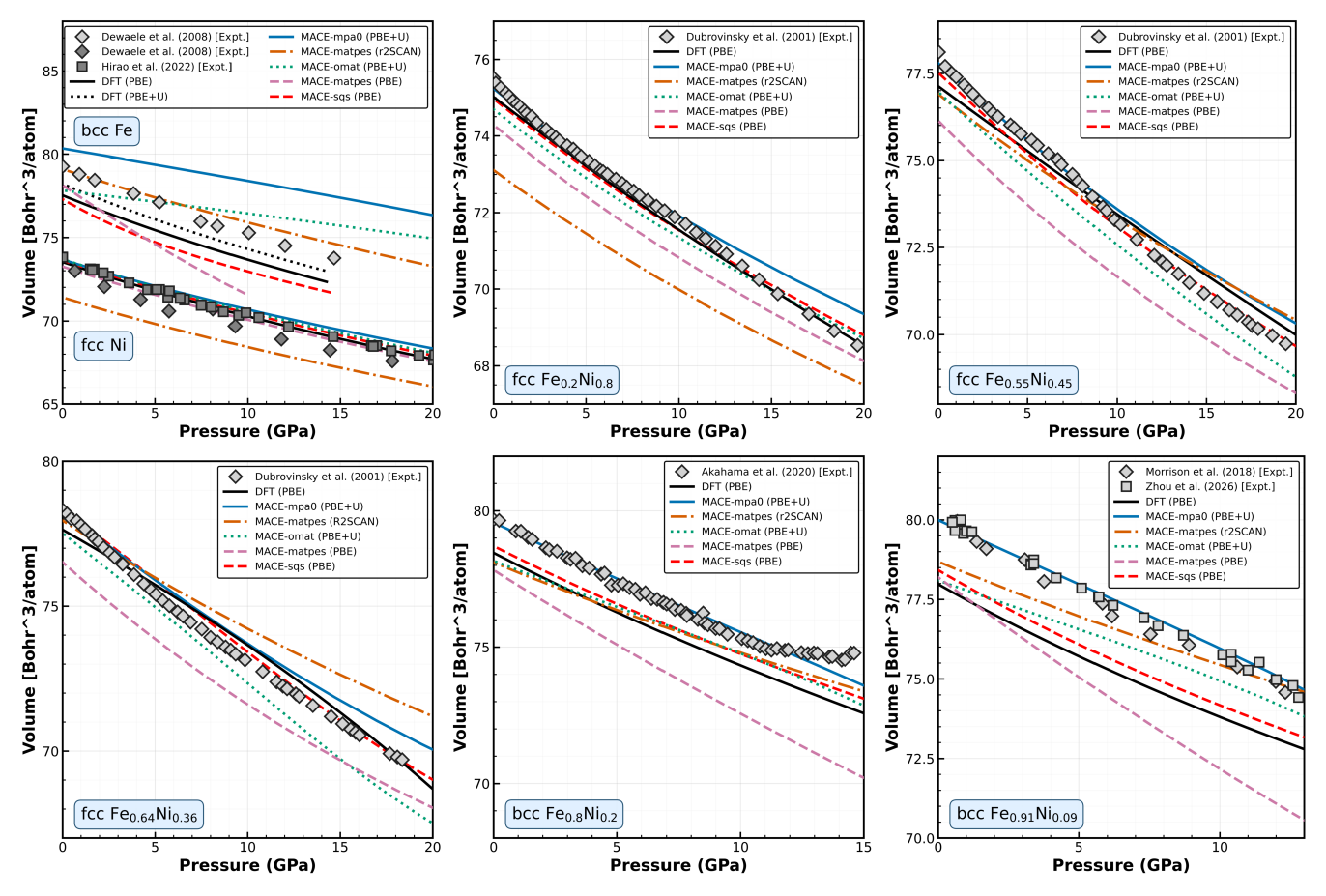}
  \caption{Equation of state (EOS) for \FeNi compositions, comparing MACE models with DFT and experiments~\cite{morrison2018equations,ZHOU2026107515,akahama2020pressure,PhysRevLett.86.4851,hirao2022equations,PhysRevB.78.104102,Ramakrishna_2023}. Note that the EOS results of Zhou \textit{\textit{et al.}}~\cite{ZHOU2026107515} correspond to a slightly sulfur--bearing alloy (Fe$_{0.89}$Ni$_{0.09}$S$_{0.02}$).}
  \label{fig:EOS_flavors_array}
\end{figure*}

The EOS benchmark tests whether the MACE models, and in particular the MACE--sqs model, reproduce the pressure--volume response of bcc and fcc \FeNi alloys across composition. This comparison is central because equilibrium volumes and compressibilities are sensitive to the same bonding, chemical-disorder, and magnetovolume effects that influence the elastic and phase-stability trends discussed below. Static zero-temperature EOS curves were computed over a range of volumes and fitted using the third--order Birch--Murnaghan equation of state~\cite{PhysRev.71.809,doi:10.1073/pnas.30.9.244}. When comparing with experimental data in Fig.~\ref{fig:EOS_flavors_array}, we interpret agreement in terms of the pressure--volume trend, where the pressure is written as

\begin{align*}
P(V) = \frac{3}{2} K_0 \Bigg[ \left( \frac{V_0}{V} \right)^{7/3} - \left( \frac{V_0}{V} \right)^{5/3} \Bigg] \\
\qquad \times \Bigg\{ 1 + \frac{3}{4} (K_0' - 4) \Bigg[ \left( \frac{V_0}{V} \right)^{2/3} - 1 \Bigg] \Bigg\}
\end{align*}

with $V_0$ denoting the equilibrium volume, $K_0$ the zero--pressure bulk modulus, and $K_0'$ its pressure derivative.

For pure bcc Fe, Fig.~\ref{fig:EOS_flavors_array} shows that MACE--sqs follows the experimental EOS of Dewaele \textit{et al.}~\cite{PhysRevB.78.104102} closely over the plotted pressure range. Among the foundation models, mpa0--medium gives the nearest visual agreement but slightly underestimates the equilibrium volume, while omat and matpes--r2SCAN show systematic volume underestimation and matpes--pbe gives the largest deviation. Thus, for the pure bcc Fe EOS, the system-specific model provides the closest agreement among the models compared here.

For bcc \FeNi alloys, Fig.~\ref{fig:EOS_flavors_array} compares the model EOS curves with experimental data for Fe$_{0.91}$Ni$_{0.09}$ and Fe$_{0.8}$Ni$_{0.2}$ from Morrison \textit{et al.}~\cite{morrison2018equations}, Zhou \textit{et al.}~\cite{ZHOU2026107515}, and Akahama \textit{et al.}~\cite{akahama2020pressure}. The Zhou \textit{et al.} data correspond to a slightly sulfur-bearing alloy, as noted in the caption, and should therefore be interpreted as a compositionally close comparison rather than an exact binary \FeNi benchmark. Across these bcc alloy compositions, MACE--sqs gives the closest overall agreement with the experimental pressure--volume trends in the plotted range. 
The foundation models show increasing deviations with 
decreasing Fe content: MACE--mpa0 (PBE$+U$) remains the best-performing foundation model but exhibits a residual volume offset, while MACE--omat (PBE$+U$) and MACE--matpes (r2SCAN) systematically underestimate the volume, and MACE--matpes (PBE) overestimates it throughout. MACE--sqs consistently reduces these offsets and reproduces the pressure dependence of the volume with near--DFT accuracy for both compositions.

For fcc \FeNi alloys, the experimental EOS data from Dubrovinsky 
\textit{et al.}~\cite{PhysRevLett.86.4851} and Dewaele \textit{et al.}~\cite{PhysRevB.78.104102} 
provide well--constrained benchmarks across a broad pressure range. 
At Fe--rich fcc compositions (Fe$_{0.64}$Ni$_{0.36}$ and 
Fe$_{0.55}$Ni$_{0.45}$), which encompass the Invar region, 
MACE--sqs achieves the best agreement with experiment, correctly 
capturing the equilibrium volume and its pressure dependence. 
The foundation models again show systematic deviations, with 
MACE--matpes (PBE) and MACE--matpes (r2SCAN) showing the 
largest spread. At Ni--rich fcc compositions (Fe$_{0.2}$Ni$_{0.8}$ 
and pure fcc Ni), MACE--sqs continues to outperform the foundation 
models, reproducing the experimental volumes from Dubrovinsky 
\textit{et al.}~\cite{PhysRevLett.86.4851} and Dewaele \textit{et al.}~\cite{PhysRevB.78.104102} 
with quantitative accuracy, while the foundation models show 
persistent over-- or underestimation depending on the 
XC functional employed.
    
Across all compositions and crystal structures, MACE--sqs 
achieves the most consistent agreement with experimental EOS 
data, outperforming all foundation models including 
MACE--mpa0 (PBE$+U$). This improvement arises from two 
complementary factors: targeted training on SQS--based DFT 
data that captures the interplay between chemical disorder, 
magnetism, and volume in \FeNi alloys, and the avoidance 
of DFT$+U$ artifacts that systematically distort the 
equilibrium volume and compressibility in itinerant 
ferromagnets~\cite{warford2026betteruimpactselective}.

\subsection{Lattice Volume across Compositions}

\begin{figure}
  \centering
  \includegraphics[width=1.0\columnwidth]{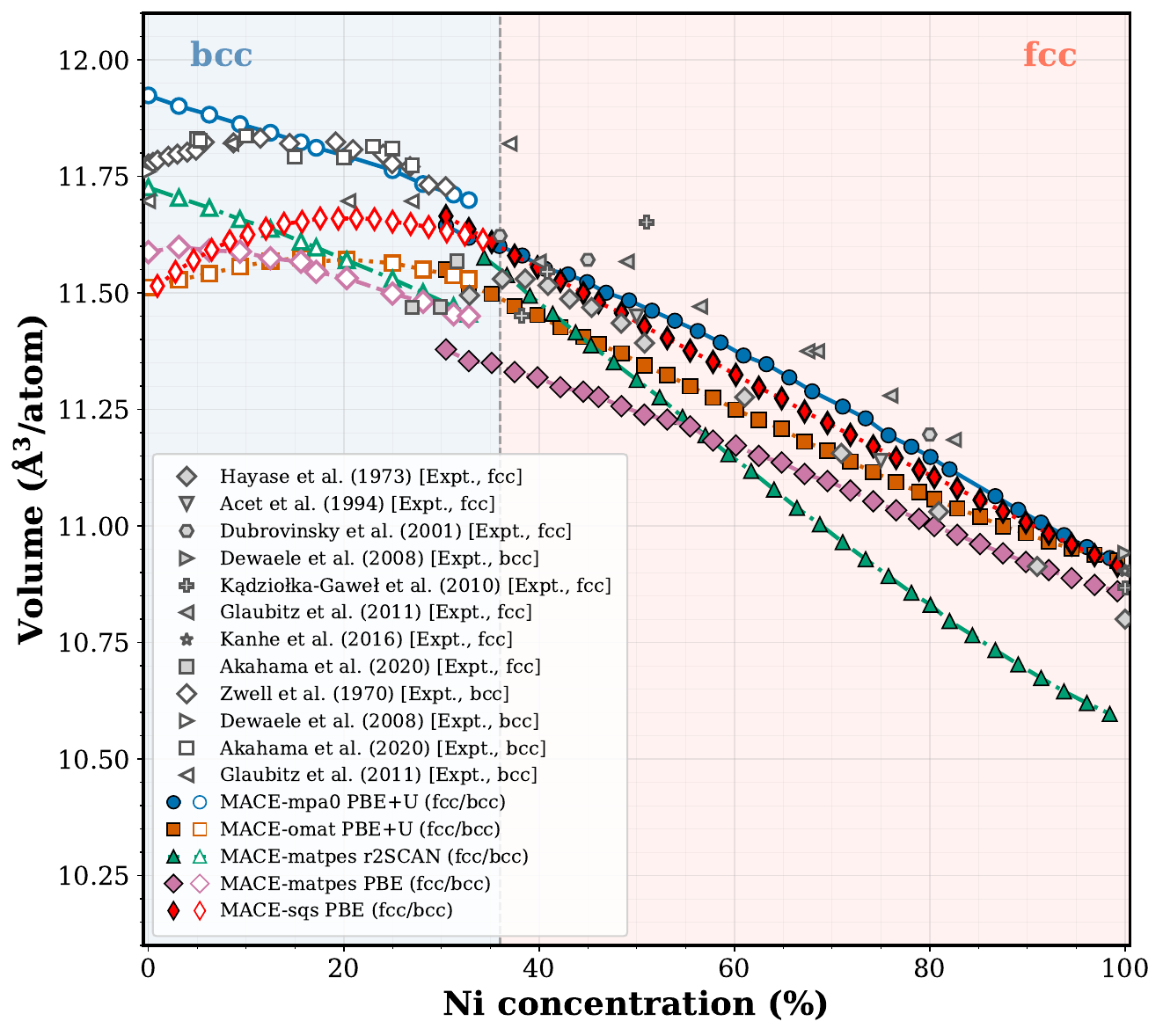}
  \caption{Volume per atom of \FeNi alloys for bcc (left) and fcc (right) structures as a function of nickel concentration, comparing MACE models with experimental data~\cite{Zwell1970,hayase1973spontaneous,PhysRevB.49.6012,PhysRevLett.86.4851,Glaubitz_2011,kkadziolka2010crystal,akahama2020pressure,KANHE201630}.}\label{fig:volume_composition}
\end{figure}

We next examine the equilibrium atomic volume as a function of nickel concentration for bcc and fcc \FeNi alloys. This benchmark complements the EOS analysis by focusing on the zero-pressure compositional trend, which is sensitive to chemical disorder, magnetic state, and finite-size effects in the SQS representation. We compare the MACE foundation models and MACE--sqs using large SQS supercells to reduce finite-size effects. The results are shown in Fig.~\ref{fig:volume_composition}.

For bcc \FeNi alloys, the MACE predictions reveal a clear decreasing trend in atomic volume with increasing nickel concentration. This behavior is consistent with the expectation that Ni atoms, being smaller than Fe atoms, reduce the average atomic volume as their fraction increases. The different MACE models yield closely similar results, indicating that the model is well--converged with respect to the choice of XC functional and training protocol. Compared with experimental data from Zwell \textit{\textit{et al.}}~\cite{Zwell1970} and Akahama \textit{\textit{et al.}}~\cite{akahama2020pressure}, the MACE results exhibit excellent agreement across the entire composition range. Small discrepancies at intermediate nickel concentrations may arise from experimental uncertainties, sample inhomogeneity, or limitations in the available training data for the model. However, the overall correspondence between theory and experiment demonstrates the reliability of the MACE approach for predicting volumetric properties in bcc alloys.

For fcc \FeNi alloys, Fig.~\ref{fig:volume_composition} shows a less linear composition dependence than in the bcc phase, with the largest sensitivity near Fe-rich and intermediate compositions where magnetovolume effects are expected to be important. MACE--sqs follows the experimental trends reported by Hayase~\cite{hayase1973spontaneous}, Acet~\cite{PhysRevB.49.6012}, Dubrovinsky \textit{et al.}~\cite{PhysRevLett.86.4851}, K{\k{a}}dzio{\l}ka--Gawe{\l} \textit{et al.}~\cite{kkadziolka2010crystal}, Glaubitz \textit{et al.}~\cite{Glaubitz_2011}, and Akahama \textit{et al.}~\cite{akahama2020pressure}. The comparison with multiple experimental datasets is useful because the measured volumes depend on composition, sample state, and experimental conditions. The spread among the MACE models provides a measure of model sensitivity, while the use of large SQS supercells reduces finite-size effects in the calculated volume trend. 

Taken together, the volume-composition results show that MACE--sqs gives the most consistent description of the zero-pressure equilibrium volumes for the bcc and fcc \FeNi compositions considered here. The benchmark also shows that agreement at the level of EOS curves is reflected in the composition-dependent equilibrium volumes, while residual model-dependent offsets remain visible in some composition ranges. These offsets motivate the more stringent elastic and phase-stability tests discussed below.

\subsection{Pressure--Induced Phase Transition}

\begin{figure}[t]
  \centering
  \includegraphics[width=1.0\columnwidth]{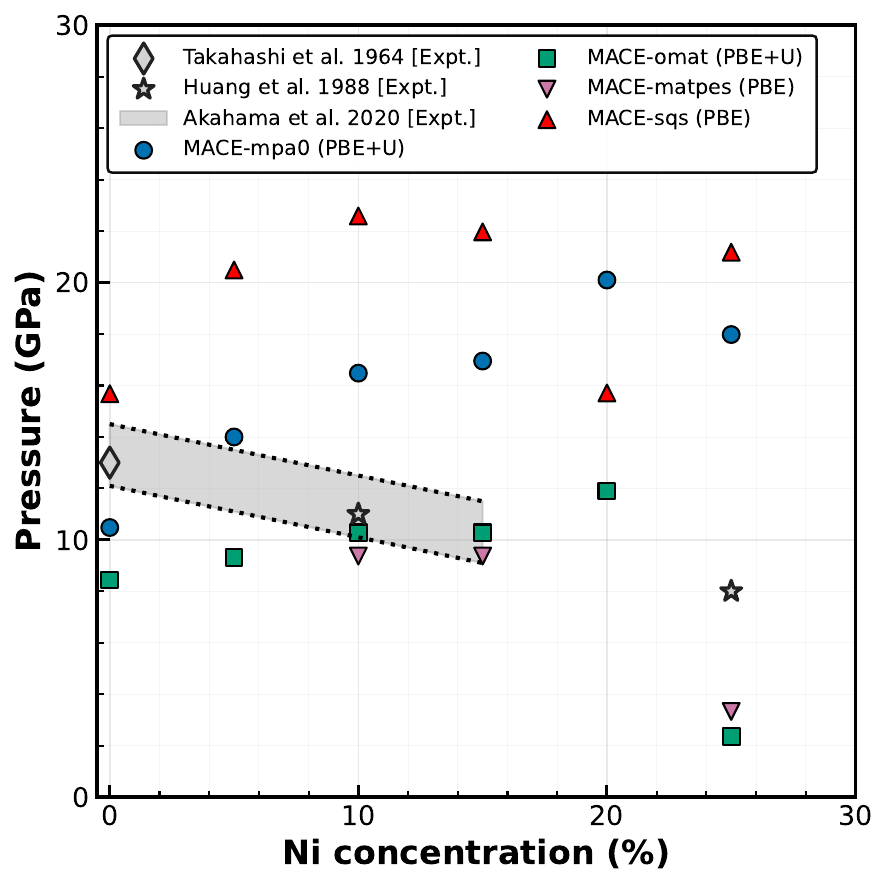}
  \caption{Transition pressure from the bcc to hcp structure as a function of nickel concentration, comparing MACE models with experiments~\cite{huang1988,akahama2020pressure,takahashi1964}.}
  \label{fig:transition_bcc}
\end{figure}

The pressure-induced transformation between the body-centered cubic and hexagonal close-packed structures is a key test of relative phase stability in Fe and Fe-rich \FeNi alloys. At ambient temperature, pure bcc Fe transforms to the hexagonal close-packed structure at approximately 13~GPa~\cite{takahashi1964}. Experimental studies have shown that the transition pressure decreases with increasing nickel concentration, reaching approximately 11~GPa and 8~GPa for \FeNi alloys with 10\% and 25\% Ni, respectively~\cite{huang1988,wei2020}. Using monochromatic synchrotron X-ray radiation, Akahama \textit{et al.}~\cite{akahama2020pressure} observed that the onset pressure decreases by about 0.2~GPa per atomic percent of Ni up to 15\% Ni, beyond which it remains approximately constant. This experimentally observed composition dependence provides the reference trend for the model comparison in Fig.~\ref{fig:transition_bcc}.

MACE--sqs was trained on bcc and fcc SQS configurations and does not include hcp structures in its training data. The hcp EOS and the bcc--to--hcp transition pressure therefore test extrapolation beyond the structural domain used for system-specific training. The foundation models considered here were trained on broader materials datasets and include hexagonal close-packed environments from those datasets~\cite{Deng2023,Schmidt2022,barrosoluque2024openmaterials2024omat24,kaplan2025foundationalpotentialenergysurface}, although not necessarily chemically disordered \FeNi configurations. The transition pressures should therefore be interpreted primarily as a test of cross-phase transferability rather than interpolation within the MACE--sqs training distribution.

To assess the capability of the MACE potentials to reproduce these trends, and to provide a stringent measure of transferability beyond the structures and thermodynamic conditions represented in the training set, we computed the zero-temperature enthalpies of bcc and hcp \FeNi alloys as a function of pressure. For each composition, static EOS curves were generated by relaxing structures at fixed volumes using the respective MACE model. The enthalpy was evaluated as $H(P) = E(V) + PV$, where $E(V)$ is the relaxed internal energy at volume $V$ and $P$ is the pressure. By interpolating $H(P)$ for both phases, the transition pressure is identified as the crossing point where the enthalpies of bcc and hcp become equal.

The predicted transition pressures differ substantially among the models. The MACE--mpa0 (PBE+$U$) model yields 10.5~GPa for pure Fe, significantly below the experimental value, consistent with the known tendency of PBE+$U$ to overstabilize localized magnetic moments in itinerant ferromagnets and thereby reduce the bcc--hcp energy difference. The MACE--omat (PBE+$U$) model predicts an even lower transition pressure of 8.4~GPa, reflecting a stronger destabilization of bcc Fe relative to hcp and highlighting the limited transferability of DFT+$U$-based training data for metallic \FeNi alloys, where the appropriate Hubbard $U$ is both ambiguous and strongly environment-dependent. In contrast, the MACE--sqs model (PBE) produces a transition pressure of 15.7~GPa for pure Fe. The most natural reference for a PBE-trained model is the PBE result of 10.3~GPa~\cite{PhysRevB.58.5296}, which itself undershoots the experimental value of $\sim$13~GPa~\cite{takahashi1964} due to the well-known tendency of PBE to underestimate the bcc--hcp energy 
difference in Fe; our PBE+$U$ calculation gives 12.7~GPa~\cite{Ramakrishna_2023}, closer to experiment owing to partial cancellation between PBE's underestimation of the bcc--hcp energy difference and the Hubbard correction's tendency to overstabilize the bcc phase. The MACE--sqs value of 15.7~GPa overshoots the experimental value of $\sim$13~GPa, in contrast to the PBE reference which undershoots it at 10.3~GPa~\cite{PhysRevB.58.5296}, 
consistent with the somewhat stiffer EOS obtained from the SQS-trained potential; nonetheless, MACE--sqs remains substantially closer to experiment than the PBE+$U$-based MACE foundation models. The improved performance of 
MACE--sqs arises from two key factors. First, training on chemically disordered SQS structures ensures that the model captures the interplay between local chemical environment, magnetism, and volume collapse in the bcc phase---an important ingredient for the onset of the bcc--to--hcp 
transition. Second, avoiding DFT+$U$ artifacts is essential: \FeNi alloys exhibit itinerant 3$d$ electrons with strong hybridization, and the use of DFT+$U$ introduces environment-dependent and non-transferable corrections that distort the relative energetics of bcc and hcp phases. The MACE--sqs model avoids both of these pitfalls and yields a smoother, more physically consistent pressure dependence.

The composition dependence in Fig.~\ref{fig:transition_bcc} reveals a more important limitation: all tested MACE models predict an increasing bcc--to--hcp transition pressure with increasing nickel concentration, whereas the experimental data show a decrease. This incorrect trend indicates that the models do not fully capture the composition-dependent magnetic and magnetoelastic effects that destabilize the bcc phase in Fe-rich \FeNi under compression. A plausible interpretation is that the potentials do not adequately represent the coupling between Ni content, Fe magnetic moments, and pressure-induced magnetic collapse. We attribute this primarily to the absence of hcp training configurations under high-pressure magnetic-collapse conditions, with a secondary contribution from the collinear fixed-spin framework that cannot represent longitudinal spin fluctuations. The MACE--sqs model reduces several EOS errors relative to the foundation models  (Fig.~\ref{fig:EOS_flavors_array}), but its out-of-distribution hcp prediction and residual EOS stiffness are not sufficient to recover the experimental transition-pressure trend. Correcting this failure will likely require reference data that explicitly sample high-pressure magnetic collapse, hcp environments, and composition-dependent disorder effects in Fe-rich \FeNi alloys.

\begin{figure*}
  \centering
  \includegraphics[width=2.0\columnwidth]{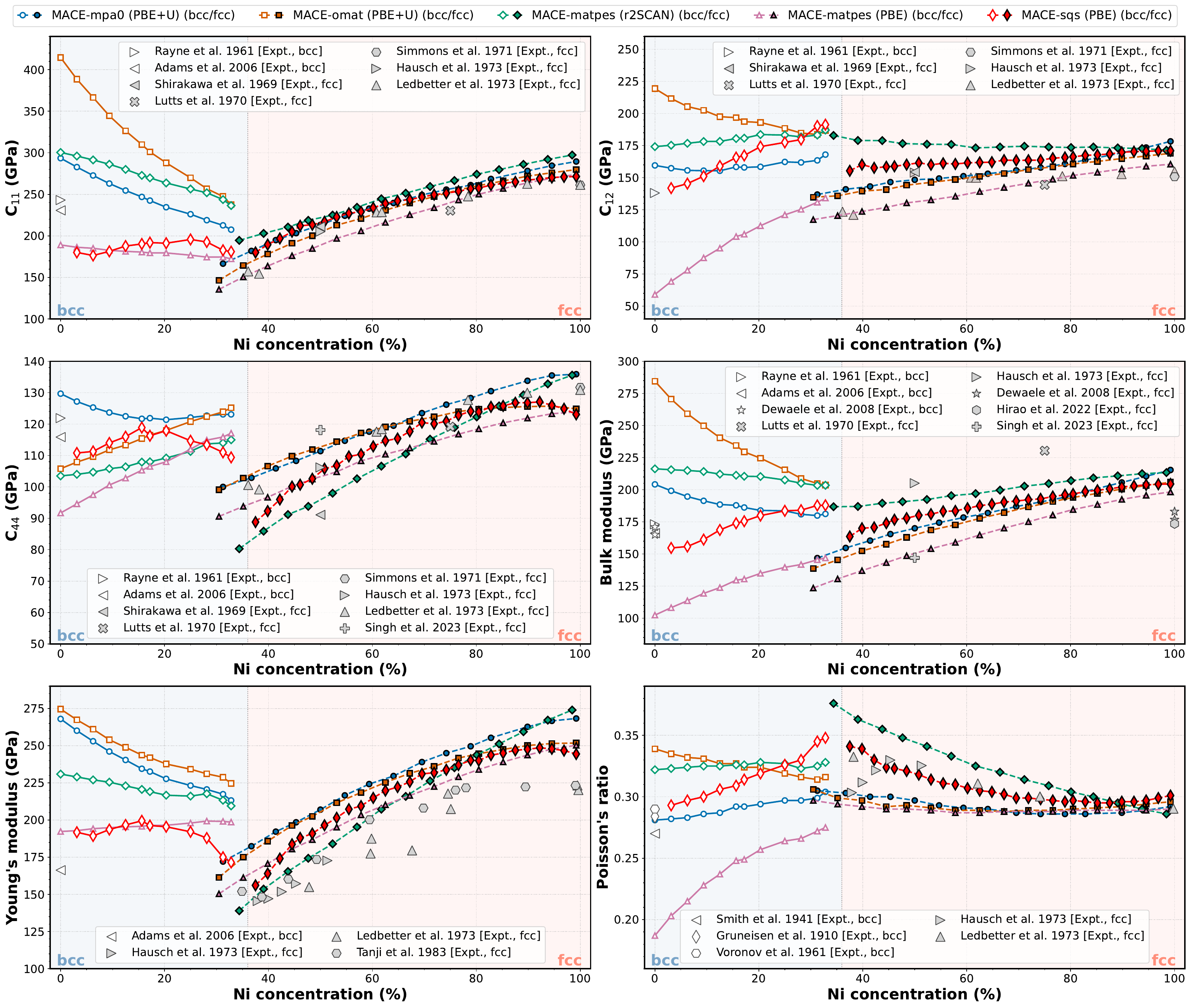}
  \caption{Elastic properties of \FeNi alloys for bcc and fcc structures as a function of nickel concentration, comparing MACE models with experiments~\cite{Adams2006,Rayne1961,Lutts_FeNi3,simmons1971single,shirakawa1969elastic,10.1063/5.0174535,hirao2022equations,PhysRevB.78.104102,10.1063/1.3253127,hausch1973401,Tanji1983,Voronov1961,Smith1941,Gruneisen1910}. }
  \label{fig:elastic_bcc}
\end{figure*}

\subsection{Elastic Properties}

The elastic properties of crystalline materials are fundamental for understanding their mechanical stability and response to external forces. These properties are quantified by the elastic stiffness tensor $C_{ijkl}$, which relates the induced stress tensor $\sigma_{ij}$ to the applied strain tensor $\epsilon_{kl}$ through Hooke's law $\sigma_{ij} = \sum_{kl} C_{ijkl} \epsilon_{kl}$. In practice, the evaluation of elastic constants begins with a full structural optimization to obtain equilibrium lattice parameters and atomic positions, ensuring that the system is free of residual stresses. Subsequently, a series of small, symmetry--adapted strain tensors is applied to the relaxed structure. For each strained configuration, the stress tensor is computed using the MLIP. The elastic constants are then extracted by fitting the linear relationship between the applied strains and the calculated stresses, with particular attention paid to the crystal symmetry. For cubic systems, the independent elastic constants $C_{11}$, $C_{12}$, and $C_{44}$ are determined by analyzing the stress response to uniaxial and shear stresses. The Born stability criteria for cubic crystals require the elastic stiffness tensor to be positive definite, ensuring mechanical stability against all infinitesimal homogeneous deformations~\cite{born1954dynamical}. For a cubic crystal characterized by three independent elastic constants, $C_{11}$, $C_{12}$, and $C_{44}$, mechanical stability requires resistance to shear deformation, tetragonal distortion, and hydrostatic compression. This is ensured when the following Born stability criteria are simultaneously satisfied:
\begin{align}
    C_{11} - C_{12} &> 0, \qquad C_{11} + 2C_{12} > 0, \notag\\
    C_{44} &> 0, \qquad  C_{11} > K > C_{12}.
\end{align}
The bulk modulus $K$ is calculated from the elastic constants using the standard relation for cubic crystals:
\begin{equation}
  K = \frac{C_{11} + 2C_{12}}{3}.
\end{equation}
Similarly, the shear modulus $G$ is given by:
\begin{equation}
  G = \frac{C_{11} - C_{12} + 3C_{44}}{5}.
\end{equation}
Young's modulus $E$ and Poisson's ratio $\nu$ are then obtained as:
\begin{equation}
  E = \frac{9KG}{3K + G},
\end{equation}
\begin{equation}
  \nu = \frac{3K - 2G}{2(3K + G)}.
\end{equation}
The strain magnitude is kept below 1\% to ensure the validity of the linear elastic approximation. This methodology enables a robust and systematic evaluation of the full elastic tensor and derived mechanical moduli, providing essential insight into the mechanical behavior of the material and facilitating direct comparison with experimental measurements.

For bcc \FeNi alloys, Fig.~\ref{fig:elastic_bcc} shows that MACE--sqs achieves the best overall agreement with the experimental elastic constants across all compositions. The predicted values of $C_{11}$, $C_{12}$, and $C_{44}$ from MACE--sqs closely reproduce the compositional trends measured by Rayne~\textit{et al.} and Adams~\textit{et al.}, with only minor deviations at intermediate nickel concentrations. The foundation models reproduce the qualitative trends but show systematic offsets in several elastic constants and in the derived bulk modulus: MACE--mpa0 (PBE$+U$) performs best but still exhibits residual offsets in $C_{11}$ and the derived bulk modulus, while MACE--matpes (PBE) and MACE--matpes (r2SCAN) show the largest deviations, particularly at low nickel content. The bulk modulus $K$ predicted by MACE--sqs decreases smoothly with increasing Ni content, consistent with experimental observations, whereas the PBE$+U$ foundation models tend to overestimate $K$ due to the artificial stiffening of the bcc phase introduced by the Hubbard correction. The agreement between MACE--sqs and experiment for Young's modulus and Poisson's ratio further validates its accuracy in capturing the full mechanical response of bcc \FeNi alloys.

For fcc \FeNi alloys, MACE--sqs best reproduces the nonmonotonic compositional dependence of the elastic constants reported by Lutts~\textit{et al.}~\cite{Lutts_FeNi3}, Shirakawa~\textit{et al.}~\cite{shirakawa1969elastic}, Hausch~\textit{et al.}~\cite{hausch1973401}, Singh~\textit{et al.}, and others. In particular, MACE--sqs correctly captures subtle features such as the local extrema in $C_{11}$, $C_{12}$, and $C_{44}$ near the Invar composition range ($\sim$36\% Ni), which are absent or strongly smeared out in most foundation model predictions. The bulk modulus $K$ and Young's modulus $E$ derived from MACE--sqs are in quantitative agreement with experimental measurements across the full range of Ni concentrations, while MACE--omat (PBE$+U$) and MACE--matpes (r2SCAN) exhibit systematic over-- or underestimation depending on the XC functional employed. The Poisson ratio $\nu$ predicted by MACE--sqs matches the experimental values of Hausch~\textit{et al.} and Ledbetter~\textit{et al.}, confirming the model's ability to simultaneously capture both the compressibility and the shear response of fcc \FeNi alloys.

The elastic-property benchmark therefore supports the same main conclusion as the EOS and volume-composition tests: MACE--sqs provides the most consistent description of bcc and fcc \FeNi among the models compared here, particularly for composition-dependent trends in the elastic constants. This behavior is consistent with the targeted SQS-based training strategy, which samples local chemical environments and lattice distortions directly relevant to disordered \FeNi alloys. At the same time, the remaining model-dependent offsets show that elastic transferability is not guaranteed by broad foundation-model training alone. This is consistent with broader MLIP benchmarks showing that Fe, with partially filled $d$ states, presents a more complex potential-energy surface than late transition metals such as Ni~\cite{Owen2024}.

\begin{figure*}[t]
\centering
\includegraphics[width=2.0\columnwidth]{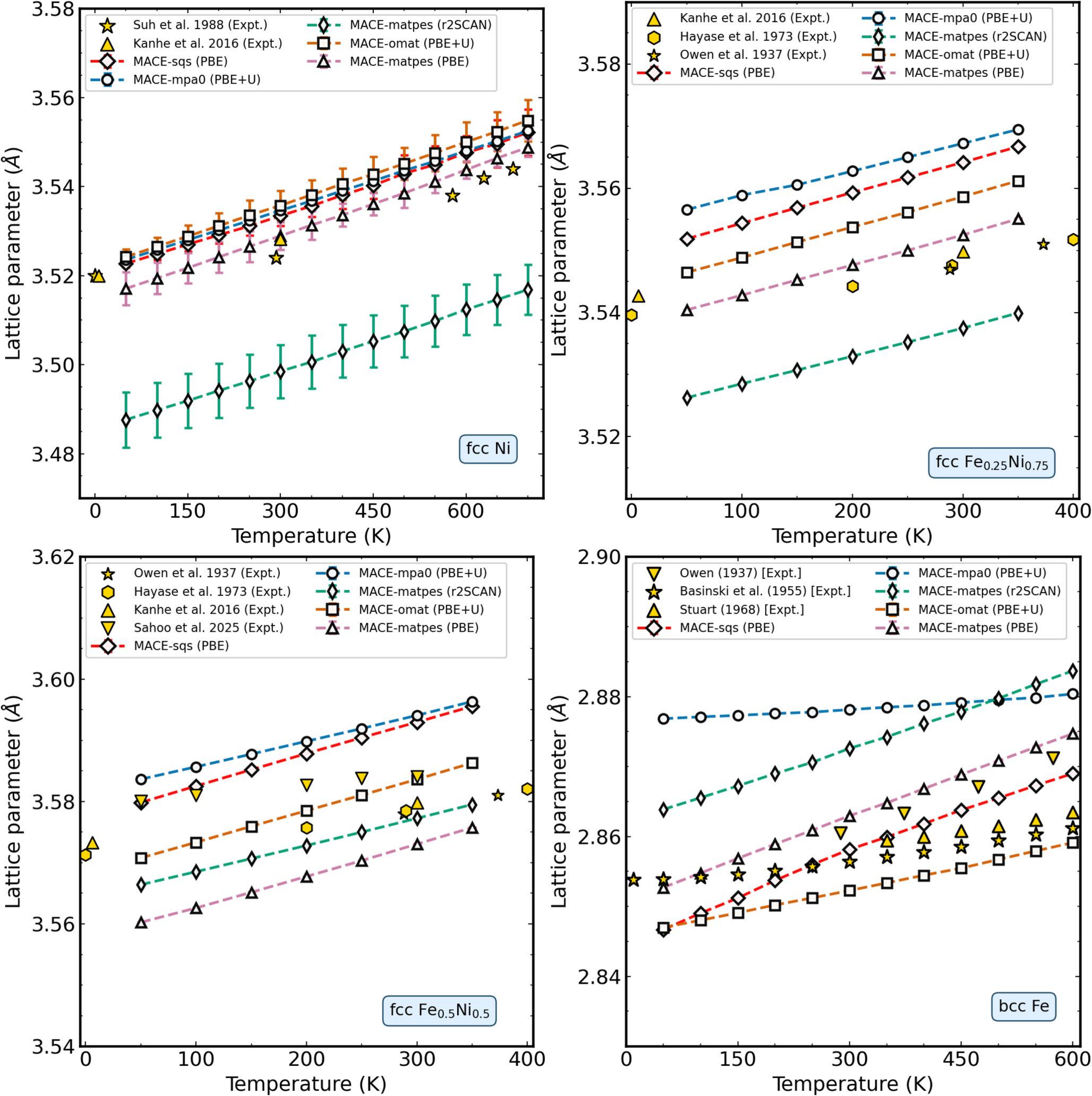}
\caption{Lattice thermal expansion of Fe–Ni compositions, comparing MACE models with experiments~\cite{Suh1988,KANHE201630,Owen_1937,10.1098/rspa.1955.0102,Ridley_1968,sahoo2025anomalieselectronicmagneticthermal,hayase1973spontaneous}.}
\label{fig:latt_temp}
\end{figure*}

\subsection{Transferability: Finite--Temperature Properties}

Assessing finite-temperature lattice expansion provides an additional transferability test because MACE--sqs was trained on static zero-temperature SQS-based DFT configurations and was not exposed to finite-temperature trajectories during training. Any accurate reproduction of finite-temperature lattice expansion therefore reflects genuine transferability of the learned potential-energy surface beyond the training distribution, rather than interpolation within it. The foundation models, while trained on substantially larger and more diverse datasets, face the complementary challenge of capturing the composition- and structure-dependent magnetovolume coupling specific to \FeNi alloys without system-specific training data.

For the finite-temperature calculations, molecular dynamics simulations were performed in the isothermal--isobaric (NPT) ensemble at zero external pressure. For each temperature, three independent replicas were initialized with different velocity seeds sampled from the Maxwell--Boltzmann distribution. Each replica used 10,000 equilibration steps followed by 20,000 production steps with a 1.0~fs timestep. Temperature and pressure were controlled with Nos\'e--Hoover~\cite{PhysRevA.31.1695} and Parrinello--Rahman~\cite{10.1063/1.328693} schemes. The lattice constant was computed from the mean lattice-vector magnitude sampled every 10~fs during production, and uncertainties were estimated from the standard error across the three replicas. Thermal expansion coefficients in Table~\ref{alpha_table} were obtained from $\alpha = \left( \frac{\partial a}{\partial T} \right) / a(T)$.

\begin{table*}
\caption{
Thermal expansion coefficients (in units of
$10^{-5}\,\mathrm{K}^{-1}$) of \FeNi compositions at
$T=300$~K obtained using MACE and compared with experimental values.
}
\begin{ruledtabular}
\small
\begin{tabular}{llcccc}
Model & XC & Ni & Fe & Fe$_{0.5}$Ni$_{0.5}$ & Fe$_{0.25}$Ni$_{0.75}$ \\
\hline
mpa0-medium   & PBE+\textit{U} & 1.25 & 0.24 & 1.23 & 1.25 \\
omat          & PBE+\textit{U} & 1.29 & 0.75 & 1.47 & 1.42 \\
matpes-pbe    & PBE            & 1.36 & 1.36 & 1.51 & 1.43 \\
matpes-r2SCAN & r2SCAN         & 1.23 & 1.26 & 1.26 & 1.33 \\
sqs           & PBE            & 1.27 & 1.38 & 1.42 & 1.40 \\
Experiment &
&
1.29~\cite{PhysRevB.16.4872}
&
1.18~\cite{Kozlovskii_2019}
&
1.60~\cite{sahoo2025anomalieselectronicmagneticthermal}
& -
\\
& 
& 
1.27~\cite{PhysRev.60.597}
&
1.18~\cite{PhysRev.60.597}
& -
& -
\\
\end{tabular}
\end{ruledtabular}
\label{alpha_table}
\end{table*}

Figure~\ref{fig:latt_temp} shows the temperature dependence of the lattice parameter for fcc Ni, bcc Fe, fcc Fe$_{0.5}$Ni$_{0.5}$, and fcc Fe$_{0.25}$Ni$_{0.75}$, with corresponding thermal expansion coefficients at T=300~K summarized in Table~\ref{alpha_table}. For fcc Ni, the MACE--sqs model reproduces the experimental lattice parameter with excellent accuracy below the Curie temperature ($\approx 630$~K), where phonon-driven thermal expansion dominates. The systematic underestimation above 630~K is attributed to the absence of spin--disorder contributions in the nonmagnetic NPT--MD framework, consistent with the known magnetovolume coupling in nickel above its Curie temperature. This deviation provides a direct quantitative estimate of the magnetic contribution to thermal expansion; its correct description would require spin--lattice coupled dynamics. The inability of the nonmagnetic NPT--MD framework to capture the Invar anomaly further underscores that accurate finite-temperature modeling of \FeNi alloys across all compositions ultimately requires explicit treatment of spin--lattice coupling, representing a natural and necessary extension of the present work. Quantitatively, MACE--sqs yields $\alpha = 1.27 \times 10^{-5}$~K$^{-1}$ for Ni at 300~K, in excellent agreement with the experimental values of $1.29 \times 10^{-5}$~K$^{-1}$~\cite{PhysRevB.16.4872} and $1.27 \times 10^{-5}$~K$^{-1}$~\cite{PhysRev.60.597}. For bcc Fe, MACE--sqs yields $\alpha = 1.38 \times 10^{-5}$~K$^{-1}$ at 300~K, in reasonable agreement with the experimental values of $1.18 \times 10^{-5}$~K$^{-1}$~\cite{Kozlovskii_2019,PhysRev.60.597}. The modest overestimation is consistent with the known tendency of collinear spin-polarized PBE to slightly overestimate the thermal compliance of bcc Fe, and with the absence of longitudinal spin fluctuations in the fixed-moment NPT framework at elevated temperatures. Across the full temperature range shown in Fig.~\ref{fig:latt_temp}, the lattice parameter trajectory lies systematically above experiment, the slope and hence $\alpha$ are well captured up to $\sim$400~K. For fcc Fe$_{0.5}$Ni$_{0.5}$, MACE--sqs predicts $\alpha = 1.42 \times 10^{-5}$~K$^{-1}$, compared to the experimental value of $1.60 \times 10^{-5}$~K$^{-1}$~\cite{sahoo2025anomalieselectronicmagneticthermal}. The underestimation at this composition is attributable in part to Invar-proximal anomalies in the magnetovolume coupling that are not captured by the fixed-spin, nonmagnetic MD protocol. The lattice parameter trajectory is in reasonable quantitative agreement with experiment below 300~K, with the deviation growing at higher temperatures where spin-disorder effects become nonnegligible. Experimental data for Fe$_{0.25}$Ni$_{0.75}$ are not available for direct comparison of $\alpha$; MACE--sqs predicts $\alpha = 1.40 \times 10^{-5}$~K$^{-1}$ for this composition, and the temperature dependence of $a(T)$ is smooth and physically consistent, lying between the values predicted for fcc Fe$_{0.5}$Ni$_{0.5}$ and fcc Ni, consistent with the expected compositional dependence.

Among the foundation models, MACE--mpa0 (PBE$+U$) systematically underestimates $\alpha$ for Fe ($0.24 \times 10^{-5}$~K$^{-1}$), reflecting the artificial over-stiffening of the bcc phase introduced by the Hubbard correction. MACE--omat (PBE$+U$) partially corrects this but overestimates $\alpha$ for the alloy compositions. MACE--matpes (PBE) and MACE--matpes (r2SCAN) yield thermal expansion coefficients for the pure elements that bracket the experimental range but exhibit less consistent behavior across alloy compositions. Overall, no single foundation model reproduces the experimental thermal expansion consistently across all four compositions, whereas MACE--sqs achieves the most balanced agreement, correctly capturing both the magnitude and the temperature dependence of $a(T)$ across the full composition range studied.


\section{Conclusion}

We have presented a systematic benchmark of MACE potentials for \FeNi alloys, assessing the structural, elastic, magnetic, phase-stability, and finite-temperature lattice-expansion properties  across composition. The main result is that the system-specific MACE--sqs model, trained on SQS-based spin-polarized DFT data, gives the most consistent agreement with experiment for bcc and fcc equations of state, equilibrium volumes, elastic constants, and thermal expansion trends (Figs.~\ref{fig:EOS_flavors_array}, \ref{fig:elastic_bcc}, and \ref{fig:latt_temp}). At the same time, the bcc--to--hcp transition benchmark exposes a remaining limitation: all tested models fail to reproduce the experimentally observed decrease of transition pressure with Ni content (Fig.~\ref{fig:transition_bcc}). Targeted SQS-based training thus substantially improves bcc and fcc property predictions, while cross-phase transferability under high-pressure magnetic collapse remains an open challenge.

For the EOS and zero-pressure volume trends, MACE--sqs most closely follows the experimental pressure--volume and composition-dependent volume data for the bcc and fcc phases considered here (Figs.~\ref{fig:EOS_flavors_array} and \ref{fig:volume_composition}). For elastic properties, MACE--sqs gives the most consistent composition-dependent trends in the cubic elastic constants and derived moduli, including the stronger nonmonotonic behavior observed in fcc alloys near intermediate Ni concentrations (Fig.~\ref{fig:elastic_bcc}). For finite-temperature lattice expansion, the model gives reasonable 300~K thermal expansion coefficients for Ni, Fe, and the alloy compositions considered, while deviations at higher temperature and near Invar compositions are consistent with the absence of explicit spin-disorder and spin-lattice effects (Fig.~\ref{fig:latt_temp} and Table~\ref{alpha_table}).

The improved bcc/fcc performance of MACE--sqs is consistent with its targeted training data. The SQS-based reference structures sample chemically disordered local environments, volume changes, and lattice distortions that are directly relevant to the benchmarks in this work. In addition, using spin-polarized PBE as a consistent reference level avoids mixing environment-dependent Hubbard corrections into the MACE--sqs training data. The training and validation errors of 2.2 and 2.0~meV/atom for energies, together with the corresponding force errors shown in Fig.~\ref{fig:training}, indicate accurate interpolation within the chosen dataset split. Extrapolative transferability, by contrast, remains property-dependent, as demonstrated by the hcp transition-pressure benchmark.

The present benchmark identifies MACE--sqs as a useful starting point for large-scale simulations of chemically disordered bcc and fcc \FeNi alloys, particularly for structural, elastic, and finite-temperature lattice-expansion properties within the validated composition and pressure ranges. Applications such as finite-temperature phase stability~\cite{KUWAYAMA2008379}, thermal expansion~\cite{Suh1988}, defect energetics~\cite{4q1c-97bp}, and grain-boundary behavior~\cite{Ito2024} remain natural next targets. Extending the training data to include high-pressure magnetic collapse, hcp environments, and composition-dependent disorder effects will be necessary to correct the remaining discrepancy in the bcc--to--hcp transition-pressure trend. More broadly, targeted SQS-based training provides a practical route to bridging the accuracy–efficiency gap for chemically disordered transition-metal alloys. At the same time, reliable phase transferability demands reference data that explicitly cover the relevant structural and magnetic regimes, a requirement that system-specific training alone cannot fully satisfy.

\begin{acknowledgments}
K.R. acknowledges funding from the HPC Gateway project (Grant Agreement No.~1077561005) through the HHPAKT fund. This work was supported by the Center for Advanced Systems Understanding (CASUS), which is financed by Germany’s Federal Ministry of Research, Technology and Space (BMFTR) and by the Saxon State government through the state budget approved by the Saxon State Parliament.\\
Computations were performed on the Hemera and ROSI clusters at 
Helmholtz--Zentrum Dresden--Rossendorf (HZDR), as well as on the Barnard and Capella clusters at the Center for Information Services and High Performance Computing (ZIH) at Technische Universit\"at Dresden. We thank Henrik Schulz and Jens Lasch at HZDR for the support in the area of High Performance Computing. This work was additionally supported by Helmholtz AI computing resources (HAICORE) of the Helmholtz Association's Initiative and Networking Fund through Helmholtz AI (Project No.~hai\_1036). Additional computations were carried out on the JUWELS Booster at the J\"ulich Supercomputing Center (JSC).
\end{acknowledgments}

\pagebreak


\bibliography{bibliography}

\begin{thebibliography}{103}%
\makeatletter
\providecommand \@ifxundefined [1]{%
 \@ifx{#1\undefined}
}%
\providecommand \@ifnum [1]{%
 \ifnum #1\expandafter \@firstoftwo
 \else \expandafter \@secondoftwo
 \fi
}%
\providecommand \@ifx [1]{%
 \ifx #1\expandafter \@firstoftwo
 \else \expandafter \@secondoftwo
 \fi
}%
\providecommand \natexlab [1]{#1}%
\providecommand \enquote  [1]{``#1''}%
\providecommand \bibnamefont  [1]{#1}%
\providecommand \bibfnamefont [1]{#1}%
\providecommand \citenamefont [1]{#1}%
\providecommand \href@noop [0]{\@secondoftwo}%
\providecommand \href [0]{\begingroup \@sanitize@url \@href}%
\providecommand \@href[1]{\@@startlink{#1}\@@href}%
\providecommand \@@href[1]{\endgroup#1\@@endlink}%
\providecommand \@sanitize@url [0]{\catcode `\\12\catcode `\$12\catcode
  `\&12\catcode `\#12\catcode `\^12\catcode `\_12\catcode `\%12\relax}%
\providecommand \@@startlink[1]{}%
\providecommand \@@endlink[0]{}%
\providecommand \url  [0]{\begingroup\@sanitize@url \@url }%
\providecommand \@url [1]{\endgroup\@href {#1}{\urlprefix }}%
\providecommand \urlprefix  [0]{URL }%
\providecommand \Eprint [0]{\href }%
\providecommand \doibase [0]{https://doi.org/}%
\providecommand \selectlanguage [0]{\@gobble}%
\providecommand \bibinfo  [0]{\@secondoftwo}%
\providecommand \bibfield  [0]{\@secondoftwo}%
\providecommand \translation [1]{[#1]}%
\providecommand \BibitemOpen [0]{}%
\providecommand \bibitemStop [0]{}%
\providecommand \bibitemNoStop [0]{.\EOS\space}%
\providecommand \EOS [0]{\spacefactor3000\relax}%
\providecommand \BibitemShut  [1]{\csname bibitem#1\endcsname}%
\let\auto@bib@innerbib\@empty
\bibitem [{\citenamefont {Kuwayama}\ \emph {et~al.}(2008)\citenamefont
  {Kuwayama}, \citenamefont {Hirose}, \citenamefont {Sata},\ and\ \citenamefont
  {Ohishi}}]{KUWAYAMA2008379}%
  \BibitemOpen
  \bibfield  {author} {\bibinfo {author} {\bibfnamefont {Y.}~\bibnamefont
  {Kuwayama}}, \bibinfo {author} {\bibfnamefont {K.}~\bibnamefont {Hirose}},
  \bibinfo {author} {\bibfnamefont {N.}~\bibnamefont {Sata}},\ and\ \bibinfo
  {author} {\bibfnamefont {Y.}~\bibnamefont {Ohishi}},\ }\bibfield  {title}
  {\bibinfo {title} {Phase relations of iron and iron--nickel alloys up to
  300~{GPa}: Implications for composition and structure of the earth's inner
  core},\ }\href {https://doi.org/https://doi.org/10.1016/j.epsl.2008.07.001}
  {\bibfield  {journal} {\bibinfo  {journal} {Earth and Planetary Science
  Letters}\ }\textbf {\bibinfo {volume} {273}},\ \bibinfo {pages} {379}
  (\bibinfo {year} {2008})}\BibitemShut {NoStop}%
\bibitem [{\citenamefont {Mao}\ \emph {et~al.}(2006)\citenamefont {Mao},
  \citenamefont {Campbell}, \citenamefont {Heinz},\ and\ \citenamefont
  {Shen}}]{MAO2006146}%
  \BibitemOpen
  \bibfield  {author} {\bibinfo {author} {\bibfnamefont {W.~L.}\ \bibnamefont
  {Mao}}, \bibinfo {author} {\bibfnamefont {A.~J.}\ \bibnamefont {Campbell}},
  \bibinfo {author} {\bibfnamefont {D.~L.}\ \bibnamefont {Heinz}},\ and\
  \bibinfo {author} {\bibfnamefont {G.}~\bibnamefont {Shen}},\ }\bibfield
  {title} {\bibinfo {title} {Phase relations of {Fe--Ni} alloys at high
  pressure and temperature},\ }\href
  {https://doi.org/https://doi.org/10.1016/j.pepi.2005.11.002} {\bibfield
  {journal} {\bibinfo  {journal} {Physics of the Earth and Planetary
  Interiors}\ }\textbf {\bibinfo {volume} {155}},\ \bibinfo {pages} {146}
  (\bibinfo {year} {2006})}\BibitemShut {NoStop}%
\bibitem [{\citenamefont {van Schilfgaarde}\ \emph {et~al.}(1999)\citenamefont
  {van Schilfgaarde}, \citenamefont {Abrikosov},\ and\ \citenamefont
  {Johansson}}]{Schilfgaarde:1999aa}%
  \BibitemOpen
  \bibfield  {author} {\bibinfo {author} {\bibfnamefont {M.}~\bibnamefont {van
  Schilfgaarde}}, \bibinfo {author} {\bibfnamefont {I.~A.}\ \bibnamefont
  {Abrikosov}},\ and\ \bibinfo {author} {\bibfnamefont {B.}~\bibnamefont
  {Johansson}},\ }\bibfield  {title} {\bibinfo {title} {Origin of the invar
  effect in iron--nickel alloys},\ }\href {https://doi.org/10.1038/21848}
  {\bibfield  {journal} {\bibinfo  {journal} {Nature}\ }\textbf {\bibinfo
  {volume} {400}},\ \bibinfo {pages} {46} (\bibinfo {year} {1999})}\BibitemShut
  {NoStop}%
\bibitem [{\citenamefont {Dubrovinsky}\ \emph {et~al.}(2001)\citenamefont
  {Dubrovinsky}, \citenamefont {Dubrovinskaia}, \citenamefont {Abrikosov},
  \citenamefont {Vennstr\"om}, \citenamefont {Westman}, \citenamefont
  {Carlson}, \citenamefont {van Schilfgaarde},\ and\ \citenamefont
  {Johansson}}]{PhysRevLett.86.4851}%
  \BibitemOpen
  \bibfield  {author} {\bibinfo {author} {\bibfnamefont {L.}~\bibnamefont
  {Dubrovinsky}}, \bibinfo {author} {\bibfnamefont {N.}~\bibnamefont
  {Dubrovinskaia}}, \bibinfo {author} {\bibfnamefont {I.~A.}\ \bibnamefont
  {Abrikosov}}, \bibinfo {author} {\bibfnamefont {M.}~\bibnamefont
  {Vennstr\"om}}, \bibinfo {author} {\bibfnamefont {F.}~\bibnamefont
  {Westman}}, \bibinfo {author} {\bibfnamefont {S.}~\bibnamefont {Carlson}},
  \bibinfo {author} {\bibfnamefont {M.}~\bibnamefont {van Schilfgaarde}},\ and\
  \bibinfo {author} {\bibfnamefont {B.}~\bibnamefont {Johansson}},\ }\bibfield
  {title} {\bibinfo {title} {Pressure-induced invar effect in {Fe}-{Ni}
  alloys},\ }\href {https://doi.org/10.1103/PhysRevLett.86.4851} {\bibfield
  {journal} {\bibinfo  {journal} {Phys. Rev. Lett.}\ }\textbf {\bibinfo
  {volume} {86}},\ \bibinfo {pages} {4851} (\bibinfo {year}
  {2001})}\BibitemShut {NoStop}%
\bibitem [{\citenamefont {Mushnikov}\ \emph {et~al.}(2022)\citenamefont
  {Mushnikov}, \citenamefont {Popov}, \citenamefont {Gaviko}, \citenamefont
  {Protasov}, \citenamefont {Kleinerman}, \citenamefont {Golovnya},\ and\
  \citenamefont {Naumov}}]{MUSHNIKOV2022118330}%
  \BibitemOpen
  \bibfield  {author} {\bibinfo {author} {\bibfnamefont {N.}~\bibnamefont
  {Mushnikov}}, \bibinfo {author} {\bibfnamefont {A.}~\bibnamefont {Popov}},
  \bibinfo {author} {\bibfnamefont {V.}~\bibnamefont {Gaviko}}, \bibinfo
  {author} {\bibfnamefont {A.}~\bibnamefont {Protasov}}, \bibinfo {author}
  {\bibfnamefont {N.}~\bibnamefont {Kleinerman}}, \bibinfo {author}
  {\bibfnamefont {O.}~\bibnamefont {Golovnya}},\ and\ \bibinfo {author}
  {\bibfnamefont {S.}~\bibnamefont {Naumov}},\ }\bibfield  {title} {\bibinfo
  {title} {Peculiarities of phase diagram of {Fe--Ni} system at {Ni}
  concentrations 0 – 20 at.\%.},\ }\href
  {https://doi.org/https://doi.org/10.1016/j.actamat.2022.118330} {\bibfield
  {journal} {\bibinfo  {journal} {Acta Materialia}\ }\textbf {\bibinfo {volume}
  {240}},\ \bibinfo {pages} {118330} (\bibinfo {year} {2022})}\BibitemShut
  {NoStop}%
\bibitem [{\citenamefont {Ahles}\ \emph {et~al.}(2021)\citenamefont {Ahles},
  \citenamefont {Emery},\ and\ \citenamefont {Dunand}}]{AHLES2021465}%
  \BibitemOpen
  \bibfield  {author} {\bibinfo {author} {\bibfnamefont {A.~A.}\ \bibnamefont
  {Ahles}}, \bibinfo {author} {\bibfnamefont {J.~D.}\ \bibnamefont {Emery}},\
  and\ \bibinfo {author} {\bibfnamefont {D.~C.}\ \bibnamefont {Dunand}},\
  }\bibfield  {title} {\bibinfo {title} {Mechanical properties of meteoritic
  {Fe--Ni} alloys for in-situ extraterrestrial structures},\ }\href
  {https://doi.org/https://doi.org/10.1016/j.actaastro.2021.09.001} {\bibfield
  {journal} {\bibinfo  {journal} {Acta Astronautica}\ }\textbf {\bibinfo
  {volume} {189}},\ \bibinfo {pages} {465} (\bibinfo {year}
  {2021})}\BibitemShut {NoStop}%
\bibitem [{\citenamefont {Huang}\ \emph {et~al.}(1988)\citenamefont {Huang},
  \citenamefont {Bassett},\ and\ \citenamefont {Weathers}}]{huang1988}%
  \BibitemOpen
  \bibfield  {author} {\bibinfo {author} {\bibfnamefont {E.}~\bibnamefont
  {Huang}}, \bibinfo {author} {\bibfnamefont {W.~A.}\ \bibnamefont {Bassett}},\
  and\ \bibinfo {author} {\bibfnamefont {M.~S.}\ \bibnamefont {Weathers}},\
  }\bibfield  {title} {\bibinfo {title} {Phase relationships in {Fe--Ni} alloys
  at high pressures and temperatures},\ }\href
  {https://doi.org/https://doi.org/10.1029/JB093iB07p07741} {\bibfield
  {journal} {\bibinfo  {journal} {Journal of Geophysical Research: Solid
  Earth}\ }\textbf {\bibinfo {volume} {93}},\ \bibinfo {pages} {7741} (\bibinfo
  {year} {1988})}\BibitemShut {NoStop}%
\bibitem [{\citenamefont {Akahama}\ \emph {et~al.}(2020)\citenamefont
  {Akahama}, \citenamefont {Fujimoto}, \citenamefont {Terai}, \citenamefont
  {Fukuda}, \citenamefont {Kawaguchi}, \citenamefont {Hirao}, \citenamefont
  {Ohishi},\ and\ \citenamefont {Kakeshita}}]{akahama2020pressure}%
  \BibitemOpen
  \bibfield  {author} {\bibinfo {author} {\bibfnamefont {Y.}~\bibnamefont
  {Akahama}}, \bibinfo {author} {\bibfnamefont {Y.}~\bibnamefont {Fujimoto}},
  \bibinfo {author} {\bibfnamefont {T.}~\bibnamefont {Terai}}, \bibinfo
  {author} {\bibfnamefont {T.}~\bibnamefont {Fukuda}}, \bibinfo {author}
  {\bibfnamefont {S.}~\bibnamefont {Kawaguchi}}, \bibinfo {author}
  {\bibfnamefont {N.}~\bibnamefont {Hirao}}, \bibinfo {author} {\bibfnamefont
  {Y.}~\bibnamefont {Ohishi}},\ and\ \bibinfo {author} {\bibfnamefont
  {T.}~\bibnamefont {Kakeshita}},\ }\bibfield  {title} {\bibinfo {title}
  {Pressure--composition phase diagram of {Fe--Ni} alloy},\ }\href
  {https://doi.org/10.2320/matertrans.MT-M2020047} {\bibfield  {journal}
  {\bibinfo  {journal} {MATERIALS TRANSACTIONS}\ }\textbf {\bibinfo {volume}
  {61}},\ \bibinfo {pages} {1058} (\bibinfo {year} {2020})}\BibitemShut
  {NoStop}%
\bibitem [{\citenamefont {Ramakrishna}\ \emph
  {et~al.}(2023{\natexlab{a}})\citenamefont {Ramakrishna}, \citenamefont
  {Lokamani}, \citenamefont {Baczewski}, \citenamefont {Vorberger},\ and\
  \citenamefont {Cangi}}]{PhysRevB.107.115131}%
  \BibitemOpen
  \bibfield  {author} {\bibinfo {author} {\bibfnamefont {K.}~\bibnamefont
  {Ramakrishna}}, \bibinfo {author} {\bibfnamefont {M.}~\bibnamefont
  {Lokamani}}, \bibinfo {author} {\bibfnamefont {A.}~\bibnamefont {Baczewski}},
  \bibinfo {author} {\bibfnamefont {J.}~\bibnamefont {Vorberger}},\ and\
  \bibinfo {author} {\bibfnamefont {A.}~\bibnamefont {Cangi}},\ }\bibfield
  {title} {\bibinfo {title} {Electrical conductivity of iron in earth's core
  from microscopic ohm's law},\ }\href
  {https://doi.org/10.1103/PhysRevB.107.115131} {\bibfield  {journal} {\bibinfo
   {journal} {Phys. Rev. B}\ }\textbf {\bibinfo {volume} {107}},\ \bibinfo
  {pages} {115131} (\bibinfo {year} {2023}{\natexlab{a}})}\BibitemShut
  {NoStop}%
\bibitem [{\citenamefont {Takahashi}\ and\ \citenamefont
  {Bassett}(1964)}]{takahashi1964}%
  \BibitemOpen
  \bibfield  {author} {\bibinfo {author} {\bibfnamefont {T.}~\bibnamefont
  {Takahashi}}\ and\ \bibinfo {author} {\bibfnamefont {W.~A.}\ \bibnamefont
  {Bassett}},\ }\bibfield  {title} {\bibinfo {title} {High-pressure polymorph
  of iron},\ }\href {https://doi.org/10.1126/science.145.3631.483} {\bibfield
  {journal} {\bibinfo  {journal} {Science}\ }\textbf {\bibinfo {volume}
  {145}},\ \bibinfo {pages} {483} (\bibinfo {year} {1964})}\BibitemShut
  {NoStop}%
\bibitem [{\citenamefont {Mathon}\ \emph {et~al.}(2004)\citenamefont {Mathon},
  \citenamefont {Baudelet}, \citenamefont {Iti\'e}, \citenamefont {Polian},
  \citenamefont {d'Astuto}, \citenamefont {Chervin},\ and\ \citenamefont
  {Pascarelli}}]{PhysRevLett.93.255503}%
  \BibitemOpen
  \bibfield  {author} {\bibinfo {author} {\bibfnamefont {O.}~\bibnamefont
  {Mathon}}, \bibinfo {author} {\bibfnamefont {F.}~\bibnamefont {Baudelet}},
  \bibinfo {author} {\bibfnamefont {J.~P.}\ \bibnamefont {Iti\'e}}, \bibinfo
  {author} {\bibfnamefont {A.}~\bibnamefont {Polian}}, \bibinfo {author}
  {\bibfnamefont {M.}~\bibnamefont {d'Astuto}}, \bibinfo {author}
  {\bibfnamefont {J.~C.}\ \bibnamefont {Chervin}},\ and\ \bibinfo {author}
  {\bibfnamefont {S.}~\bibnamefont {Pascarelli}},\ }\bibfield  {title}
  {\bibinfo {title} {Dynamics of the magnetic and structural
  $\ensuremath{\alpha}\mathrm{\text{\ensuremath{-}}}\ensuremath{\epsilon}$
  phase transition in iron},\ }\href
  {https://doi.org/10.1103/PhysRevLett.93.255503} {\bibfield  {journal}
  {\bibinfo  {journal} {Phys. Rev. Lett.}\ }\textbf {\bibinfo {volume} {93}},\
  \bibinfo {pages} {255503} (\bibinfo {year} {2004})}\BibitemShut {NoStop}%
\bibitem [{\citenamefont {Ramakrishna}\ \emph
  {et~al.}(2023{\natexlab{b}})\citenamefont {Ramakrishna}, \citenamefont
  {Lokamani}, \citenamefont {Baczewski}, \citenamefont {Vorberger},\ and\
  \citenamefont {Cangi}}]{Ramakrishna_2023}%
  \BibitemOpen
  \bibfield  {author} {\bibinfo {author} {\bibfnamefont {K.}~\bibnamefont
  {Ramakrishna}}, \bibinfo {author} {\bibfnamefont {M.}~\bibnamefont
  {Lokamani}}, \bibinfo {author} {\bibfnamefont {A.}~\bibnamefont {Baczewski}},
  \bibinfo {author} {\bibfnamefont {J.}~\bibnamefont {Vorberger}},\ and\
  \bibinfo {author} {\bibfnamefont {A.}~\bibnamefont {Cangi}},\ }\bibfield
  {title} {\bibinfo {title} {Impact of electronic correlations on high-pressure
  iron: insights from time-dependent density functional theory},\ }\href
  {https://doi.org/10.1088/2516-1075/acfd75} {\bibfield  {journal} {\bibinfo
  {journal} {Electronic Structure}\ }\textbf {\bibinfo {volume} {5}},\ \bibinfo
  {pages} {045002} (\bibinfo {year} {2023}{\natexlab{b}})}\BibitemShut
  {NoStop}%
\bibitem [{\citenamefont {Behler}\ and\ \citenamefont
  {Parrinello}(2007)}]{PhysRevLett.98.146401}%
  \BibitemOpen
  \bibfield  {author} {\bibinfo {author} {\bibfnamefont {J.}~\bibnamefont
  {Behler}}\ and\ \bibinfo {author} {\bibfnamefont {M.}~\bibnamefont
  {Parrinello}},\ }\bibfield  {title} {\bibinfo {title} {Generalized
  neural-network representation of high-dimensional potential-energy
  surfaces},\ }\href {https://doi.org/10.1103/PhysRevLett.98.146401} {\bibfield
   {journal} {\bibinfo  {journal} {Phys. Rev. Lett.}\ }\textbf {\bibinfo
  {volume} {98}},\ \bibinfo {pages} {146401} (\bibinfo {year}
  {2007})}\BibitemShut {NoStop}%
\bibitem [{\citenamefont {Behler}(2021)}]{doi:10.1021/acs.chemrev.0c00868}%
  \BibitemOpen
  \bibfield  {author} {\bibinfo {author} {\bibfnamefont {J.}~\bibnamefont
  {Behler}},\ }\bibfield  {title} {\bibinfo {title} {Four generations of
  high-dimensional neural network potentials},\ }\href
  {https://doi.org/10.1021/acs.chemrev.0c00868} {\bibfield  {journal} {\bibinfo
   {journal} {Chemical Reviews}\ }\textbf {\bibinfo {volume} {121}},\ \bibinfo
  {pages} {10037} (\bibinfo {year} {2021})}\BibitemShut {NoStop}%
\bibitem [{\citenamefont {Thompson}\ \emph {et~al.}(2015)\citenamefont
  {Thompson}, \citenamefont {Swiler}, \citenamefont {Trott}, \citenamefont
  {Foiles},\ and\ \citenamefont {Tucker}}]{thompson2015spectral}%
  \BibitemOpen
  \bibfield  {author} {\bibinfo {author} {\bibfnamefont {A.}~\bibnamefont
  {Thompson}}, \bibinfo {author} {\bibfnamefont {L.}~\bibnamefont {Swiler}},
  \bibinfo {author} {\bibfnamefont {C.}~\bibnamefont {Trott}}, \bibinfo
  {author} {\bibfnamefont {S.}~\bibnamefont {Foiles}},\ and\ \bibinfo {author}
  {\bibfnamefont {G.}~\bibnamefont {Tucker}},\ }\bibfield  {title} {\bibinfo
  {title} {Spectral neighbor analysis method for automated generation of
  quantum-accurate interatomic potentials},\ }\href
  {https://doi.org/https://doi.org/10.1016/j.jcp.2014.12.018} {\bibfield
  {journal} {\bibinfo  {journal} {Journal of Computational Physics}\ }\textbf
  {\bibinfo {volume} {285}},\ \bibinfo {pages} {316} (\bibinfo {year}
  {2015})}\BibitemShut {NoStop}%
\bibitem [{\citenamefont {Zuo}\ \emph {et~al.}(2020)\citenamefont {Zuo},
  \citenamefont {Chen}, \citenamefont {Li}, \citenamefont {Deng}, \citenamefont
  {Chen}, \citenamefont {Behler}, \citenamefont {Cs{\'a}nyi}, \citenamefont
  {Shapeev}, \citenamefont {Thompson}, \citenamefont {Wood},\ and\
  \citenamefont {Ong}}]{doi:10.1021/acs.jpca.9b08723}%
  \BibitemOpen
  \bibfield  {author} {\bibinfo {author} {\bibfnamefont {Y.}~\bibnamefont
  {Zuo}}, \bibinfo {author} {\bibfnamefont {C.}~\bibnamefont {Chen}}, \bibinfo
  {author} {\bibfnamefont {X.}~\bibnamefont {Li}}, \bibinfo {author}
  {\bibfnamefont {Z.}~\bibnamefont {Deng}}, \bibinfo {author} {\bibfnamefont
  {Y.}~\bibnamefont {Chen}}, \bibinfo {author} {\bibfnamefont {J.}~\bibnamefont
  {Behler}}, \bibinfo {author} {\bibfnamefont {G.}~\bibnamefont {Cs{\'a}nyi}},
  \bibinfo {author} {\bibfnamefont {A.~V.}\ \bibnamefont {Shapeev}}, \bibinfo
  {author} {\bibfnamefont {A.~P.}\ \bibnamefont {Thompson}}, \bibinfo {author}
  {\bibfnamefont {M.~A.}\ \bibnamefont {Wood}},\ and\ \bibinfo {author}
  {\bibfnamefont {S.~P.}\ \bibnamefont {Ong}},\ }\bibfield  {title} {\bibinfo
  {title} {Performance and cost assessment of machine learning interatomic
  potentials},\ }\href {https://doi.org/10.1021/acs.jpca.9b08723} {\bibfield
  {journal} {\bibinfo  {journal} {The Journal of Physical Chemistry A}\
  }\textbf {\bibinfo {volume} {124}},\ \bibinfo {pages} {731} (\bibinfo {year}
  {2020})}\BibitemShut {NoStop}%
\bibitem [{\citenamefont {Nikolov}\ \emph {et~al.}(2021)\citenamefont
  {Nikolov}, \citenamefont {Wood}, \citenamefont {Cangi}, \citenamefont
  {Maillet}, \citenamefont {Marinica}, \citenamefont {Thompson}, \citenamefont
  {Desjarlais},\ and\ \citenamefont {Tranchida}}]{Nikolov2021}%
  \BibitemOpen
  \bibfield  {author} {\bibinfo {author} {\bibfnamefont {S.}~\bibnamefont
  {Nikolov}}, \bibinfo {author} {\bibfnamefont {M.~A.}\ \bibnamefont {Wood}},
  \bibinfo {author} {\bibfnamefont {A.}~\bibnamefont {Cangi}}, \bibinfo
  {author} {\bibfnamefont {J.-B.}\ \bibnamefont {Maillet}}, \bibinfo {author}
  {\bibfnamefont {M.-C.}\ \bibnamefont {Marinica}}, \bibinfo {author}
  {\bibfnamefont {A.~P.}\ \bibnamefont {Thompson}}, \bibinfo {author}
  {\bibfnamefont {M.~P.}\ \bibnamefont {Desjarlais}},\ and\ \bibinfo {author}
  {\bibfnamefont {J.}~\bibnamefont {Tranchida}},\ }\bibfield  {title} {\bibinfo
  {title} {Data-driven magneto-elastic predictions with scalable classical
  spin-lattice dynamics},\ }\href {https://doi.org/10.1038/s41524-021-00617-2}
  {\bibfield  {journal} {\bibinfo  {journal} {npj Computational Materials}\
  }\textbf {\bibinfo {volume} {7}},\ \bibinfo {pages} {153} (\bibinfo {year}
  {2021})}\BibitemShut {NoStop}%
\bibitem [{\citenamefont {Nikolov}\ \emph {et~al.}(2022)\citenamefont
  {Nikolov}, \citenamefont {Tranchida}, \citenamefont {Ramakrishna},
  \citenamefont {Lokamani}, \citenamefont {Cangi},\ and\ \citenamefont
  {Wood}}]{Nikolov2022}%
  \BibitemOpen
  \bibfield  {author} {\bibinfo {author} {\bibfnamefont {S.}~\bibnamefont
  {Nikolov}}, \bibinfo {author} {\bibfnamefont {J.}~\bibnamefont {Tranchida}},
  \bibinfo {author} {\bibfnamefont {K.}~\bibnamefont {Ramakrishna}}, \bibinfo
  {author} {\bibfnamefont {M.}~\bibnamefont {Lokamani}}, \bibinfo {author}
  {\bibfnamefont {A.}~\bibnamefont {Cangi}},\ and\ \bibinfo {author}
  {\bibfnamefont {M.~A.}\ \bibnamefont {Wood}},\ }\bibfield  {title} {\bibinfo
  {title} {Dissociating the phononic, magnetic and electronic contributions to
  thermal conductivity: a computational study in alpha-iron},\ }\href
  {https://doi.org/10.1007/s10853-021-06865-3} {\bibfield  {journal} {\bibinfo
  {journal} {Journal of Materials Science}\ }\textbf {\bibinfo {volume} {57}},\
  \bibinfo {pages} {10535} (\bibinfo {year} {2022})}\BibitemShut {NoStop}%
\bibitem [{\citenamefont {Nikolov}\ \emph {et~al.}(2024)\citenamefont
  {Nikolov}, \citenamefont {Ramakrishna}, \citenamefont {Rohskopf},
  \citenamefont {Lokamani}, \citenamefont {Tranchida}, \citenamefont
  {Carpenter}, \citenamefont {Cangi},\ and\ \citenamefont
  {Wood}}]{Nikolov2024}%
  \BibitemOpen
  \bibfield  {author} {\bibinfo {author} {\bibfnamefont {S.}~\bibnamefont
  {Nikolov}}, \bibinfo {author} {\bibfnamefont {K.}~\bibnamefont
  {Ramakrishna}}, \bibinfo {author} {\bibfnamefont {A.}~\bibnamefont
  {Rohskopf}}, \bibinfo {author} {\bibfnamefont {M.}~\bibnamefont {Lokamani}},
  \bibinfo {author} {\bibfnamefont {J.}~\bibnamefont {Tranchida}}, \bibinfo
  {author} {\bibfnamefont {J.}~\bibnamefont {Carpenter}}, \bibinfo {author}
  {\bibfnamefont {A.}~\bibnamefont {Cangi}},\ and\ \bibinfo {author}
  {\bibfnamefont {M.~A.}\ \bibnamefont {Wood}},\ }\bibfield  {title} {\bibinfo
  {title} {Probing iron in earth’s core with molecular-spin dynamics},\
  }\href {https://doi.org/10.1073/pnas.2408897121} {\bibfield  {journal}
  {\bibinfo  {journal} {Proceedings of the National Academy of Sciences}\
  }\textbf {\bibinfo {volume} {121}},\ \bibinfo {pages} {e2408897121} (\bibinfo
  {year} {2024})}\BibitemShut {NoStop}%
\bibitem [{\citenamefont {Bart\'ok}\ \emph {et~al.}(2010)\citenamefont
  {Bart\'ok}, \citenamefont {Payne}, \citenamefont {Kondor},\ and\
  \citenamefont {Cs\'anyi}}]{bartok2010gaussian}%
  \BibitemOpen
  \bibfield  {author} {\bibinfo {author} {\bibfnamefont {A.~P.}\ \bibnamefont
  {Bart\'ok}}, \bibinfo {author} {\bibfnamefont {M.~C.}\ \bibnamefont {Payne}},
  \bibinfo {author} {\bibfnamefont {R.}~\bibnamefont {Kondor}},\ and\ \bibinfo
  {author} {\bibfnamefont {G.}~\bibnamefont {Cs\'anyi}},\ }\bibfield  {title}
  {\bibinfo {title} {Gaussian approximation potentials: The accuracy of quantum
  mechanics, without the electrons},\ }\href
  {https://doi.org/10.1103/PhysRevLett.104.136403} {\bibfield  {journal}
  {\bibinfo  {journal} {Phys. Rev. Lett.}\ }\textbf {\bibinfo {volume} {104}},\
  \bibinfo {pages} {136403} (\bibinfo {year} {2010})}\BibitemShut {NoStop}%
\bibitem [{\citenamefont {Dragoni}\ \emph {et~al.}(2018)\citenamefont
  {Dragoni}, \citenamefont {Daff}, \citenamefont {Cs\'anyi},\ and\
  \citenamefont {Marzari}}]{PhysRevMaterials.2.013808}%
  \BibitemOpen
  \bibfield  {author} {\bibinfo {author} {\bibfnamefont {D.}~\bibnamefont
  {Dragoni}}, \bibinfo {author} {\bibfnamefont {T.~D.}\ \bibnamefont {Daff}},
  \bibinfo {author} {\bibfnamefont {G.}~\bibnamefont {Cs\'anyi}},\ and\
  \bibinfo {author} {\bibfnamefont {N.}~\bibnamefont {Marzari}},\ }\bibfield
  {title} {\bibinfo {title} {Achieving {DFT} accuracy with a machine-learning
  interatomic potential: Thermomechanics and defects in bcc ferromagnetic
  iron},\ }\href {https://doi.org/10.1103/PhysRevMaterials.2.013808} {\bibfield
   {journal} {\bibinfo  {journal} {Phys. Rev. Mater.}\ }\textbf {\bibinfo
  {volume} {2}},\ \bibinfo {pages} {013808} (\bibinfo {year}
  {2018})}\BibitemShut {NoStop}%
\bibitem [{\citenamefont {Shenoy}\ \emph {et~al.}(2024)\citenamefont {Shenoy},
  \citenamefont {Woodgate}, \citenamefont {Staunton}, \citenamefont {Bart\'ok},
  \citenamefont {Becquart}, \citenamefont {Domain},\ and\ \citenamefont
  {Kermode}}]{PhysRevMaterials.8.033804}%
  \BibitemOpen
  \bibfield  {author} {\bibinfo {author} {\bibfnamefont {L.}~\bibnamefont
  {Shenoy}}, \bibinfo {author} {\bibfnamefont {C.~D.}\ \bibnamefont
  {Woodgate}}, \bibinfo {author} {\bibfnamefont {J.~B.}\ \bibnamefont
  {Staunton}}, \bibinfo {author} {\bibfnamefont {A.~P.}\ \bibnamefont
  {Bart\'ok}}, \bibinfo {author} {\bibfnamefont {C.~S.}\ \bibnamefont
  {Becquart}}, \bibinfo {author} {\bibfnamefont {C.}~\bibnamefont {Domain}},\
  and\ \bibinfo {author} {\bibfnamefont {J.~R.}\ \bibnamefont {Kermode}},\
  }\bibfield  {title} {\bibinfo {title} {Collinear-spin machine learned
  interatomic potential for {Fe$_{7}$Cr$_{2}$Ni} alloy},\ }\href
  {https://doi.org/10.1103/PhysRevMaterials.8.033804} {\bibfield  {journal}
  {\bibinfo  {journal} {Phys. Rev. Mater.}\ }\textbf {\bibinfo {volume} {8}},\
  \bibinfo {pages} {033804} (\bibinfo {year} {2024})}\BibitemShut {NoStop}%
\bibitem [{\citenamefont {Fellman}\ \emph {et~al.}(2025)\citenamefont
  {Fellman}, \citenamefont {Byggm\"astar}, \citenamefont {Granberg},
  \citenamefont {Nordlund},\ and\ \citenamefont
  {Djurabekova}}]{PhysRevMaterials.9.053807}%
  \BibitemOpen
  \bibfield  {author} {\bibinfo {author} {\bibfnamefont {A.}~\bibnamefont
  {Fellman}}, \bibinfo {author} {\bibfnamefont {J.}~\bibnamefont
  {Byggm\"astar}}, \bibinfo {author} {\bibfnamefont {F.}~\bibnamefont
  {Granberg}}, \bibinfo {author} {\bibfnamefont {K.}~\bibnamefont {Nordlund}},\
  and\ \bibinfo {author} {\bibfnamefont {F.}~\bibnamefont {Djurabekova}},\
  }\bibfield  {title} {\bibinfo {title} {Fast and accurate machine-learned
  interatomic potentials for large-scale simulations of {Cu}, {Al}, and {Ni}},\
  }\href {https://doi.org/10.1103/PhysRevMaterials.9.053807} {\bibfield
  {journal} {\bibinfo  {journal} {Phys. Rev. Mater.}\ }\textbf {\bibinfo
  {volume} {9}},\ \bibinfo {pages} {053807} (\bibinfo {year}
  {2025})}\BibitemShut {NoStop}%
\bibitem [{\citenamefont {Shapeev}(2016)}]{Shapeev2016}%
  \BibitemOpen
  \bibfield  {author} {\bibinfo {author} {\bibfnamefont {A.~V.}\ \bibnamefont
  {Shapeev}},\ }\bibfield  {title} {\bibinfo {title} {Moment tensor potentials:
  A class of systematically improvable interatomic potentials},\ }\href
  {https://doi.org/10.1137/15M1054183} {\bibfield  {journal} {\bibinfo
  {journal} {Multiscale Modeling \& Simulation}\ }\textbf {\bibinfo {volume}
  {14}},\ \bibinfo {pages} {1153} (\bibinfo {year} {2016})}\BibitemShut
  {NoStop}%
\bibitem [{\citenamefont {Novikov}\ \emph {et~al.}(2022)\citenamefont
  {Novikov}, \citenamefont {Grabowski}, \citenamefont {K{\"o}rmann},\ and\
  \citenamefont {Shapeev}}]{Novikov2022}%
  \BibitemOpen
  \bibfield  {author} {\bibinfo {author} {\bibfnamefont {I.}~\bibnamefont
  {Novikov}}, \bibinfo {author} {\bibfnamefont {B.}~\bibnamefont {Grabowski}},
  \bibinfo {author} {\bibfnamefont {F.}~\bibnamefont {K{\"o}rmann}},\ and\
  \bibinfo {author} {\bibfnamefont {A.}~\bibnamefont {Shapeev}},\ }\bibfield
  {title} {\bibinfo {title} {Magnetic moment tensor potentials for collinear
  spin-polarized materials reproduce different magnetic states of bcc {Fe}},\
  }\href {https://doi.org/10.1038/s41524-022-00696-9} {\bibfield  {journal}
  {\bibinfo  {journal} {npj Computational Materials}\ }\textbf {\bibinfo
  {volume} {8}},\ \bibinfo {pages} {13} (\bibinfo {year} {2022})}\BibitemShut
  {NoStop}%
\bibitem [{\citenamefont {Hao}\ \emph {et~al.}(2025)\citenamefont {Hao},
  \citenamefont {Singh}, \citenamefont {Smirnov}, \citenamefont {Johnson},
  \citenamefont {Alman},\ and\ \citenamefont {Gao}}]{10.1063/5.0280935}%
  \BibitemOpen
  \bibfield  {author} {\bibinfo {author} {\bibfnamefont {S.}~\bibnamefont
  {Hao}}, \bibinfo {author} {\bibfnamefont {P.}~\bibnamefont {Singh}}, \bibinfo
  {author} {\bibfnamefont {A.~V.}\ \bibnamefont {Smirnov}}, \bibinfo {author}
  {\bibfnamefont {D.~D.}\ \bibnamefont {Johnson}}, \bibinfo {author}
  {\bibfnamefont {D.~E.}\ \bibnamefont {Alman}},\ and\ \bibinfo {author}
  {\bibfnamefont {M.~C.}\ \bibnamefont {Gao}},\ }\bibfield  {title} {\bibinfo
  {title} {Developing reliable machine learning interatomic potential for
  {Fe–Cr–Ni} austenitic alloys},\ }\href
  {https://doi.org/10.1063/5.0280935} {\bibfield  {journal} {\bibinfo
  {journal} {Journal of Applied Physics}\ }\textbf {\bibinfo {volume} {138}},\
  \bibinfo {pages} {085103} (\bibinfo {year} {2025})}\BibitemShut {NoStop}%
\bibitem [{\citenamefont {Drautz}(2019)}]{drautz2019atomic}%
  \BibitemOpen
  \bibfield  {author} {\bibinfo {author} {\bibfnamefont {R.}~\bibnamefont
  {Drautz}},\ }\bibfield  {title} {\bibinfo {title} {Atomic cluster expansion
  for accurate and transferable interatomic potentials},\ }\href
  {https://doi.org/10.1103/PhysRevB.99.014104} {\bibfield  {journal} {\bibinfo
  {journal} {Phys. Rev. B}\ }\textbf {\bibinfo {volume} {99}},\ \bibinfo
  {pages} {014104} (\bibinfo {year} {2019})}\BibitemShut {NoStop}%
\bibitem [{\citenamefont {Rinaldi}\ \emph {et~al.}(2024)\citenamefont
  {Rinaldi}, \citenamefont {Mrovec}, \citenamefont {Bochkarev}, \citenamefont
  {Lysogorskiy},\ and\ \citenamefont {Drautz}}]{Rinaldi2024}%
  \BibitemOpen
  \bibfield  {author} {\bibinfo {author} {\bibfnamefont {M.}~\bibnamefont
  {Rinaldi}}, \bibinfo {author} {\bibfnamefont {M.}~\bibnamefont {Mrovec}},
  \bibinfo {author} {\bibfnamefont {A.}~\bibnamefont {Bochkarev}}, \bibinfo
  {author} {\bibfnamefont {Y.}~\bibnamefont {Lysogorskiy}},\ and\ \bibinfo
  {author} {\bibfnamefont {R.}~\bibnamefont {Drautz}},\ }\bibfield  {title}
  {\bibinfo {title} {Non-collinear magnetic atomic cluster expansion for
  iron},\ }\href {https://doi.org/10.1038/s41524-024-01196-8} {\bibfield
  {journal} {\bibinfo  {journal} {npj Computational Materials}\ }\textbf
  {\bibinfo {volume} {10}},\ \bibinfo {pages} {12} (\bibinfo {year}
  {2024})}\BibitemShut {NoStop}%
\bibitem [{\citenamefont {Owen}\ \emph {et~al.}(2024)\citenamefont {Owen},
  \citenamefont {Torrisi}, \citenamefont {Xie}, \citenamefont {Batzner},
  \citenamefont {Bystrom}, \citenamefont {Coulter}, \citenamefont {Musaelian},
  \citenamefont {Sun},\ and\ \citenamefont {Kozinsky}}]{Owen2024}%
  \BibitemOpen
  \bibfield  {author} {\bibinfo {author} {\bibfnamefont {C.~J.}\ \bibnamefont
  {Owen}}, \bibinfo {author} {\bibfnamefont {S.~B.}\ \bibnamefont {Torrisi}},
  \bibinfo {author} {\bibfnamefont {Y.}~\bibnamefont {Xie}}, \bibinfo {author}
  {\bibfnamefont {S.}~\bibnamefont {Batzner}}, \bibinfo {author} {\bibfnamefont
  {K.}~\bibnamefont {Bystrom}}, \bibinfo {author} {\bibfnamefont
  {J.}~\bibnamefont {Coulter}}, \bibinfo {author} {\bibfnamefont
  {A.}~\bibnamefont {Musaelian}}, \bibinfo {author} {\bibfnamefont
  {L.}~\bibnamefont {Sun}},\ and\ \bibinfo {author} {\bibfnamefont
  {B.}~\bibnamefont {Kozinsky}},\ }\bibfield  {title} {\bibinfo {title}
  {Complexity of many-body interactions in transition metals via
  machine-learned force fields from the {TM23} data set},\ }\href
  {https://doi.org/10.1038/s41524-024-01264-z} {\bibfield  {journal} {\bibinfo
  {journal} {npj Computational Materials}\ }\textbf {\bibinfo {volume} {10}},\
  \bibinfo {pages} {92} (\bibinfo {year} {2024})}\BibitemShut {NoStop}%
\bibitem [{\citenamefont {Zhang}\ \emph {et~al.}(2018)\citenamefont {Zhang},
  \citenamefont {Han}, \citenamefont {Wang}, \citenamefont {Car},\ and\
  \citenamefont {E}}]{zhang2018deep}%
  \BibitemOpen
  \bibfield  {author} {\bibinfo {author} {\bibfnamefont {L.}~\bibnamefont
  {Zhang}}, \bibinfo {author} {\bibfnamefont {J.}~\bibnamefont {Han}}, \bibinfo
  {author} {\bibfnamefont {H.}~\bibnamefont {Wang}}, \bibinfo {author}
  {\bibfnamefont {R.}~\bibnamefont {Car}},\ and\ \bibinfo {author}
  {\bibfnamefont {W.}~\bibnamefont {E}},\ }\bibfield  {title} {\bibinfo {title}
  {Deep potential molecular dynamics: A scalable model with the accuracy of
  quantum mechanics},\ }\href {https://doi.org/10.1103/PhysRevLett.120.143001}
  {\bibfield  {journal} {\bibinfo  {journal} {Phys. Rev. Lett.}\ }\textbf
  {\bibinfo {volume} {120}},\ \bibinfo {pages} {143001} (\bibinfo {year}
  {2018})}\BibitemShut {NoStop}%
\bibitem [{\citenamefont {Zhang}\ \emph {et~al.}(2020)\citenamefont {Zhang},
  \citenamefont {Wang}, \citenamefont {Chen}, \citenamefont {Zeng},
  \citenamefont {Zhang}, \citenamefont {Wang},\ and\ \citenamefont
  {E}}]{ZHANG2020107206}%
  \BibitemOpen
  \bibfield  {author} {\bibinfo {author} {\bibfnamefont {Y.}~\bibnamefont
  {Zhang}}, \bibinfo {author} {\bibfnamefont {H.}~\bibnamefont {Wang}},
  \bibinfo {author} {\bibfnamefont {W.}~\bibnamefont {Chen}}, \bibinfo {author}
  {\bibfnamefont {J.}~\bibnamefont {Zeng}}, \bibinfo {author} {\bibfnamefont
  {L.}~\bibnamefont {Zhang}}, \bibinfo {author} {\bibfnamefont
  {H.}~\bibnamefont {Wang}},\ and\ \bibinfo {author} {\bibfnamefont
  {W.}~\bibnamefont {E}},\ }\bibfield  {title} {\bibinfo {title} {{DP-GEN}: A
  concurrent learning platform for the generation of reliable deep learning
  based potential energy models},\ }\href
  {https://doi.org/https://doi.org/10.1016/j.cpc.2020.107206} {\bibfield
  {journal} {\bibinfo  {journal} {Computer Physics Communications}\ }\textbf
  {\bibinfo {volume} {253}},\ \bibinfo {pages} {107206} (\bibinfo {year}
  {2020})}\BibitemShut {NoStop}%
\bibitem [{\citenamefont {Gong}\ \emph {et~al.}(2024)\citenamefont {Gong},
  \citenamefont {Li}, \citenamefont {Pattamatta}, \citenamefont {Wen},\ and\
  \citenamefont {Srolovitz}}]{Gong2024}%
  \BibitemOpen
  \bibfield  {author} {\bibinfo {author} {\bibfnamefont {X.}~\bibnamefont
  {Gong}}, \bibinfo {author} {\bibfnamefont {Z.}~\bibnamefont {Li}}, \bibinfo
  {author} {\bibfnamefont {A.~S. L.~S.}\ \bibnamefont {Pattamatta}}, \bibinfo
  {author} {\bibfnamefont {T.}~\bibnamefont {Wen}},\ and\ \bibinfo {author}
  {\bibfnamefont {D.~J.}\ \bibnamefont {Srolovitz}},\ }\bibfield  {title}
  {\bibinfo {title} {An accurate and transferable machine learning interatomic
  potential for nickel},\ }\href {https://doi.org/10.1038/s43246-024-00603-3}
  {\bibfield  {journal} {\bibinfo  {journal} {Communications Materials}\
  }\textbf {\bibinfo {volume} {5}},\ \bibinfo {pages} {157} (\bibinfo {year}
  {2024})}\BibitemShut {NoStop}%
\bibitem [{\citenamefont {Khazieva}\ \emph {et~al.}(2025)\citenamefont
  {Khazieva}, \citenamefont {Chtchelkatchev}, \citenamefont {Katkov},\ and\
  \citenamefont {Ryltsev}}]{913y-p6qf}%
  \BibitemOpen
  \bibfield  {author} {\bibinfo {author} {\bibfnamefont {E.~O.}\ \bibnamefont
  {Khazieva}}, \bibinfo {author} {\bibfnamefont {N.~M.}\ \bibnamefont
  {Chtchelkatchev}}, \bibinfo {author} {\bibfnamefont {N.~N.}\ \bibnamefont
  {Katkov}},\ and\ \bibinfo {author} {\bibfnamefont {R.~E.}\ \bibnamefont
  {Ryltsev}},\ }\bibfield  {title} {\bibinfo {title} {Accuracy and limitations
  of machine-learned interatomic potentials for magnetic systems: A case study
  on {Fe}-{Cr}-{C}},\ }\href {https://doi.org/10.1103/913y-p6qf} {\bibfield
  {journal} {\bibinfo  {journal} {Phys. Rev. E}\ }\textbf {\bibinfo {volume}
  {112}},\ \bibinfo {pages} {055302} (\bibinfo {year} {2025})}\BibitemShut
  {NoStop}%
\bibitem [{\citenamefont {Batzner}\ \emph {et~al.}(2022)\citenamefont
  {Batzner}, \citenamefont {Musaelian}, \citenamefont {Sun}, \citenamefont
  {Geiger}, \citenamefont {Mailoa}, \citenamefont {Kornbluth}, \citenamefont
  {Molinari}, \citenamefont {Smidt},\ and\ \citenamefont
  {Kozinsky}}]{batzner2022e3equivariant}%
  \BibitemOpen
  \bibfield  {author} {\bibinfo {author} {\bibfnamefont {S.}~\bibnamefont
  {Batzner}}, \bibinfo {author} {\bibfnamefont {A.}~\bibnamefont {Musaelian}},
  \bibinfo {author} {\bibfnamefont {L.}~\bibnamefont {Sun}}, \bibinfo {author}
  {\bibfnamefont {M.}~\bibnamefont {Geiger}}, \bibinfo {author} {\bibfnamefont
  {J.~P.}\ \bibnamefont {Mailoa}}, \bibinfo {author} {\bibfnamefont
  {M.}~\bibnamefont {Kornbluth}}, \bibinfo {author} {\bibfnamefont
  {N.}~\bibnamefont {Molinari}}, \bibinfo {author} {\bibfnamefont {T.~E.}\
  \bibnamefont {Smidt}},\ and\ \bibinfo {author} {\bibfnamefont
  {B.}~\bibnamefont {Kozinsky}},\ }\bibfield  {title} {\bibinfo {title}
  {E(3)-equivariant graph neural networks for data-efficient and accurate
  interatomic potentials},\ }\href {https://doi.org/10.1038/s41467-022-29939-5}
  {\bibfield  {journal} {\bibinfo  {journal} {Nature Communications}\ }\textbf
  {\bibinfo {volume} {13}},\ \bibinfo {pages} {2453} (\bibinfo {year}
  {2022})}\BibitemShut {NoStop}%
\bibitem [{\citenamefont {Batatia}\ \emph {et~al.}(2022)\citenamefont
  {Batatia}, \citenamefont {Kov\'acs}, \citenamefont {Simm}, \citenamefont
  {Ortner},\ and\ \citenamefont {Cs\'anyi}}]{batatia2022mace}%
  \BibitemOpen
  \bibfield  {author} {\bibinfo {author} {\bibfnamefont {I.}~\bibnamefont
  {Batatia}}, \bibinfo {author} {\bibfnamefont {D.~P.}\ \bibnamefont
  {Kov\'acs}}, \bibinfo {author} {\bibfnamefont {G.}~\bibnamefont {Simm}},
  \bibinfo {author} {\bibfnamefont {C.}~\bibnamefont {Ortner}},\ and\ \bibinfo
  {author} {\bibfnamefont {G.}~\bibnamefont {Cs\'anyi}},\ }\bibfield  {title}
  {\bibinfo {title} {{MACE}: Higher order equivariant message passing neural
  networks for fast and accurate force fields},\ }in\ \href
  {https://proceedings.neurips.cc/paper_files/paper/2022/file/4a36c3c51af11ed9f34615b81edb5bbc-Paper-Conference.pdf}
  {\emph {\bibinfo {booktitle} {Advances in Neural Information Processing
  Systems}}},\ Vol.~\bibinfo {volume} {35},\ \bibinfo {editor} {edited by\
  \bibinfo {editor} {\bibfnamefont {S.}~\bibnamefont {Koyejo}}, \bibinfo
  {editor} {\bibfnamefont {S.}~\bibnamefont {Mohamed}}, \bibinfo {editor}
  {\bibfnamefont {A.}~\bibnamefont {Agarwal}}, \bibinfo {editor} {\bibfnamefont
  {D.}~\bibnamefont {Belgrave}}, \bibinfo {editor} {\bibfnamefont
  {K.}~\bibnamefont {Cho}},\ and\ \bibinfo {editor} {\bibfnamefont
  {A.}~\bibnamefont {Oh}}}\ (\bibinfo  {publisher} {Curran Associates, Inc.},\
  \bibinfo {year} {2022})\ pp.\ \bibinfo {pages} {11423--11436}\BibitemShut
  {NoStop}%
\bibitem [{\citenamefont {Deng}\ \emph {et~al.}(2023)\citenamefont {Deng},
  \citenamefont {Zhong}, \citenamefont {Jun}, \citenamefont {Riebesell},
  \citenamefont {Han}, \citenamefont {Bartel},\ and\ \citenamefont
  {Ceder}}]{Deng2023}%
  \BibitemOpen
  \bibfield  {author} {\bibinfo {author} {\bibfnamefont {B.}~\bibnamefont
  {Deng}}, \bibinfo {author} {\bibfnamefont {P.}~\bibnamefont {Zhong}},
  \bibinfo {author} {\bibfnamefont {K.}~\bibnamefont {Jun}}, \bibinfo {author}
  {\bibfnamefont {J.}~\bibnamefont {Riebesell}}, \bibinfo {author}
  {\bibfnamefont {K.}~\bibnamefont {Han}}, \bibinfo {author} {\bibfnamefont
  {C.~J.}\ \bibnamefont {Bartel}},\ and\ \bibinfo {author} {\bibfnamefont
  {G.}~\bibnamefont {Ceder}},\ }\bibfield  {title} {\bibinfo {title} {Chgnet as
  a pretrained universal neural network potential for charge-informed atomistic
  modelling},\ }\href {https://doi.org/10.1038/s42256-023-00716-3} {\bibfield
  {journal} {\bibinfo  {journal} {Nature Machine Intelligence}\ }\textbf
  {\bibinfo {volume} {5}},\ \bibinfo {pages} {1031} (\bibinfo {year}
  {2023})}\BibitemShut {NoStop}%
\bibitem [{\citenamefont {Schmidt}\ \emph {et~al.}(2022)\citenamefont
  {Schmidt}, \citenamefont {Wang}, \citenamefont {Cerqueira}, \citenamefont
  {Botti},\ and\ \citenamefont {Marques}}]{Schmidt2022}%
  \BibitemOpen
  \bibfield  {author} {\bibinfo {author} {\bibfnamefont {J.}~\bibnamefont
  {Schmidt}}, \bibinfo {author} {\bibfnamefont {H.-C.}\ \bibnamefont {Wang}},
  \bibinfo {author} {\bibfnamefont {T.~F.~T.}\ \bibnamefont {Cerqueira}},
  \bibinfo {author} {\bibfnamefont {S.}~\bibnamefont {Botti}},\ and\ \bibinfo
  {author} {\bibfnamefont {M.~A.~L.}\ \bibnamefont {Marques}},\ }\bibfield
  {title} {\bibinfo {title} {A dataset of 175{K} stable and metastable
  materials calculated with the {PBEsol} and {SCAN} functionals},\ }\href
  {https://doi.org/10.1038/s41597-022-01177-w} {\bibfield  {journal} {\bibinfo
  {journal} {Scientific Data}\ }\textbf {\bibinfo {volume} {9}},\ \bibinfo
  {pages} {64} (\bibinfo {year} {2022})}\BibitemShut {NoStop}%
\bibitem [{\citenamefont {Barroso-Luque}\ \emph {et~al.}(2024)\citenamefont
  {Barroso-Luque}, \citenamefont {Shuaibi}, \citenamefont {Fu}, \citenamefont
  {Wood}, \citenamefont {Dzamba}, \citenamefont {Gao}, \citenamefont {Rizvi},
  \citenamefont {Zitnick},\ and\ \citenamefont
  {Ulissi}}]{barrosoluque2024openmaterials2024omat24}%
  \BibitemOpen
  \bibfield  {author} {\bibinfo {author} {\bibfnamefont {L.}~\bibnamefont
  {Barroso-Luque}}, \bibinfo {author} {\bibfnamefont {M.}~\bibnamefont
  {Shuaibi}}, \bibinfo {author} {\bibfnamefont {X.}~\bibnamefont {Fu}},
  \bibinfo {author} {\bibfnamefont {B.~M.}\ \bibnamefont {Wood}}, \bibinfo
  {author} {\bibfnamefont {M.}~\bibnamefont {Dzamba}}, \bibinfo {author}
  {\bibfnamefont {M.}~\bibnamefont {Gao}}, \bibinfo {author} {\bibfnamefont
  {A.}~\bibnamefont {Rizvi}}, \bibinfo {author} {\bibfnamefont {C.~L.}\
  \bibnamefont {Zitnick}},\ and\ \bibinfo {author} {\bibfnamefont {Z.~W.}\
  \bibnamefont {Ulissi}},\ }\href {https://arxiv.org/abs/2410.12771} {\bibinfo
  {title} {Open materials 2024 (omat24) inorganic materials dataset and
  models}} (\bibinfo {year} {2024}),\ \Eprint
  {https://arxiv.org/abs/2410.12771} {arXiv:2410.12771 [cond-mat.mtrl-sci]}
  \BibitemShut {NoStop}%
\bibitem [{\citenamefont {Kaplan}\ \emph {et~al.}(2025)\citenamefont {Kaplan},
  \citenamefont {Liu}, \citenamefont {Qi}, \citenamefont {Ko}, \citenamefont
  {Deng}, \citenamefont {Riebesell}, \citenamefont {Ceder}, \citenamefont
  {Persson},\ and\ \citenamefont
  {Ong}}]{kaplan2025foundationalpotentialenergysurface}%
  \BibitemOpen
  \bibfield  {author} {\bibinfo {author} {\bibfnamefont {A.~D.}\ \bibnamefont
  {Kaplan}}, \bibinfo {author} {\bibfnamefont {R.}~\bibnamefont {Liu}},
  \bibinfo {author} {\bibfnamefont {J.}~\bibnamefont {Qi}}, \bibinfo {author}
  {\bibfnamefont {T.~W.}\ \bibnamefont {Ko}}, \bibinfo {author} {\bibfnamefont
  {B.}~\bibnamefont {Deng}}, \bibinfo {author} {\bibfnamefont {J.}~\bibnamefont
  {Riebesell}}, \bibinfo {author} {\bibfnamefont {G.}~\bibnamefont {Ceder}},
  \bibinfo {author} {\bibfnamefont {K.~A.}\ \bibnamefont {Persson}},\ and\
  \bibinfo {author} {\bibfnamefont {S.~P.}\ \bibnamefont {Ong}},\ }\href
  {https://arxiv.org/abs/2503.04070} {\bibinfo {title} {A foundational
  potential energy surface dataset for materials}} (\bibinfo {year} {2025}),\
  \Eprint {https://arxiv.org/abs/2503.04070} {arXiv:2503.04070
  [cond-mat.mtrl-sci]} \BibitemShut {NoStop}%
\bibitem [{\citenamefont {van~de Walle}\ \emph {et~al.}(2002)\citenamefont
  {van~de Walle}, \citenamefont {Asta},\ and\ \citenamefont
  {Ceder}}]{avdw:atat}%
  \BibitemOpen
  \bibfield  {author} {\bibinfo {author} {\bibfnamefont {A.}~\bibnamefont
  {van~de Walle}}, \bibinfo {author} {\bibfnamefont {M.~D.}\ \bibnamefont
  {Asta}},\ and\ \bibinfo {author} {\bibfnamefont {G.}~\bibnamefont {Ceder}},\
  }\bibfield  {title} {\bibinfo {title} {{T}he {A}lloy {T}heoretic {A}utomated
  {T}oolkit: {A} user guide},\ }\href
  {https://doi.org/10.1016/S0364-5916(02)80006-2} {\bibfield  {journal}
  {\bibinfo  {journal} {Calphad}\ }\textbf {\bibinfo {volume} {26}},\ \bibinfo
  {pages} {539} (\bibinfo {year} {2002})}\BibitemShut {NoStop}%
\bibitem [{\citenamefont {{van de Walle}}\ \emph {et~al.}(2013)\citenamefont
  {{van de Walle}}, \citenamefont {Tiwary}, \citenamefont {{de Jong}},
  \citenamefont {Olmsted}, \citenamefont {Asta}, \citenamefont {Dick},
  \citenamefont {Shin}, \citenamefont {Wang}, \citenamefont {Chen},\ and\
  \citenamefont {Liu}}]{VANDEWALLE201313}%
  \BibitemOpen
  \bibfield  {author} {\bibinfo {author} {\bibfnamefont {A.}~\bibnamefont {{van
  de Walle}}}, \bibinfo {author} {\bibfnamefont {P.}~\bibnamefont {Tiwary}},
  \bibinfo {author} {\bibfnamefont {M.}~\bibnamefont {{de Jong}}}, \bibinfo
  {author} {\bibfnamefont {D.}~\bibnamefont {Olmsted}}, \bibinfo {author}
  {\bibfnamefont {M.}~\bibnamefont {Asta}}, \bibinfo {author} {\bibfnamefont
  {A.}~\bibnamefont {Dick}}, \bibinfo {author} {\bibfnamefont {D.}~\bibnamefont
  {Shin}}, \bibinfo {author} {\bibfnamefont {Y.}~\bibnamefont {Wang}}, \bibinfo
  {author} {\bibfnamefont {L.-Q.}\ \bibnamefont {Chen}},\ and\ \bibinfo
  {author} {\bibfnamefont {Z.-K.}\ \bibnamefont {Liu}},\ }\bibfield  {title}
  {\bibinfo {title} {Efficient stochastic generation of special quasirandom
  structures},\ }\href
  {https://doi.org/https://doi.org/10.1016/j.calphad.2013.06.006} {\bibfield
  {journal} {\bibinfo  {journal} {Calphad}\ }\textbf {\bibinfo {volume} {42}},\
  \bibinfo {pages} {13} (\bibinfo {year} {2013})}\BibitemShut {NoStop}%
\bibitem [{\citenamefont {Cowley}(1965)}]{PhysRev.138.A1384}%
  \BibitemOpen
  \bibfield  {author} {\bibinfo {author} {\bibfnamefont {J.~M.}\ \bibnamefont
  {Cowley}},\ }\bibfield  {title} {\bibinfo {title} {Short-range order and
  long-range order parameters},\ }\href
  {https://doi.org/10.1103/PhysRev.138.A1384} {\bibfield  {journal} {\bibinfo
  {journal} {Phys. Rev.}\ }\textbf {\bibinfo {volume} {138}},\ \bibinfo {pages}
  {A1384} (\bibinfo {year} {1965})}\BibitemShut {NoStop}%
\bibitem [{\citenamefont {Cowley}(1950)}]{PhysRev.77.669}%
  \BibitemOpen
  \bibfield  {author} {\bibinfo {author} {\bibfnamefont {J.~M.}\ \bibnamefont
  {Cowley}},\ }\bibfield  {title} {\bibinfo {title} {An approximate theory of
  order in alloys},\ }\href {https://doi.org/10.1103/PhysRev.77.669} {\bibfield
   {journal} {\bibinfo  {journal} {Phys. Rev.}\ }\textbf {\bibinfo {volume}
  {77}},\ \bibinfo {pages} {669} (\bibinfo {year} {1950})}\BibitemShut
  {NoStop}%
\bibitem [{\citenamefont {Kresse}\ and\ \citenamefont
  {Hafner}(1993)}]{PhysRevB.47.558}%
  \BibitemOpen
  \bibfield  {author} {\bibinfo {author} {\bibfnamefont {G.}~\bibnamefont
  {Kresse}}\ and\ \bibinfo {author} {\bibfnamefont {J.}~\bibnamefont
  {Hafner}},\ }\bibfield  {title} {\bibinfo {title} {Ab initio molecular
  dynamics for liquid metals},\ }\href
  {https://doi.org/10.1103/PhysRevB.47.558} {\bibfield  {journal} {\bibinfo
  {journal} {Phys. Rev. B}\ }\textbf {\bibinfo {volume} {47}},\ \bibinfo
  {pages} {558} (\bibinfo {year} {1993})}\BibitemShut {NoStop}%
\bibitem [{\citenamefont {Kresse}\ and\ \citenamefont
  {Joubert}(1999)}]{PhysRevB.59.1758}%
  \BibitemOpen
  \bibfield  {author} {\bibinfo {author} {\bibfnamefont {G.}~\bibnamefont
  {Kresse}}\ and\ \bibinfo {author} {\bibfnamefont {D.}~\bibnamefont
  {Joubert}},\ }\bibfield  {title} {\bibinfo {title} {From ultrasoft
  pseudopotentials to the projector augmented-wave method},\ }\href
  {https://doi.org/10.1103/PhysRevB.59.1758} {\bibfield  {journal} {\bibinfo
  {journal} {Phys. Rev. B}\ }\textbf {\bibinfo {volume} {59}},\ \bibinfo
  {pages} {1758} (\bibinfo {year} {1999})}\BibitemShut {NoStop}%
\bibitem [{\citenamefont {Kresse}\ and\ \citenamefont
  {Furthm{\"u}ller}(1996{\natexlab{a}})}]{KRESSE199615}%
  \BibitemOpen
  \bibfield  {author} {\bibinfo {author} {\bibfnamefont {G.}~\bibnamefont
  {Kresse}}\ and\ \bibinfo {author} {\bibfnamefont {J.}~\bibnamefont
  {Furthm{\"u}ller}},\ }\bibfield  {title} {\bibinfo {title} {Efficiency of
  ab-initio total energy calculations for metals and semiconductors using a
  plane-wave basis set},\ }\href
  {https://doi.org/https://doi.org/10.1016/0927-0256(96)00008-0} {\bibfield
  {journal} {\bibinfo  {journal} {Computational Materials Science}\ }\textbf
  {\bibinfo {volume} {6}},\ \bibinfo {pages} {15 } (\bibinfo {year}
  {1996}{\natexlab{a}})}\BibitemShut {NoStop}%
\bibitem [{\citenamefont {Kresse}\ and\ \citenamefont
  {Furthm{\"u}ller}(1996{\natexlab{b}})}]{PhysRevB.54.11169}%
  \BibitemOpen
  \bibfield  {author} {\bibinfo {author} {\bibfnamefont {G.}~\bibnamefont
  {Kresse}}\ and\ \bibinfo {author} {\bibfnamefont {J.}~\bibnamefont
  {Furthm{\"u}ller}},\ }\bibfield  {title} {\bibinfo {title} {Efficient
  iterative schemes for ab initio total-energy calculations using a plane-wave
  basis set},\ }\href {https://doi.org/10.1103/PhysRevB.54.11169} {\bibfield
  {journal} {\bibinfo  {journal} {Phys. Rev. B}\ }\textbf {\bibinfo {volume}
  {54}},\ \bibinfo {pages} {11169} (\bibinfo {year}
  {1996}{\natexlab{b}})}\BibitemShut {NoStop}%
\bibitem [{\citenamefont {Perdew}\ \emph {et~al.}(1996)\citenamefont {Perdew},
  \citenamefont {Burke},\ and\ \citenamefont
  {Ernzerhof}}]{perdew1996generalized}%
  \BibitemOpen
  \bibfield  {author} {\bibinfo {author} {\bibfnamefont {J.~P.}\ \bibnamefont
  {Perdew}}, \bibinfo {author} {\bibfnamefont {K.}~\bibnamefont {Burke}},\ and\
  \bibinfo {author} {\bibfnamefont {M.}~\bibnamefont {Ernzerhof}},\ }\bibfield
  {title} {\bibinfo {title} {Generalized gradient approximation made simple},\
  }\href {https://doi.org/10.1103/PhysRevLett.77.3865} {\bibfield  {journal}
  {\bibinfo  {journal} {Phys. Rev. Lett.}\ }\textbf {\bibinfo {volume} {77}},\
  \bibinfo {pages} {3865} (\bibinfo {year} {1996})}\BibitemShut {NoStop}%
\bibitem [{\citenamefont {Tesch}\ and\ \citenamefont
  {Kowalski}(2022)}]{PhysRevB.105.195153}%
  \BibitemOpen
  \bibfield  {author} {\bibinfo {author} {\bibfnamefont {R.}~\bibnamefont
  {Tesch}}\ and\ \bibinfo {author} {\bibfnamefont {P.~M.}\ \bibnamefont
  {Kowalski}},\ }\bibfield  {title} {\bibinfo {title} {Hubbard $u$ parameters
  for transition metals from first principles},\ }\href
  {https://doi.org/10.1103/PhysRevB.105.195153} {\bibfield  {journal} {\bibinfo
   {journal} {Phys. Rev. B}\ }\textbf {\bibinfo {volume} {105}},\ \bibinfo
  {pages} {195153} (\bibinfo {year} {2022})}\BibitemShut {NoStop}%
\bibitem [{\citenamefont {Hausoel}\ \emph {et~al.}(2017)\citenamefont
  {Hausoel}, \citenamefont {Karolak}, \citenamefont
  {{\c{S}}a{\c{s}}$\iota$o{\u{g}}lu}, \citenamefont {Lichtenstein},
  \citenamefont {Held}, \citenamefont {Katanin}, \citenamefont {Toschi},\ and\
  \citenamefont {Sangiovanni}}]{Hausoel2017}%
  \BibitemOpen
  \bibfield  {author} {\bibinfo {author} {\bibfnamefont {A.}~\bibnamefont
  {Hausoel}}, \bibinfo {author} {\bibfnamefont {M.}~\bibnamefont {Karolak}},
  \bibinfo {author} {\bibfnamefont {E.}~\bibnamefont
  {{\c{S}}a{\c{s}}$\iota$o{\u{g}}lu}}, \bibinfo {author} {\bibfnamefont
  {A.}~\bibnamefont {Lichtenstein}}, \bibinfo {author} {\bibfnamefont
  {K.}~\bibnamefont {Held}}, \bibinfo {author} {\bibfnamefont {A.}~\bibnamefont
  {Katanin}}, \bibinfo {author} {\bibfnamefont {A.}~\bibnamefont {Toschi}},\
  and\ \bibinfo {author} {\bibfnamefont {G.}~\bibnamefont {Sangiovanni}},\
  }\bibfield  {title} {\bibinfo {title} {Local magnetic moments in iron and
  nickel at ambient and earth's core conditions},\ }\href
  {https://doi.org/10.1038/ncomms16062} {\bibfield  {journal} {\bibinfo
  {journal} {Nature Communications}\ }\textbf {\bibinfo {volume} {8}},\
  \bibinfo {pages} {16062} (\bibinfo {year} {2017})}\BibitemShut {NoStop}%
\bibitem [{\citenamefont {Warford}\ \emph {et~al.}(2026)\citenamefont
  {Warford}, \citenamefont {Thiemann},\ and\ \citenamefont
  {Csányi}}]{warford2026betteruimpactselective}%
  \BibitemOpen
  \bibfield  {author} {\bibinfo {author} {\bibfnamefont {T.~D.}\ \bibnamefont
  {Warford}}, \bibinfo {author} {\bibfnamefont {F.~L.}\ \bibnamefont
  {Thiemann}},\ and\ \bibinfo {author} {\bibfnamefont {G.}~\bibnamefont
  {Csányi}},\ }\bibfield  {title} {\bibinfo {title} {Better without {U}:
  Impact of selective {Hubbard U} correction on foundational {MLIPs}},\
  }\bibfield  {journal} {\bibinfo  {journal} {Machine Learning: Science and
  Technology}\ }\href {https://doi.org/10.1088/2632-2153/ae6be5}
  {10.1088/2632-2153/ae6be5} (\bibinfo {year} {2026})\BibitemShut {NoStop}%
\bibitem [{\citenamefont {Xiong}\ \emph {et~al.}(2011)\citenamefont {Xiong},
  \citenamefont {Zhang}, \citenamefont {Vitos},\ and\ \citenamefont
  {Selleby}}]{xiong2011magnetic}%
  \BibitemOpen
  \bibfield  {author} {\bibinfo {author} {\bibfnamefont {W.}~\bibnamefont
  {Xiong}}, \bibinfo {author} {\bibfnamefont {H.}~\bibnamefont {Zhang}},
  \bibinfo {author} {\bibfnamefont {L.}~\bibnamefont {Vitos}},\ and\ \bibinfo
  {author} {\bibfnamefont {M.}~\bibnamefont {Selleby}},\ }\bibfield  {title}
  {\bibinfo {title} {Magnetic phase diagram of the {Fe--Ni} system},\ }\href
  {https://doi.org/https://doi.org/10.1016/j.actamat.2010.09.055} {\bibfield
  {journal} {\bibinfo  {journal} {Acta Materialia}\ }\textbf {\bibinfo {volume}
  {59}},\ \bibinfo {pages} {521} (\bibinfo {year} {2011})}\BibitemShut
  {NoStop}%
\bibitem [{\citenamefont {Peschard}(1925)}]{peschard1925contribution}%
  \BibitemOpen
  \bibfield  {author} {\bibinfo {author} {\bibnamefont {Peschard}},\ }\bibfield
   {title} {\bibinfo {title} {Contribution {\`a} l’{\'e}tude des
  ferro-nickels},\ }\href@noop {} {\bibfield  {journal} {\bibinfo  {journal}
  {Revue de M{\'e}tallurgie}\ }\textbf {\bibinfo {volume} {22}},\ \bibinfo
  {pages} {490} (\bibinfo {year} {1925})}\BibitemShut {NoStop}%
\bibitem [{\citenamefont {Crangle}\ \emph {et~al.}(1963)\citenamefont
  {Crangle}, \citenamefont {Hallam},\ and\ \citenamefont
  {Sucksmith}}]{crangle1963magnetization}%
  \BibitemOpen
  \bibfield  {author} {\bibinfo {author} {\bibfnamefont {J.}~\bibnamefont
  {Crangle}}, \bibinfo {author} {\bibfnamefont {G.~C.}\ \bibnamefont
  {Hallam}},\ and\ \bibinfo {author} {\bibfnamefont {W.}~\bibnamefont
  {Sucksmith}},\ }\bibfield  {title} {\bibinfo {title} {The magnetization of
  face-centred cubic and body-centred cubic iron + nickel alloys},\ }\href
  {https://doi.org/10.1098/rspa.1963.0045} {\bibfield  {journal} {\bibinfo
  {journal} {Proceedings of the Royal Society of London. Series A. Mathematical
  and Physical Sciences}\ }\textbf {\bibinfo {volume} {272}},\ \bibinfo {pages}
  {119} (\bibinfo {year} {1963})}\BibitemShut {NoStop}%
\bibitem [{\citenamefont {Tian}\ \emph {et~al.}(2005)\citenamefont {Tian},
  \citenamefont {Qian}, \citenamefont {Wu}, \citenamefont {He}, \citenamefont
  {Wu}, \citenamefont {Tang}, \citenamefont {Yin}, \citenamefont {Shi},
  \citenamefont {Dong}, \citenamefont {Jin}, \citenamefont {Jiang},
  \citenamefont {Liu}, \citenamefont {Qian}, \citenamefont {Sun}, \citenamefont
  {Wang}, \citenamefont {Rossi}, \citenamefont {Qiu},\ and\ \citenamefont
  {Shi}}]{PhysRevLett.94.137210}%
  \BibitemOpen
  \bibfield  {author} {\bibinfo {author} {\bibfnamefont {C.~S.}\ \bibnamefont
  {Tian}}, \bibinfo {author} {\bibfnamefont {D.}~\bibnamefont {Qian}}, \bibinfo
  {author} {\bibfnamefont {D.}~\bibnamefont {Wu}}, \bibinfo {author}
  {\bibfnamefont {R.~H.}\ \bibnamefont {He}}, \bibinfo {author} {\bibfnamefont
  {Y.~Z.}\ \bibnamefont {Wu}}, \bibinfo {author} {\bibfnamefont {W.~X.}\
  \bibnamefont {Tang}}, \bibinfo {author} {\bibfnamefont {L.~F.}\ \bibnamefont
  {Yin}}, \bibinfo {author} {\bibfnamefont {Y.~S.}\ \bibnamefont {Shi}},
  \bibinfo {author} {\bibfnamefont {G.~S.}\ \bibnamefont {Dong}}, \bibinfo
  {author} {\bibfnamefont {X.~F.}\ \bibnamefont {Jin}}, \bibinfo {author}
  {\bibfnamefont {X.~M.}\ \bibnamefont {Jiang}}, \bibinfo {author}
  {\bibfnamefont {F.~Q.}\ \bibnamefont {Liu}}, \bibinfo {author} {\bibfnamefont
  {H.~J.}\ \bibnamefont {Qian}}, \bibinfo {author} {\bibfnamefont
  {K.}~\bibnamefont {Sun}}, \bibinfo {author} {\bibfnamefont {L.~M.}\
  \bibnamefont {Wang}}, \bibinfo {author} {\bibfnamefont {G.}~\bibnamefont
  {Rossi}}, \bibinfo {author} {\bibfnamefont {Z.~Q.}\ \bibnamefont {Qiu}},\
  and\ \bibinfo {author} {\bibfnamefont {J.}~\bibnamefont {Shi}},\ }\bibfield
  {title} {\bibinfo {title} {Body-centered-cubic {Ni} and its magnetic
  properties},\ }\href {https://doi.org/10.1103/PhysRevLett.94.137210}
  {\bibfield  {journal} {\bibinfo  {journal} {Phys. Rev. Lett.}\ }\textbf
  {\bibinfo {volume} {94}},\ \bibinfo {pages} {137210} (\bibinfo {year}
  {2005})}\BibitemShut {NoStop}%
\bibitem [{\citenamefont {Sumiyama}\ \emph {et~al.}(1983)\citenamefont
  {Sumiyama}, \citenamefont {Kadono},\ and\ \citenamefont
  {Nakamura}}]{sumiyama1983metastable}%
  \BibitemOpen
  \bibfield  {author} {\bibinfo {author} {\bibfnamefont {K.}~\bibnamefont
  {Sumiyama}}, \bibinfo {author} {\bibfnamefont {M.}~\bibnamefont {Kadono}},\
  and\ \bibinfo {author} {\bibfnamefont {Y.}~\bibnamefont {Nakamura}},\
  }\bibfield  {title} {\bibinfo {title} {Metastable bcc phase in sputtered
  {Fe}--{Ni} alloys},\ }\href {https://doi.org/10.2320/matertrans1960.24.190}
  {\bibfield  {journal} {\bibinfo  {journal} {Transactions of the Japan
  institute of metals}\ }\textbf {\bibinfo {volume} {24}},\ \bibinfo {pages}
  {190} (\bibinfo {year} {1983})}\BibitemShut {NoStop}%
\bibitem [{\citenamefont {Asano}(1969)}]{asano1969magnetism}%
  \BibitemOpen
  \bibfield  {author} {\bibinfo {author} {\bibfnamefont {H.}~\bibnamefont
  {Asano}},\ }\bibfield  {title} {\bibinfo {title} {Magnetism of $\gamma$
  {Fe}-{Ni} invar alloys with low nickel concentration},\ }\href
  {https://doi.org/10.1143/JPSJ.27.542} {\bibfield  {journal} {\bibinfo
  {journal} {Journal of the Physical Society of Japan}\ }\textbf {\bibinfo
  {volume} {27}},\ \bibinfo {pages} {542} (\bibinfo {year} {1969})}\BibitemShut
  {NoStop}%
\bibitem [{\citenamefont {Bando}(1964)}]{bando1964magnetization}%
  \BibitemOpen
  \bibfield  {author} {\bibinfo {author} {\bibfnamefont {Y.}~\bibnamefont
  {Bando}},\ }\bibfield  {title} {\bibinfo {title} {The magnetization of face
  centered cubic iron-nickel alloys in the vicinity of invar region},\ }\href
  {https://doi.org/10.1143/JPSJ.19.237} {\bibfield  {journal} {\bibinfo
  {journal} {Journal of the Physical Society of Japan}\ }\textbf {\bibinfo
  {volume} {19}},\ \bibinfo {pages} {237} (\bibinfo {year} {1964})}\BibitemShut
  {NoStop}%
\bibitem [{\citenamefont {Glaubitz}\ \emph {et~al.}(2011)\citenamefont
  {Glaubitz}, \citenamefont {Buschhorn}, \citenamefont {Br{\"u}ssing},
  \citenamefont {Abrudan},\ and\ \citenamefont {Zabel}}]{Glaubitz_2011}%
  \BibitemOpen
  \bibfield  {author} {\bibinfo {author} {\bibfnamefont {B.}~\bibnamefont
  {Glaubitz}}, \bibinfo {author} {\bibfnamefont {S.}~\bibnamefont {Buschhorn}},
  \bibinfo {author} {\bibfnamefont {F.}~\bibnamefont {Br{\"u}ssing}}, \bibinfo
  {author} {\bibfnamefont {R.}~\bibnamefont {Abrudan}},\ and\ \bibinfo {author}
  {\bibfnamefont {H.}~\bibnamefont {Zabel}},\ }\bibfield  {title} {\bibinfo
  {title} {Development of magnetic moments in {Fe$_{1-x}$}{Ni$_{x}$}-alloys},\
  }\href {https://doi.org/10.1088/0953-8984/23/25/254210} {\bibfield  {journal}
  {\bibinfo  {journal} {Journal of Physics: Condensed Matter}\ }\textbf
  {\bibinfo {volume} {23}},\ \bibinfo {pages} {254210} (\bibinfo {year}
  {2011})}\BibitemShut {NoStop}%
\bibitem [{\citenamefont {Tsoukalas}\ and\ \citenamefont
  {Melidis}(1987)}]{Tsoukalas1987}%
  \BibitemOpen
  \bibfield  {author} {\bibinfo {author} {\bibfnamefont {I.~A.}\ \bibnamefont
  {Tsoukalas}}\ and\ \bibinfo {author} {\bibfnamefont {K.~G.}\ \bibnamefont
  {Melidis}},\ }\bibfield  {title} {\bibinfo {title} {Measurements on the
  magnetic properties of {Fe--Ni} alloys},\ }\href
  {https://doi.org/doi:10.1515/ijmr-1987-780705} {\bibfield  {journal}
  {\bibinfo  {journal} {International Journal of Materials Research}\ }\textbf
  {\bibinfo {volume} {78}},\ \bibinfo {pages} {498} (\bibinfo {year}
  {1987})}\BibitemShut {NoStop}%
\bibitem [{\citenamefont {Kalinin}\ \emph {et~al.}(1972)\citenamefont
  {Kalinin}, \citenamefont {Kornyakov}, \citenamefont {Khomenko}, \citenamefont
  {Dunaev},\ and\ \citenamefont {Yurchikov}}]{Kalinin1972En}%
  \BibitemOpen
  \bibfield  {author} {\bibinfo {author} {\bibfnamefont {V.~M.}\ \bibnamefont
  {Kalinin}}, \bibinfo {author} {\bibfnamefont {V.~A.}\ \bibnamefont
  {Kornyakov}}, \bibinfo {author} {\bibfnamefont {O.~A.}\ \bibnamefont
  {Khomenko}}, \bibinfo {author} {\bibfnamefont {F.~N.}\ \bibnamefont
  {Dunaev}},\ and\ \bibinfo {author} {\bibfnamefont {E.~E.}\ \bibnamefont
  {Yurchikov}},\ }\bibfield  {title} {\bibinfo {title} {M\"ossbauer
  investigation of iron-nickel invar alloys},\ }\href@noop {} {\bibfield
  {journal} {\bibinfo  {journal} {Bulletin of the Academy of Sciences of the
  USSR, Physical Series}\ }\textbf {\bibinfo {volume} {36}},\ \bibinfo {pages}
  {1420} (\bibinfo {year} {1972})}\BibitemShut {NoStop}%
\bibitem [{\citenamefont {Biewald}(2020)}]{wandb}%
  \BibitemOpen
  \bibfield  {author} {\bibinfo {author} {\bibfnamefont {L.}~\bibnamefont
  {Biewald}},\ }\href {https://www.wandb.com/} {\bibinfo {title} {Experiment
  tracking with weights and biases}} (\bibinfo {year} {2020}),\ \bibinfo {note}
  {software available from wandb.com}\BibitemShut {NoStop}%
\bibitem [{\citenamefont {Riebesell}\ \emph {et~al.}(2025)\citenamefont
  {Riebesell}, \citenamefont {Goodall}, \citenamefont {Benner}, \citenamefont
  {Chiang}, \citenamefont {Deng}, \citenamefont {Ceder}, \citenamefont {Asta},
  \citenamefont {Lee}, \citenamefont {Jain},\ and\ \citenamefont
  {Persson}}]{riebesell2025}%
  \BibitemOpen
  \bibfield  {author} {\bibinfo {author} {\bibfnamefont {J.}~\bibnamefont
  {Riebesell}}, \bibinfo {author} {\bibfnamefont {R.~E.~A.}\ \bibnamefont
  {Goodall}}, \bibinfo {author} {\bibfnamefont {P.}~\bibnamefont {Benner}},
  \bibinfo {author} {\bibfnamefont {Y.}~\bibnamefont {Chiang}}, \bibinfo
  {author} {\bibfnamefont {B.}~\bibnamefont {Deng}}, \bibinfo {author}
  {\bibfnamefont {G.}~\bibnamefont {Ceder}}, \bibinfo {author} {\bibfnamefont
  {M.}~\bibnamefont {Asta}}, \bibinfo {author} {\bibfnamefont {A.~A.}\
  \bibnamefont {Lee}}, \bibinfo {author} {\bibfnamefont {A.}~\bibnamefont
  {Jain}},\ and\ \bibinfo {author} {\bibfnamefont {K.~A.}\ \bibnamefont
  {Persson}},\ }\bibfield  {title} {\bibinfo {title} {A framework to evaluate
  machine learning crystal stability predictions},\ }\href
  {https://doi.org/10.1038/s42256-025-01055-1} {\bibfield  {journal} {\bibinfo
  {journal} {Nature Machine Intelligence}\ }\textbf {\bibinfo {volume} {7}},\
  \bibinfo {pages} {836} (\bibinfo {year} {2025})}\BibitemShut {NoStop}%
\bibitem [{\citenamefont {Hjorth~Larsen}\ \emph {et~al.}(2017)\citenamefont
  {Hjorth~Larsen}, \citenamefont {J{\o}rgen~Mortensen}, \citenamefont
  {Blomqvist}, \citenamefont {Castelli}, \citenamefont {Christensen},
  \citenamefont {Du{\l}ak}, \citenamefont {Friis}, \citenamefont {Groves},
  \citenamefont {Hammer}, \citenamefont {Hargus}, \citenamefont {Hermes},
  \citenamefont {Jennings}, \citenamefont {Bjerre~Jensen}, \citenamefont
  {Kermode}, \citenamefont {Kitchin}, \citenamefont {Leonhard~Kolsbjerg},
  \citenamefont {Kubal}, \citenamefont {Kaasbjerg}, \citenamefont {Lysgaard},
  \citenamefont {Bergmann~Maronsson}, \citenamefont {Maxson}, \citenamefont
  {Olsen}, \citenamefont {Pastewka}, \citenamefont {Peterson}, \citenamefont
  {Rostgaard}, \citenamefont {Schi{\o}tz}, \citenamefont {Sch{\"u}tt},
  \citenamefont {Strange}, \citenamefont {Thygesen}, \citenamefont {Vegge},
  \citenamefont {Vilhelmsen}, \citenamefont {Walter}, \citenamefont {Zeng},\
  and\ \citenamefont {Jacobsen}}]{Hjorth_Larsen_2017}%
  \BibitemOpen
  \bibfield  {author} {\bibinfo {author} {\bibfnamefont {A.}~\bibnamefont
  {Hjorth~Larsen}}, \bibinfo {author} {\bibfnamefont {J.}~\bibnamefont
  {J{\o}rgen~Mortensen}}, \bibinfo {author} {\bibfnamefont {J.}~\bibnamefont
  {Blomqvist}}, \bibinfo {author} {\bibfnamefont {I.~E.}\ \bibnamefont
  {Castelli}}, \bibinfo {author} {\bibfnamefont {R.}~\bibnamefont
  {Christensen}}, \bibinfo {author} {\bibfnamefont {M.}~\bibnamefont
  {Du{\l}ak}}, \bibinfo {author} {\bibfnamefont {J.}~\bibnamefont {Friis}},
  \bibinfo {author} {\bibfnamefont {M.~N.}\ \bibnamefont {Groves}}, \bibinfo
  {author} {\bibfnamefont {B.}~\bibnamefont {Hammer}}, \bibinfo {author}
  {\bibfnamefont {C.}~\bibnamefont {Hargus}}, \bibinfo {author} {\bibfnamefont
  {E.~D.}\ \bibnamefont {Hermes}}, \bibinfo {author} {\bibfnamefont {P.~C.}\
  \bibnamefont {Jennings}}, \bibinfo {author} {\bibfnamefont {P.}~\bibnamefont
  {Bjerre~Jensen}}, \bibinfo {author} {\bibfnamefont {J.}~\bibnamefont
  {Kermode}}, \bibinfo {author} {\bibfnamefont {J.~R.}\ \bibnamefont
  {Kitchin}}, \bibinfo {author} {\bibfnamefont {E.}~\bibnamefont
  {Leonhard~Kolsbjerg}}, \bibinfo {author} {\bibfnamefont {J.}~\bibnamefont
  {Kubal}}, \bibinfo {author} {\bibfnamefont {K.}~\bibnamefont {Kaasbjerg}},
  \bibinfo {author} {\bibfnamefont {S.}~\bibnamefont {Lysgaard}}, \bibinfo
  {author} {\bibfnamefont {J.}~\bibnamefont {Bergmann~Maronsson}}, \bibinfo
  {author} {\bibfnamefont {T.}~\bibnamefont {Maxson}}, \bibinfo {author}
  {\bibfnamefont {T.}~\bibnamefont {Olsen}}, \bibinfo {author} {\bibfnamefont
  {L.}~\bibnamefont {Pastewka}}, \bibinfo {author} {\bibfnamefont
  {A.}~\bibnamefont {Peterson}}, \bibinfo {author} {\bibfnamefont
  {C.}~\bibnamefont {Rostgaard}}, \bibinfo {author} {\bibfnamefont
  {J.}~\bibnamefont {Schi{\o}tz}}, \bibinfo {author} {\bibfnamefont
  {O.}~\bibnamefont {Sch{\"u}tt}}, \bibinfo {author} {\bibfnamefont
  {M.}~\bibnamefont {Strange}}, \bibinfo {author} {\bibfnamefont {K.~S.}\
  \bibnamefont {Thygesen}}, \bibinfo {author} {\bibfnamefont {T.}~\bibnamefont
  {Vegge}}, \bibinfo {author} {\bibfnamefont {L.}~\bibnamefont {Vilhelmsen}},
  \bibinfo {author} {\bibfnamefont {M.}~\bibnamefont {Walter}}, \bibinfo
  {author} {\bibfnamefont {Z.}~\bibnamefont {Zeng}},\ and\ \bibinfo {author}
  {\bibfnamefont {K.~W.}\ \bibnamefont {Jacobsen}},\ }\bibfield  {title}
  {\bibinfo {title} {The atomic simulation environment—a python library for
  working with atoms},\ }\href {https://doi.org/10.1088/1361-648X/aa680e}
  {\bibfield  {journal} {\bibinfo  {journal} {Journal of Physics: Condensed
  Matter}\ }\textbf {\bibinfo {volume} {29}},\ \bibinfo {pages} {273002}
  (\bibinfo {year} {2017})}\BibitemShut {NoStop}%
\bibitem [{\citenamefont {Cohen}\ \emph {et~al.}(2025)\citenamefont {Cohen},
  \citenamefont {Riebesell}, \citenamefont {Goodall}, \citenamefont {Kolluru},
  \citenamefont {Falletta}, \citenamefont {Krause}, \citenamefont {Colindres},
  \citenamefont {Ceder},\ and\ \citenamefont {Gangan}}]{Cohen_2025}%
  \BibitemOpen
  \bibfield  {author} {\bibinfo {author} {\bibfnamefont {O.}~\bibnamefont
  {Cohen}}, \bibinfo {author} {\bibfnamefont {J.}~\bibnamefont {Riebesell}},
  \bibinfo {author} {\bibfnamefont {R.}~\bibnamefont {Goodall}}, \bibinfo
  {author} {\bibfnamefont {A.}~\bibnamefont {Kolluru}}, \bibinfo {author}
  {\bibfnamefont {S.}~\bibnamefont {Falletta}}, \bibinfo {author}
  {\bibfnamefont {J.}~\bibnamefont {Krause}}, \bibinfo {author} {\bibfnamefont
  {J.}~\bibnamefont {Colindres}}, \bibinfo {author} {\bibfnamefont
  {G.}~\bibnamefont {Ceder}},\ and\ \bibinfo {author} {\bibfnamefont {A.~S.}\
  \bibnamefont {Gangan}},\ }\bibfield  {title} {\bibinfo {title} {Torchsim: an
  efficient atomistic simulation engine in pytorch},\ }\href
  {https://doi.org/10.1088/3050-287X/ae1799} {\bibfield  {journal} {\bibinfo
  {journal} {AI for Science}\ }\textbf {\bibinfo {volume} {1}},\ \bibinfo
  {pages} {025003} (\bibinfo {year} {2025})}\BibitemShut {NoStop}%
\bibitem [{\citenamefont {Birch}(1947)}]{PhysRev.71.809}%
  \BibitemOpen
  \bibfield  {author} {\bibinfo {author} {\bibfnamefont {F.}~\bibnamefont
  {Birch}},\ }\bibfield  {title} {\bibinfo {title} {Finite elastic strain of
  cubic crystals},\ }\href {https://doi.org/10.1103/PhysRev.71.809} {\bibfield
  {journal} {\bibinfo  {journal} {Phys. Rev.}\ }\textbf {\bibinfo {volume}
  {71}},\ \bibinfo {pages} {809} (\bibinfo {year} {1947})}\BibitemShut
  {NoStop}%
\bibitem [{\citenamefont {Murnaghan}(1944)}]{doi:10.1073/pnas.30.9.244}%
  \BibitemOpen
  \bibfield  {author} {\bibinfo {author} {\bibfnamefont {F.~D.}\ \bibnamefont
  {Murnaghan}},\ }\bibfield  {title} {\bibinfo {title} {The compressibility of
  media under extreme pressures},\ }\href
  {https://doi.org/10.1073/pnas.30.9.244} {\bibfield  {journal} {\bibinfo
  {journal} {Proceedings of the National Academy of Sciences}\ }\textbf
  {\bibinfo {volume} {30}},\ \bibinfo {pages} {244} (\bibinfo {year}
  {1944})}\BibitemShut {NoStop}%
\bibitem [{\citenamefont {Dewaele}\ \emph {et~al.}(2008)\citenamefont
  {Dewaele}, \citenamefont {Torrent}, \citenamefont {Loubeyre},\ and\
  \citenamefont {Mezouar}}]{PhysRevB.78.104102}%
  \BibitemOpen
  \bibfield  {author} {\bibinfo {author} {\bibfnamefont {A.}~\bibnamefont
  {Dewaele}}, \bibinfo {author} {\bibfnamefont {M.}~\bibnamefont {Torrent}},
  \bibinfo {author} {\bibfnamefont {P.}~\bibnamefont {Loubeyre}},\ and\
  \bibinfo {author} {\bibfnamefont {M.}~\bibnamefont {Mezouar}},\ }\bibfield
  {title} {\bibinfo {title} {Compression curves of transition metals in the
  mbar range: Experiments and projector augmented-wave calculations},\ }\href
  {https://doi.org/10.1103/PhysRevB.78.104102} {\bibfield  {journal} {\bibinfo
  {journal} {Phys. Rev. B}\ }\textbf {\bibinfo {volume} {78}},\ \bibinfo
  {pages} {104102} (\bibinfo {year} {2008})}\BibitemShut {NoStop}%
\bibitem [{\citenamefont {Morrison}\ \emph {et~al.}(2018)\citenamefont
  {Morrison}, \citenamefont {Jackson}, \citenamefont {Sturhahn}, \citenamefont
  {Zhang},\ and\ \citenamefont {Greenberg}}]{morrison2018equations}%
  \BibitemOpen
  \bibfield  {author} {\bibinfo {author} {\bibfnamefont {R.~A.}\ \bibnamefont
  {Morrison}}, \bibinfo {author} {\bibfnamefont {J.~M.}\ \bibnamefont
  {Jackson}}, \bibinfo {author} {\bibfnamefont {W.}~\bibnamefont {Sturhahn}},
  \bibinfo {author} {\bibfnamefont {D.}~\bibnamefont {Zhang}},\ and\ \bibinfo
  {author} {\bibfnamefont {E.}~\bibnamefont {Greenberg}},\ }\bibfield  {title}
  {\bibinfo {title} {Equations of state and anisotropy of {Fe-Ni-Si} alloys},\
  }\href {https://doi.org/https://doi.org/10.1029/2017JB015343} {\bibfield
  {journal} {\bibinfo  {journal} {Journal of Geophysical Research: Solid
  Earth}\ }\textbf {\bibinfo {volume} {123}},\ \bibinfo {pages} {4647}
  (\bibinfo {year} {2018})}\BibitemShut {NoStop}%
\bibitem [{\citenamefont {Zhou}\ \emph {et~al.}(2026)\citenamefont {Zhou},
  \citenamefont {Buchen}, \citenamefont {Dobrosavljevic}, \citenamefont
  {Strozewski}, \citenamefont {Pardo}, \citenamefont {Sturhahn}, \citenamefont
  {Ishii}, \citenamefont {Toellner}, \citenamefont {Wilding}, \citenamefont
  {Chariton}, \citenamefont {Kalkan}, \citenamefont {Kunz},\ and\ \citenamefont
  {Jackson}}]{ZHOU2026107515}%
  \BibitemOpen
  \bibfield  {author} {\bibinfo {author} {\bibfnamefont {C.}~\bibnamefont
  {Zhou}}, \bibinfo {author} {\bibfnamefont {J.}~\bibnamefont {Buchen}},
  \bibinfo {author} {\bibfnamefont {V.~V.}\ \bibnamefont {Dobrosavljevic}},
  \bibinfo {author} {\bibfnamefont {B.}~\bibnamefont {Strozewski}}, \bibinfo
  {author} {\bibfnamefont {O.~S.}\ \bibnamefont {Pardo}}, \bibinfo {author}
  {\bibfnamefont {W.}~\bibnamefont {Sturhahn}}, \bibinfo {author}
  {\bibfnamefont {T.}~\bibnamefont {Ishii}}, \bibinfo {author} {\bibfnamefont
  {T.~S.}\ \bibnamefont {Toellner}}, \bibinfo {author} {\bibfnamefont {J.~D.}\
  \bibnamefont {Wilding}}, \bibinfo {author} {\bibfnamefont {S.}~\bibnamefont
  {Chariton}}, \bibinfo {author} {\bibfnamefont {B.}~\bibnamefont {Kalkan}},
  \bibinfo {author} {\bibfnamefont {M.}~\bibnamefont {Kunz}},\ and\ \bibinfo
  {author} {\bibfnamefont {J.~M.}\ \bibnamefont {Jackson}},\ }\bibfield
  {title} {\bibinfo {title} {Equation of state of a low-sulfur iron–nickel
  alloy up to 143 {GPa}},\ }\href
  {https://doi.org/https://doi.org/10.1016/j.pepi.2026.107515} {\bibfield
  {journal} {\bibinfo  {journal} {Physics of the Earth and Planetary
  Interiors}\ }\textbf {\bibinfo {volume} {372}},\ \bibinfo {pages} {107515}
  (\bibinfo {year} {2026})}\BibitemShut {NoStop}%
\bibitem [{\citenamefont {Hirao}\ \emph {et~al.}(2022)\citenamefont {Hirao},
  \citenamefont {Akahama},\ and\ \citenamefont {Ohishi}}]{hirao2022equations}%
  \BibitemOpen
  \bibfield  {author} {\bibinfo {author} {\bibfnamefont {N.}~\bibnamefont
  {Hirao}}, \bibinfo {author} {\bibfnamefont {Y.}~\bibnamefont {Akahama}},\
  and\ \bibinfo {author} {\bibfnamefont {Y.}~\bibnamefont {Ohishi}},\
  }\bibfield  {title} {\bibinfo {title} {Equations of state of iron and nickel
  to the pressure at the center of the earth},\ }\href
  {https://doi.org/10.1063/5.0074340} {\bibfield  {journal} {\bibinfo
  {journal} {Matter and Radiation at Extremes}\ }\textbf {\bibinfo {volume}
  {7}},\ \bibinfo {pages} {038403} (\bibinfo {year} {2022})}\BibitemShut
  {NoStop}%
\bibitem [{\citenamefont {Zwell}\ \emph {et~al.}(1970)\citenamefont {Zwell},
  \citenamefont {Carnahan},\ and\ \citenamefont {Speich}}]{Zwell1970}%
  \BibitemOpen
  \bibfield  {author} {\bibinfo {author} {\bibfnamefont {L.}~\bibnamefont
  {Zwell}}, \bibinfo {author} {\bibfnamefont {D.~E.}\ \bibnamefont
  {Carnahan}},\ and\ \bibinfo {author} {\bibfnamefont {G.~R.}\ \bibnamefont
  {Speich}},\ }\bibfield  {title} {\bibinfo {title} {Lattice parameter of
  ferritic and martensitic {Fe}-{Ni} alloys},\ }\href
  {https://doi.org/10.1007/BF02811785} {\bibfield  {journal} {\bibinfo
  {journal} {Metallurgical Transactions}\ }\textbf {\bibinfo {volume} {1}},\
  \bibinfo {pages} {1007} (\bibinfo {year} {1970})}\BibitemShut {NoStop}%
\bibitem [{\citenamefont {Hayase}\ \emph {et~al.}(1973)\citenamefont {Hayase},
  \citenamefont {Shiga},\ and\ \citenamefont
  {Nakamura}}]{hayase1973spontaneous}%
  \BibitemOpen
  \bibfield  {author} {\bibinfo {author} {\bibfnamefont {M.}~\bibnamefont
  {Hayase}}, \bibinfo {author} {\bibfnamefont {M.}~\bibnamefont {Shiga}},\ and\
  \bibinfo {author} {\bibfnamefont {Y.}~\bibnamefont {Nakamura}},\ }\bibfield
  {title} {\bibinfo {title} {Spontaneous volume magnetostriction and lattice
  constant of face-centered cubic {Fe}-{Ni} and {Ni}-{Cu} alloys},\ }\href
  {https://doi.org/10.1143/JPSJ.34.925} {\bibfield  {journal} {\bibinfo
  {journal} {Journal of the Physical Society of Japan}\ }\textbf {\bibinfo
  {volume} {34}},\ \bibinfo {pages} {925} (\bibinfo {year} {1973})}\BibitemShut
  {NoStop}%
\bibitem [{\citenamefont {Acet}\ \emph {et~al.}(1994)\citenamefont {Acet},
  \citenamefont {Z\"ahres}, \citenamefont {Wassermann},\ and\ \citenamefont
  {Pepperhoff}}]{PhysRevB.49.6012}%
  \BibitemOpen
  \bibfield  {author} {\bibinfo {author} {\bibfnamefont {M.}~\bibnamefont
  {Acet}}, \bibinfo {author} {\bibfnamefont {H.}~\bibnamefont {Z\"ahres}},
  \bibinfo {author} {\bibfnamefont {E.~F.}\ \bibnamefont {Wassermann}},\ and\
  \bibinfo {author} {\bibfnamefont {W.}~\bibnamefont {Pepperhoff}},\ }\bibfield
   {title} {\bibinfo {title} {High-temperature moment-volume instability and
  anti-invar of \ensuremath{\gamma}-{Fe}},\ }\href
  {https://doi.org/10.1103/PhysRevB.49.6012} {\bibfield  {journal} {\bibinfo
  {journal} {Phys. Rev. B}\ }\textbf {\bibinfo {volume} {49}},\ \bibinfo
  {pages} {6012} (\bibinfo {year} {1994})}\BibitemShut {NoStop}%
\bibitem [{\citenamefont {K{\k{a}}dzio{\l}ka-Gawe{\l}}\ \emph
  {et~al.}(2010)\citenamefont {K{\k{a}}dzio{\l}ka-Gawe{\l}}, \citenamefont
  {Zarek}, \citenamefont {Popiel},\ and\ \citenamefont
  {Chrobak}}]{kkadziolka2010crystal}%
  \BibitemOpen
  \bibfield  {author} {\bibinfo {author} {\bibfnamefont {M.}~\bibnamefont
  {K{\k{a}}dzio{\l}ka-Gawe{\l}}}, \bibinfo {author} {\bibfnamefont
  {W.}~\bibnamefont {Zarek}}, \bibinfo {author} {\bibfnamefont
  {E.}~\bibnamefont {Popiel}},\ and\ \bibinfo {author} {\bibfnamefont
  {A.}~\bibnamefont {Chrobak}},\ }\bibfield  {title} {\bibinfo {title} {The
  crystal structure and magnetic properties of selected fcc {FeNi} and
  {Fe$_{40}$Ni$_{40}$B$_{20}$} alloys},\ }\href@noop {} {\bibfield  {journal}
  {\bibinfo  {journal} {Acta Physica Polonica A}\ }\textbf {\bibinfo {volume}
  {117}} (\bibinfo {year} {2010})}\BibitemShut {NoStop}%
\bibitem [{\citenamefont {Kanhe}\ \emph {et~al.}(2016)\citenamefont {Kanhe},
  \citenamefont {Kumar}, \citenamefont {Yusuf}, \citenamefont {Nawale},
  \citenamefont {Gaikwad}, \citenamefont {Raut}, \citenamefont {Bhoraskar},
  \citenamefont {Wu}, \citenamefont {Das},\ and\ \citenamefont
  {Mathe}}]{KANHE201630}%
  \BibitemOpen
  \bibfield  {author} {\bibinfo {author} {\bibfnamefont {N.~S.}\ \bibnamefont
  {Kanhe}}, \bibinfo {author} {\bibfnamefont {A.}~\bibnamefont {Kumar}},
  \bibinfo {author} {\bibfnamefont {S.}~\bibnamefont {Yusuf}}, \bibinfo
  {author} {\bibfnamefont {A.}~\bibnamefont {Nawale}}, \bibinfo {author}
  {\bibfnamefont {S.}~\bibnamefont {Gaikwad}}, \bibinfo {author} {\bibfnamefont
  {S.~A.}\ \bibnamefont {Raut}}, \bibinfo {author} {\bibfnamefont
  {S.}~\bibnamefont {Bhoraskar}}, \bibinfo {author} {\bibfnamefont {S.~Y.}\
  \bibnamefont {Wu}}, \bibinfo {author} {\bibfnamefont {A.}~\bibnamefont
  {Das}},\ and\ \bibinfo {author} {\bibfnamefont {V.}~\bibnamefont {Mathe}},\
  }\bibfield  {title} {\bibinfo {title} {Investigation of structural and
  magnetic properties of thermal plasma-synthesized {Fe$_{1-x}$Ni$_{x}$} alloy
  nanoparticles},\ }\href
  {https://doi.org/https://doi.org/10.1016/j.jallcom.2015.11.190} {\bibfield
  {journal} {\bibinfo  {journal} {Journal of Alloys and Compounds}\ }\textbf
  {\bibinfo {volume} {663}},\ \bibinfo {pages} {30} (\bibinfo {year}
  {2016})}\BibitemShut {NoStop}%
\bibitem [{\citenamefont {Wei}\ \emph {et~al.}(2020)\citenamefont {Wei},
  \citenamefont {Gilder}, \citenamefont {Ertel-Ingrisch}, \citenamefont
  {Guillou},\ and\ \citenamefont {Wilhelm}}]{wei2020}%
  \BibitemOpen
  \bibfield  {author} {\bibinfo {author} {\bibfnamefont {Q.}~\bibnamefont
  {Wei}}, \bibinfo {author} {\bibfnamefont {S.~A.}\ \bibnamefont {Gilder}},
  \bibinfo {author} {\bibfnamefont {W.}~\bibnamefont {Ertel-Ingrisch}},
  \bibinfo {author} {\bibfnamefont {F.}~\bibnamefont {Guillou}},\ and\ \bibinfo
  {author} {\bibfnamefont {F.}~\bibnamefont {Wilhelm}},\ }\bibfield  {title}
  {\bibinfo {title} {Magnetism of body-centered cubic {Fe--Ni} alloys under
  pressure: Strain-enhanced ferromagnetism at the phase transitions},\ }\href
  {https://doi.org/https://doi.org/10.1029/2020JB020922} {\bibfield  {journal}
  {\bibinfo  {journal} {Journal of Geophysical Research: Solid Earth}\ }\textbf
  {\bibinfo {volume} {125}},\ \bibinfo {pages} {e2020JB020922} (\bibinfo {year}
  {2020})}\BibitemShut {NoStop}%
\bibitem [{\citenamefont {Ekman}\ \emph {et~al.}(1998)\citenamefont {Ekman},
  \citenamefont {Sadigh}, \citenamefont {Einarsdotter},\ and\ \citenamefont
  {Blaha}}]{PhysRevB.58.5296}%
  \BibitemOpen
  \bibfield  {author} {\bibinfo {author} {\bibfnamefont {M.}~\bibnamefont
  {Ekman}}, \bibinfo {author} {\bibfnamefont {B.}~\bibnamefont {Sadigh}},
  \bibinfo {author} {\bibfnamefont {K.}~\bibnamefont {Einarsdotter}},\ and\
  \bibinfo {author} {\bibfnamefont {P.}~\bibnamefont {Blaha}},\ }\bibfield
  {title} {\bibinfo {title} {Ab initio study of the martensitic bcc-hcp
  transformation in iron},\ }\href {https://doi.org/10.1103/PhysRevB.58.5296}
  {\bibfield  {journal} {\bibinfo  {journal} {Phys. Rev. B}\ }\textbf {\bibinfo
  {volume} {58}},\ \bibinfo {pages} {5296} (\bibinfo {year}
  {1998})}\BibitemShut {NoStop}%
\bibitem [{\citenamefont {Adams}\ \emph {et~al.}(2006)\citenamefont {Adams},
  \citenamefont {Agosta}, \citenamefont {Leisure},\ and\ \citenamefont
  {Ledbetter}}]{Adams2006}%
  \BibitemOpen
  \bibfield  {author} {\bibinfo {author} {\bibfnamefont {J.~J.}\ \bibnamefont
  {Adams}}, \bibinfo {author} {\bibfnamefont {D.~S.}\ \bibnamefont {Agosta}},
  \bibinfo {author} {\bibfnamefont {R.~G.}\ \bibnamefont {Leisure}},\ and\
  \bibinfo {author} {\bibfnamefont {H.}~\bibnamefont {Ledbetter}},\ }\bibfield
  {title} {\bibinfo {title} {Elastic constants of monocrystal iron from
  3to500k},\ }\href {https://doi.org/10.1063/1.2365714} {\bibfield  {journal}
  {\bibinfo  {journal} {Journal of Applied Physics}\ }\textbf {\bibinfo
  {volume} {100}},\ \bibinfo {pages} {113530} (\bibinfo {year}
  {2006})}\BibitemShut {NoStop}%
\bibitem [{\citenamefont {Rayne}\ and\ \citenamefont
  {Chandrasekhar}(1961)}]{Rayne1961}%
  \BibitemOpen
  \bibfield  {author} {\bibinfo {author} {\bibfnamefont {J.~A.}\ \bibnamefont
  {Rayne}}\ and\ \bibinfo {author} {\bibfnamefont {B.~S.}\ \bibnamefont
  {Chandrasekhar}},\ }\bibfield  {title} {\bibinfo {title} {Elastic constants
  of iron from 4.2 to 300\ifmmode^\circ\else\textdegree\fi{}{K}},\ }\href
  {https://doi.org/10.1103/PhysRev.122.1714} {\bibfield  {journal} {\bibinfo
  {journal} {Phys. Rev.}\ }\textbf {\bibinfo {volume} {122}},\ \bibinfo {pages}
  {1714} (\bibinfo {year} {1961})}\BibitemShut {NoStop}%
\bibitem [{\citenamefont {Lutts}\ and\ \citenamefont
  {Gielen}(1970)}]{Lutts_FeNi3}%
  \BibitemOpen
  \bibfield  {author} {\bibinfo {author} {\bibfnamefont {A.}~\bibnamefont
  {Lutts}}\ and\ \bibinfo {author} {\bibfnamefont {P.~M.}\ \bibnamefont
  {Gielen}},\ }\bibfield  {title} {\bibinfo {title} {The order-disorder
  transformation in {FeNi$_3$}},\ }\href
  {https://doi.org/https://doi.org/10.1002/pssb.19700410169} {\bibfield
  {journal} {\bibinfo  {journal} {physica status solidi (b)}\ }\textbf
  {\bibinfo {volume} {41}},\ \bibinfo {pages} {K81} (\bibinfo {year}
  {1970})}\BibitemShut {NoStop}%
\bibitem [{\citenamefont {Simmons}(1971)}]{simmons1971single}%
  \BibitemOpen
  \bibfield  {author} {\bibinfo {author} {\bibfnamefont {G.}~\bibnamefont
  {Simmons}},\ }\bibfield  {title} {\bibinfo {title} {Single crystal elastic
  constants and caluculated aggregate properties},\ }\href@noop {} {\bibfield
  {journal} {\bibinfo  {journal} {A handbook}\ }\textbf {\bibinfo {volume} {4}}
  (\bibinfo {year} {1971})}\BibitemShut {NoStop}%
\bibitem [{\citenamefont {Shirakawa}\ \emph {et~al.}(1969)\citenamefont
  {Shirakawa}, \citenamefont {Tanji}, \citenamefont {Moriya},\ and\
  \citenamefont {Oguma}}]{shirakawa1969elastic}%
  \BibitemOpen
  \bibfield  {author} {\bibinfo {author} {\bibfnamefont {Y.}~\bibnamefont
  {Shirakawa}}, \bibinfo {author} {\bibfnamefont {Y.}~\bibnamefont {Tanji}},
  \bibinfo {author} {\bibfnamefont {H.}~\bibnamefont {Moriya}},\ and\ \bibinfo
  {author} {\bibfnamefont {I.}~\bibnamefont {Oguma}},\ }\bibfield  {title}
  {\bibinfo {title} {Elastic constants of {Ni} and {Ni}--{Fe} fcc alloys},\
  }\href {https://doi.org/10.2320/jinstmet1952.33.10_1196} {\bibfield
  {journal} {\bibinfo  {journal} {J Japan Inst Metals}\ }\textbf {\bibinfo
  {volume} {33}},\ \bibinfo {pages} {1196} (\bibinfo {year}
  {1969})}\BibitemShut {NoStop}%
\bibitem [{\citenamefont {Singh}\ and\ \citenamefont
  {Guruswamy}(2023)}]{10.1063/5.0174535}%
  \BibitemOpen
  \bibfield  {author} {\bibinfo {author} {\bibfnamefont {R.~S.}\ \bibnamefont
  {Singh}}\ and\ \bibinfo {author} {\bibfnamefont {S.}~\bibnamefont
  {Guruswamy}},\ }\bibfield  {title} {\bibinfo {title} {Elastic constants of
  equiatomic {Fe}–{Ni} invar alloy single crystal},\ }\href
  {https://doi.org/10.1063/5.0174535} {\bibfield  {journal} {\bibinfo
  {journal} {AIP Advances}\ }\textbf {\bibinfo {volume} {13}},\ \bibinfo
  {pages} {115112} (\bibinfo {year} {2023})}\BibitemShut {NoStop}%
\bibitem [{\citenamefont {Ledbetter}\ and\ \citenamefont
  {Reed}(1973)}]{10.1063/1.3253127}%
  \BibitemOpen
  \bibfield  {author} {\bibinfo {author} {\bibfnamefont {H.~M.}\ \bibnamefont
  {Ledbetter}}\ and\ \bibinfo {author} {\bibfnamefont {R.~P.}\ \bibnamefont
  {Reed}},\ }\bibfield  {title} {\bibinfo {title} {Elastic properties of metals
  and alloys, i. iron, nickel, and iron‐nickel alloys},\ }\href
  {https://doi.org/10.1063/1.3253127} {\bibfield  {journal} {\bibinfo
  {journal} {Journal of Physical and Chemical Reference Data}\ }\textbf
  {\bibinfo {volume} {2}},\ \bibinfo {pages} {531} (\bibinfo {year}
  {1973})}\BibitemShut {NoStop}%
\bibitem [{\citenamefont {Hausch}\ and\ \citenamefont
  {Warlimont}(1973)}]{hausch1973401}%
  \BibitemOpen
  \bibfield  {author} {\bibinfo {author} {\bibfnamefont {G.}~\bibnamefont
  {Hausch}}\ and\ \bibinfo {author} {\bibfnamefont {H.}~\bibnamefont
  {Warlimont}},\ }\bibfield  {title} {\bibinfo {title} {Single crystalline
  elastic constants of ferromagnetic face centered cubic {Fe}-{Ni} invar
  alloys},\ }\href
  {https://doi.org/https://doi.org/10.1016/0001-6160(73)90197-1} {\bibfield
  {journal} {\bibinfo  {journal} {Acta Metallurgica}\ }\textbf {\bibinfo
  {volume} {21}},\ \bibinfo {pages} {401} (\bibinfo {year} {1973})}\BibitemShut
  {NoStop}%
\bibitem [{\citenamefont {Tanji}\ \emph {et~al.}(1983)\citenamefont {Tanji},
  \citenamefont {Nakagawa},\ and\ \citenamefont {Steinemann}}]{Tanji1983}%
  \BibitemOpen
  \bibfield  {author} {\bibinfo {author} {\bibfnamefont {Y.}~\bibnamefont
  {Tanji}}, \bibinfo {author} {\bibfnamefont {Y.}~\bibnamefont {Nakagawa}},\
  and\ \bibinfo {author} {\bibfnamefont {S.}~\bibnamefont {Steinemann}},\
  }\bibfield  {title} {\bibinfo {title} {Anomalous elastic properties of fe-ni
  (fcc) alloys and their invar properties},\ }\href
  {https://doi.org/https://doi.org/10.1016/0378-4363(83)90174-2} {\bibfield
  {journal} {\bibinfo  {journal} {Physica B+C}\ }\textbf {\bibinfo {volume}
  {119}},\ \bibinfo {pages} {109} (\bibinfo {year} {1983})}\BibitemShut
  {NoStop}%
\bibitem [{\citenamefont {Voronov}\ and\ \citenamefont
  {Vereshchagin}(1961)}]{Voronov1961}%
  \BibitemOpen
  \bibfield  {author} {\bibinfo {author} {\bibfnamefont {F.}~\bibnamefont
  {Voronov}}\ and\ \bibinfo {author} {\bibfnamefont {L.}~\bibnamefont
  {Vereshchagin}},\ }\bibfield  {title} {\bibinfo {title} {Influence of
  hydrostatic pressure on the elastic properties of metals. i. experimental
  data},\ }\href@noop {} {\bibfield  {journal} {\bibinfo  {journal} {Phys.
  Metals and Metallography}\ }\textbf {\bibinfo {volume} {11}},\ \bibinfo
  {pages} {111} (\bibinfo {year} {1961})}\BibitemShut {NoStop}%
\bibitem [{\citenamefont {Smith}\ and\ \citenamefont {Wood}(1941)}]{Smith1941}%
  \BibitemOpen
  \bibfield  {author} {\bibinfo {author} {\bibfnamefont {S.~L.}\ \bibnamefont
  {Smith}}\ and\ \bibinfo {author} {\bibfnamefont {W.~A.}\ \bibnamefont
  {Wood}},\ }\bibfield  {title} {\bibinfo {title} {A stress-strain curve for
  the atomic lattice of iron},\ }\href {https://doi.org/10.1098/rspa.1941.0046}
  {\bibfield  {journal} {\bibinfo  {journal} {Proceedings of the Royal Society
  of London. A. Mathematical and Physical Sciences}\ }\textbf {\bibinfo
  {volume} {178}},\ \bibinfo {pages} {93} (\bibinfo {year} {1941})}\BibitemShut
  {NoStop}%
\bibitem [{\citenamefont {Gr{\"u}neisen}(1910)}]{Gruneisen1910}%
  \BibitemOpen
  \bibfield  {author} {\bibinfo {author} {\bibfnamefont {E.}~\bibnamefont
  {Gr{\"u}neisen}},\ }\bibfield  {title} {\bibinfo {title} {Einflu\ss{} der
  temperatur auf die kompressibilit{\"a}t der metalle},\ }\href
  {https://doi.org/https://doi.org/10.1002/andp.19103381611} {\bibfield
  {journal} {\bibinfo  {journal} {Annalen der Physik}\ }\textbf {\bibinfo
  {volume} {338}},\ \bibinfo {pages} {1239} (\bibinfo {year}
  {1910})}\BibitemShut {NoStop}%
\bibitem [{\citenamefont {Born}\ and\ \citenamefont
  {Huang}(1954)}]{born1954dynamical}%
  \BibitemOpen
  \bibfield  {author} {\bibinfo {author} {\bibfnamefont {M.}~\bibnamefont
  {Born}}\ and\ \bibinfo {author} {\bibfnamefont {K.}~\bibnamefont {Huang}},\
  }\href {https://archive.org/details/dynamicaltheoryo0000born} {\emph
  {\bibinfo {title} {Dynamical Theory of Crystal Lattices}}}\ (\bibinfo
  {publisher} {Oxford University Press},\ \bibinfo {address} {Oxford},\
  \bibinfo {year} {1954})\BibitemShut {NoStop}%
\bibitem [{\citenamefont {Suh}\ \emph {et~al.}(1988)\citenamefont {Suh},
  \citenamefont {Ohta},\ and\ \citenamefont {Waseda}}]{Suh1988}%
  \BibitemOpen
  \bibfield  {author} {\bibinfo {author} {\bibfnamefont {I.-K.}\ \bibnamefont
  {Suh}}, \bibinfo {author} {\bibfnamefont {H.}~\bibnamefont {Ohta}},\ and\
  \bibinfo {author} {\bibfnamefont {Y.}~\bibnamefont {Waseda}},\ }\bibfield
  {title} {\bibinfo {title} {High-temperature thermal expansion of six metallic
  elements measured by dilatation method and x-ray diffraction},\ }\href
  {https://doi.org/10.1007/BF01174717} {\bibfield  {journal} {\bibinfo
  {journal} {Journal of Materials Science}\ }\textbf {\bibinfo {volume} {23}},\
  \bibinfo {pages} {757} (\bibinfo {year} {1988})}\BibitemShut {NoStop}%
\bibitem [{\citenamefont {Owen}\ \emph {et~al.}(1937)\citenamefont {Owen},
  \citenamefont {Yates},\ and\ \citenamefont {Sully}}]{Owen_1937}%
  \BibitemOpen
  \bibfield  {author} {\bibinfo {author} {\bibfnamefont {E.~A.}\ \bibnamefont
  {Owen}}, \bibinfo {author} {\bibfnamefont {E.~L.}\ \bibnamefont {Yates}},\
  and\ \bibinfo {author} {\bibfnamefont {A.~H.}\ \bibnamefont {Sully}},\
  }\bibfield  {title} {\bibinfo {title} {An x-ray investigation of pure
  iron-nickel alloys. part 4: the variation of lattice-parameter with
  composition},\ }\href {https://doi.org/10.1088/0959-5309/49/3/313} {\bibfield
   {journal} {\bibinfo  {journal} {Proceedings of the Physical Society}\
  }\textbf {\bibinfo {volume} {49}},\ \bibinfo {pages} {315} (\bibinfo {year}
  {1937})}\BibitemShut {NoStop}%
\bibitem [{\citenamefont {Basinski}\ \emph {et~al.}(1955)\citenamefont
  {Basinski}, \citenamefont {Hume-Rothery},\ and\ \citenamefont
  {Sutton}}]{10.1098/rspa.1955.0102}%
  \BibitemOpen
  \bibfield  {author} {\bibinfo {author} {\bibfnamefont {Z.~S.}\ \bibnamefont
  {Basinski}}, \bibinfo {author} {\bibfnamefont {W.}~\bibnamefont
  {Hume-Rothery}},\ and\ \bibinfo {author} {\bibfnamefont {A.~L.}\ \bibnamefont
  {Sutton}},\ }\bibfield  {title} {\bibinfo {title} {The lattice expansion of
  iron},\ }\href {https://doi.org/10.1098/rspa.1955.0102} {\bibfield  {journal}
  {\bibinfo  {journal} {Proceedings of the Royal Society of London. A.
  Mathematical and Physical Sciences}\ }\textbf {\bibinfo {volume} {229}},\
  \bibinfo {pages} {459} (\bibinfo {year} {1955})}\BibitemShut {NoStop}%
\bibitem [{\citenamefont {Ridley}\ and\ \citenamefont
  {Stuart}(1968)}]{Ridley_1968}%
  \BibitemOpen
  \bibfield  {author} {\bibinfo {author} {\bibfnamefont {N.}~\bibnamefont
  {Ridley}}\ and\ \bibinfo {author} {\bibfnamefont {H.}~\bibnamefont
  {Stuart}},\ }\bibfield  {title} {\bibinfo {title} {Lattice parameter
  anomalies at the curie point of pure iron},\ }\href
  {https://doi.org/10.1088/0022-3727/1/10/308} {\bibfield  {journal} {\bibinfo
  {journal} {Journal of Physics D: Applied Physics}\ }\textbf {\bibinfo
  {volume} {1}},\ \bibinfo {pages} {1291} (\bibinfo {year} {1968})}\BibitemShut
  {NoStop}%
\bibitem [{\citenamefont {Sahoo}\ \emph {et~al.}(2025)\citenamefont {Sahoo},
  \citenamefont {Biswal}, \citenamefont {Parida}, \citenamefont {Medicherla},
  \citenamefont {Behera}, \citenamefont {Singh}, \citenamefont {Sagdeo},
  \citenamefont {Datta}, \citenamefont {Singh},\ and\ \citenamefont
  {Maiti}}]{sahoo2025anomalieselectronicmagneticthermal}%
  \BibitemOpen
  \bibfield  {author} {\bibinfo {author} {\bibfnamefont {A.}~\bibnamefont
  {Sahoo}}, \bibinfo {author} {\bibfnamefont {A.~A.}\ \bibnamefont {Biswal}},
  \bibinfo {author} {\bibfnamefont {S.~K.}\ \bibnamefont {Parida}}, \bibinfo
  {author} {\bibfnamefont {V.~R.~R.}\ \bibnamefont {Medicherla}}, \bibinfo
  {author} {\bibfnamefont {S.~S.}\ \bibnamefont {Behera}}, \bibinfo {author}
  {\bibfnamefont {M.~N.}\ \bibnamefont {Singh}}, \bibinfo {author}
  {\bibfnamefont {A.}~\bibnamefont {Sagdeo}}, \bibinfo {author} {\bibfnamefont
  {S.}~\bibnamefont {Datta}}, \bibinfo {author} {\bibfnamefont
  {A.}~\bibnamefont {Singh}},\ and\ \bibinfo {author} {\bibfnamefont
  {K.}~\bibnamefont {Maiti}},\ }\href {https://arxiv.org/abs/2502.10757}
  {\bibinfo {title} {Anomalies in the electronic, magnetic and thermal behavior
  near the invar compositions of {Fe-Ni} alloys}} (\bibinfo {year} {2025}),\
  \Eprint {https://arxiv.org/abs/2502.10757} {arXiv:2502.10757
  [cond-mat.str-el]} \BibitemShut {NoStop}%
\bibitem [{\citenamefont {Hoover}(1985)}]{PhysRevA.31.1695}%
  \BibitemOpen
  \bibfield  {author} {\bibinfo {author} {\bibfnamefont {W.~G.}\ \bibnamefont
  {Hoover}},\ }\bibfield  {title} {\bibinfo {title} {Canonical dynamics:
  Equilibrium phase-space distributions},\ }\href
  {https://doi.org/10.1103/PhysRevA.31.1695} {\bibfield  {journal} {\bibinfo
  {journal} {Phys. Rev. A}\ }\textbf {\bibinfo {volume} {31}},\ \bibinfo
  {pages} {1695} (\bibinfo {year} {1985})}\BibitemShut {NoStop}%
\bibitem [{\citenamefont {Parrinello}\ and\ \citenamefont
  {Rahman}(1981)}]{10.1063/1.328693}%
  \BibitemOpen
  \bibfield  {author} {\bibinfo {author} {\bibfnamefont {M.}~\bibnamefont
  {Parrinello}}\ and\ \bibinfo {author} {\bibfnamefont {A.}~\bibnamefont
  {Rahman}},\ }\bibfield  {title} {\bibinfo {title} {Polymorphic transitions in
  single crystals: A new molecular dynamics method},\ }\href
  {https://doi.org/10.1063/1.328693} {\bibfield  {journal} {\bibinfo  {journal}
  {Journal of Applied Physics}\ }\textbf {\bibinfo {volume} {52}},\ \bibinfo
  {pages} {7182} (\bibinfo {year} {1981})}\BibitemShut {NoStop}%
\bibitem [{\citenamefont {Kollie}(1977)}]{PhysRevB.16.4872}%
  \BibitemOpen
  \bibfield  {author} {\bibinfo {author} {\bibfnamefont {T.~G.}\ \bibnamefont
  {Kollie}},\ }\bibfield  {title} {\bibinfo {title} {Measurement of the
  thermal-expansion coefficient of nickel from 300 to 1000 {K} and
  determination of the power-law constants near the curie temperature},\ }\href
  {https://doi.org/10.1103/PhysRevB.16.4872} {\bibfield  {journal} {\bibinfo
  {journal} {Phys. Rev. B}\ }\textbf {\bibinfo {volume} {16}},\ \bibinfo
  {pages} {4872} (\bibinfo {year} {1977})}\BibitemShut {NoStop}%
\bibitem [{\citenamefont {Kozlovskii}\ and\ \citenamefont
  {Stankus}(2019)}]{Kozlovskii_2019}%
  \BibitemOpen
  \bibfield  {author} {\bibinfo {author} {\bibfnamefont {Y.~M.}\ \bibnamefont
  {Kozlovskii}}\ and\ \bibinfo {author} {\bibfnamefont {S.~V.}\ \bibnamefont
  {Stankus}},\ }\bibfield  {title} {\bibinfo {title} {The linear thermal
  expansion coefficient of iron in the temperature range of 130–1180 {K}},\
  }\href {https://doi.org/10.1088/1742-6596/1382/1/012181} {\bibfield
  {journal} {\bibinfo  {journal} {Journal of Physics: Conference Series}\
  }\textbf {\bibinfo {volume} {1382}},\ \bibinfo {pages} {012181} (\bibinfo
  {year} {2019})}\BibitemShut {NoStop}%
\bibitem [{\citenamefont {Nix}\ and\ \citenamefont
  {MacNair}(1941)}]{PhysRev.60.597}%
  \BibitemOpen
  \bibfield  {author} {\bibinfo {author} {\bibfnamefont {F.~C.}\ \bibnamefont
  {Nix}}\ and\ \bibinfo {author} {\bibfnamefont {D.}~\bibnamefont {MacNair}},\
  }\bibfield  {title} {\bibinfo {title} {The thermal expansion of pure metals:
  Copper, gold, aluminum, nickel, and iron},\ }\href
  {https://doi.org/10.1103/PhysRev.60.597} {\bibfield  {journal} {\bibinfo
  {journal} {Phys. Rev.}\ }\textbf {\bibinfo {volume} {60}},\ \bibinfo {pages}
  {597} (\bibinfo {year} {1941})}\BibitemShut {NoStop}%
\bibitem [{\citenamefont {Fisher}\ \emph {et~al.}(2026)\citenamefont {Fisher},
  \citenamefont {Woodgate}, \citenamefont {Zhang}, \citenamefont
  {Hadjipanayis}, \citenamefont {Lewis},\ and\ \citenamefont
  {Staunton}}]{4q1c-97bp}%
  \BibitemOpen
  \bibfield  {author} {\bibinfo {author} {\bibfnamefont {A.~M.}\ \bibnamefont
  {Fisher}}, \bibinfo {author} {\bibfnamefont {C.~D.}\ \bibnamefont
  {Woodgate}}, \bibinfo {author} {\bibfnamefont {X.}~\bibnamefont {Zhang}},
  \bibinfo {author} {\bibfnamefont {G.~C.}\ \bibnamefont {Hadjipanayis}},
  \bibinfo {author} {\bibfnamefont {L.~H.}\ \bibnamefont {Lewis}},\ and\
  \bibinfo {author} {\bibfnamefont {J.~B.}\ \bibnamefont {Staunton}},\
  }\bibfield  {title} {\bibinfo {title} {Lattice vacancy migration barriers in
  {Fe-Ni} alloys, and an indication as to why {Ni} atoms diffuse slowly: A
  first-principles study},\ }\href {https://doi.org/10.1103/4q1c-97bp}
  {\bibfield  {journal} {\bibinfo  {journal} {Phys. Rev. Mater.}\ }\textbf
  {\bibinfo {volume} {10}},\ \bibinfo {pages} {034410} (\bibinfo {year}
  {2026})}\BibitemShut {NoStop}%
\bibitem [{\citenamefont {Ito}\ \emph {et~al.}(2024)\citenamefont {Ito},
  \citenamefont {Yokoi}, \citenamefont {Hyodo},\ and\ \citenamefont
  {Mori}}]{Ito2024}%
  \BibitemOpen
  \bibfield  {author} {\bibinfo {author} {\bibfnamefont {K.}~\bibnamefont
  {Ito}}, \bibinfo {author} {\bibfnamefont {T.}~\bibnamefont {Yokoi}}, \bibinfo
  {author} {\bibfnamefont {K.}~\bibnamefont {Hyodo}},\ and\ \bibinfo {author}
  {\bibfnamefont {H.}~\bibnamefont {Mori}},\ }\bibfield  {title} {\bibinfo
  {title} {Machine learning interatomic potential with {DFT} accuracy for
  general grain boundaries in $\alpha$-{Fe}},\ }\href
  {https://doi.org/10.1038/s41524-024-01451-y} {\bibfield  {journal} {\bibinfo
  {journal} {npj Computational Materials}\ }\textbf {\bibinfo {volume} {10}},\
  \bibinfo {pages} {255} (\bibinfo {year} {2024})}\BibitemShut {NoStop}%
\end{thebibliography}%

\end{document}